\documentclass[journal=jctcce, manuscript=article]{achemso}

\usepackage{chemformula} 
\usepackage[T1]{fontenc} 
 \usepackage[colorlinks=true,linkcolor=blue,urlcolor=blue,citecolor=blue]{hyperref}
\usepackage{array}
\usepackage{lipsum}
\usepackage{bm}
\usepackage{subfigure}
\usepackage{amssymb}
\usepackage{multirow}
\usepackage{tabularx}
\usepackage{amsmath}
\usepackage{braket}
	
\usepackage{ulem}
\renewcommand{\vec}[1]{\mathbf{#1}}

\newcommand{\angstrom}{\mbox{\normalfont\AA}}
\SectionNumbersOn

\author{Zhandos Moldabekov}
\email{z.moldabekov@hzdr.de}
\affiliation{Center for Advanced Systems Understanding (CASUS), D-02826 G\"orlitz, Germany}
\alsoaffiliation{Helmholtz-Zentrum Dresden-Rossendorf (HZDR), D-01328 Dresden, Germany}

\author{Jan Vorberger}
\affiliation{Helmholtz-Zentrum Dresden-Rossendorf (HZDR), D-01328 Dresden, Germany}

\author{Tobias Dornheim}
\affiliation{Center for Advanced Systems Understanding (CASUS), D-02826 G\"orlitz, Germany}
\alsoaffiliation{Helmholtz-Zentrum Dresden-Rossendorf (HZDR), D-01328 Dresden, Germany}

\title{Density-Functional-Theory Perspective on the Non-Linear Response of Correlated Electrons Across Temperature Regimes}

\abbreviations{IR,NMR,UV}

\begin{document}







\begin{abstract}
We explore a new formalism to study the nonlinear electronic density response based on 
Kohn-Sham density functional theory (KS-DFT) at partially and strongly {quantum} degenerate regimes.
It is demonstrated that the KS-DFT calculations are able to accurately reproduce the available path integral Monte Carlo simulation results at temperatures relevant for warm dense matter research.
The existing analytical results for the quadratic and cubic response functions are {rigorously} tested. It is demonstrated that the analytical results for the quadratic response function closely agree with the KS-DFT data. Furthermore, the performed analysis reveals that currently available analytical formulas for the cubic response function are not able to describe simulation results, neither qualitatively nor quantitatively,  at small wave-numbers $q<2q_F$, with $q_F$ being the Fermi wave-number.        
The results show that KS-DFT can be used to describe warm dense matter that is strongly perturbed by an external field {with remarkable accuracy}. Furthermore, it is demonstrated that KS-DFT constitutes a valuable tool to guide the development of the non-linear response theory of correlated quantum electrons from ambient to extreme conditions. This opens up new avenues to study nonlinear effects in a gamut of different contexts at conditions that cannot be accessed with previously used path integral Monte Carlo methods [T. Dornheim \emph{et al.}, \textit{Phys.~Rev.~Lett.}~\textbf{125}, 085001 (2020)].
\end{abstract}

\section{Introduction}


Quantum linear response theory (LRT) has been actively developed since the formulation of the foundations of quantum mechanics and has become one of the most fundamental theories for the computation of various properties~\cite{nolting2009fundamentals}.
 {At the same time,} the ongoing development of technological and, along with it, experimental capabilities has resulted in the need for a theory that captures phenomena beyond the LRT.
Specific examples include plasmonics \cite{Panoiu_2018, Lee2014}, optics \cite{doi:10.1063/1.5021553, Ventura_2019},  and more recently so-called warm dense matter (WDM) \cite{Dornheim_PRL_2020, Fuchs2015} {---an extreme state that occurs in astrophysical objects~\cite{Benuzzi_Mounaix_2014,saumon1} and that is also relevant for technological applications~\cite{Hu2010,Lazicki2021,Dattelbaum2021,Luetgert2021}.} 
However, in contrast to the LRT, the foundations of a quantum theory of the non-linear response (NLRT) at finite wave numbers is far from being established even for simple model systems such as a free electron gas~\cite{review,loos}.  
In this regard, the lack of a reliable theoretical foundation makes the \textit{ab initio} simulation an indispensable tool to guide the development of the NLRT.
This was demonstrated recently for warm dense matter by performing path integral quantum Monte Carlo (PIMC) simulations \cite{PhysRevResearch.3.033231,JCP21_nonlin}.  {Yet, while these results are exact, the fermion sign problem~\cite{PhysRevE.100.023307,troyer} limits their application to moderate levels of quantum degeneracy. In contrast, the thermal Kohn-Sham density functional theory (KS-DFT) method~\cite{Mermin_DFT_1965} does not suffer from this limitation.
Indeed, it has become standard practice to study the linear electronic response~\cite{Gross_Kohn_PRL_1985,martin2004electronic} based on the KS orbitals.
In this work, we explore a new KS-DFT based approach to the nonlinear electronic response of arbitrary materials. Firstly, this methodology allows us to compute higher-order (being quadratic, cubic, etc.~with respect to the perturbation amplitude) response functions, with the only approximation being given by the choice of the exchange--correlation (XC) functional. In addition, we can straightforwardly estimate the validity range of LRT, which is highly important in its own right.}

 {As a particular example, we apply this approach to the free electron gas~\cite{loos,review}---the archetypical} model system 
 {with general relevance} for numerous applications in condensate matter physics and  high-energy-density science. 
From a many-particle physics perspective,  {we note that} it is imperative to first develop a NLRT for 
this general free electron gas model, before applying the NLRT to specific cases.

 {In this context,  {thermal} KS-DFT~\cite{Mermin_DFT_1965} constitutes the method of choice}
because it allows calculations over large temperature ranges covering the strongly to partially degenerate regimes.  {Moreover, we note that the general nature of our present NLRT approach makes it directly usable for high-$T$ DFT methods~\cite{Zhang_POP_2016,bethkenhagen2021thermodynamic}, including orbital-free formulations~\cite{wesolowski2013recent}.} 
Since the free electron gas model and the NLRT have important applications in WDM \cite{new_POP, Dornheim_PRL_2020}, we start from  {relatively} high temperatures relevant for laboratory astrophysics \cite{Kritcher2020, Booth2015,PhysRevB.101.054301}  {as well as astrophysical models~\cite{saumon1}}, inertial confinement fusion \cite{Hu2010}, and the synthesis of new materials at extreme conditions \cite{Dattelbaum2021, Lazicki2021, Luetgert2021}.
At these parameters, we can benchmark KS-DFT results against available PIMC results~\cite{PhysRevResearch.3.033231,JCP21_nonlin,Dornheim_PRL_2020}.
 {In addition, we consider lower temperatures down to the electronic ground state that are relevant for condensed matter physics.}

It is conventional to give the temperature $T$ and density $n_0$ of the free electron gas using the reduced temperature $\theta=T/T_F$ (with $T_F$ being the Fermi temperature) and the mean inter-particle distance in a.u., $r_s=(4\pi n_0/3)^{1/3}$.  
For example, a rather loose definition of the WDM regime corresponds to temperatures $0.1 \lesssim \theta \lesssim 10$ and densities $0.5\lesssim r_s \lesssim 10$.



The prospect of the observation of nonlinear phenomena in WDM has triggered an active investigation of the non-linear density response properties of the free electron gas by Dornheim et al. \cite{PhysRevResearch.3.033231, JCP21_nonlin, dornheim2021nonlinear} using the \textit{ab initio} PIMC method~\cite{PhysRevE.100.023307,Dornheim_permutation_cycles}. The focus of these PIMC studies was the static density response of WDM at temperatures $\theta\geq 1$.
The main reason for not considering lower temperatures was 
 {the aforementioned}
fermion sign problem \cite{PhysRevE.100.023307}, which results in an exponential increase in the computation time with decreasing temperature. Although there are other quantum Monte-Carlo  (QMC) methods that have different domains of applicability, such as the configuration PIMC approach \cite{JCP_Simon17, Yilmaz}, the permutation blocking PIMC method~\cite{Dornheim_2015,JCP_tobias_15}, and a phaseless auxiliary-field quantum Monte Carlo technique~\cite{JCP_Lee}, there are always parameters at which QMC methods encounter significant difficulties.
Approximately, the problematic domain for QMC methods corresponds to $\theta<1$ and $r_s\gtrsim 2$ \cite{dornheim_physrep_18, JCP_Lee, Yilmaz}.

On the other hand, the parameter range corresponding to densities $r_s\gtrsim 2$ and temperatures $0.01 \lesssim \theta<1$ is highly important for numerous applications. For example, recently it was shown that the static  non-linear density response functions of the electron gas can be used for the construction of advanced kinetic
energy functionals required for orbital-free density functional theory (OF-DFT) based simulations  \cite{PhysRevB.100.125106, PhysRevB.100.125107} with applications at {ambient} \cite{PhysRevB.104.045118} as well  as extreme conditions \cite{PhysRevB.88.195103, PhysRevB.92.115104}. Additionally, non-linear density response functions can extend quantum fluid models (quantum hydrodynamics and time-dependent orbital-free density functional theory) beyond the weak perturbation regime \cite{zhandos_pop18, PhysRevX.11.011049, graziani2021shock, Manfredi2021, PhysRevB.104.235110, zhandos_cpp17_1d}. Moreover, static non-linear density response functions are needed for the systematic improvement of effective pair interaction models for WDM \cite{cpp21, PhysRevE.98.023207, PhysRevE.99.053203, pop15} and liquid metals \cite{SENATORE1996851, PhysRevB.81.224113, PhysRevB.64.224112}. 
 {Finally, Moldabekov and co-workers~\cite{moldabekov2021thermal,JPSP21} have recently suggested to deliberately probe the nonlinear regime in X-ray Thomson scattering experiments~\cite{RevModPhys.81.1625} as an improved method for the inference of plasma parameters such as the electronic temperature.}
However, these applications 
 {remain in their infancy}
since the NLRT of correlated electrons is significantly less developed compared to the LRT \cite{dornheim_physrep_18}. One of the reasons is that the derivations in the NLRT are much more mathematically involved \cite{PhysRevB.59.10145,PhysRevB.37.9268, PhysRevE.54.3518}. 
 In fact, the NLRT is burdened with easy-to-make-mistake mathematical tasks and poorly converging integrals.
Therefore, the \textit{ab initio} calculation of the non-linear response properties across parameter ranges is required not only to describe certain phenomena, but also to guide and test new theoretical developments.

The key goal of this work is to demonstrate the  {high value of KS-DFT to study the nonlinear density response}
across temperature regimes as an alternative to much more expensive---for certain parameters even prohibitively expensive---QMC simulations. 
  {This is achieved by developing and testing the KS-DFT  based methodology  for the analysis and investigation of the higher order static density response functions.}
Therefore, first of all, we show that KS-DFT can be effectively used to compute static non-linear density response properties of correlated electrons at low temperatures ($\theta\lesssim 1$) and is able to reproduce available PIMC results at $\theta=1$. Secondly,  we provide an analysis of the available theoretical results for the diagonal parts of quadratic and cubic response functions   {by combining the KS-DFT  simulation of correlated electrons,  the KS-DFT calculations with the exchange-correlations functional set to zero, and recently developed machine learning representation of the local field correction of the free electron gas \cite{dornheim_ML,Dornheim_PRL_2020_ESA}}. This confirms the high accuracy of the analytic results for the quadratic response function and reveals the significant deficiency of the available analytical results for the cubic response function.     Finally, we 
 {are able to show}
the change in the characteristic features of the non-linear response functions on the way from moderately to strongly degenerate regimes.  


The paper is organized as the following: in Sec. 2, we provide the theoretical background of the studied non-linear response characteristics; in Sec. 3, we give the description of the performed simulations; the new results are presented and discussed in Sec. 4; the paper is concluded by summarizing the main findings and providing an outlook over future investigations in Sec. 5.

\section{Theory}

Let us start by briefly discussing the state-of-the-art theory of the static non-linear density response functions. Along with that, we establish the terminology used throughout the paper.
In general, the definition of the non-linear response functions follow from the perturbative expansion of the density $n(\vec r)$ around its unperturbed value $n_0(\vec r)$ \cite{PhysRevB.37.9268, PhysRevB.59.10145,PhysRevResearch.3.033231}. 
Specifically, we consider the response of the uniform electron gas  to an external harmonic perturbation~\cite{PhysRevLett.75.689,Dornheim_PRL_2020},  $V(\vec r)=2A\cos{\left(\vec q\cdot \vec r\right)}$, with amplitude $A$ and wave number $\vec q$. In this case, the Fourier expansion of the density distribution reads \cite{PhysRevResearch.3.033231}:
\begin{align}\label{eq:rho_tot}
n(\mathbf{r}) = n_0(\mathbf{r}) + 2 \sum_{\eta=1}^\infty \braket{\hat\rho_{\eta\mathbf{q}}}_{q,A} \textnormal{cos}\left(\eta\mathbf{q}\cdot\mathbf{r} \right),
\end{align}
where 
we have introduced the density perturbation components in Fourier space $\braket{\hat\rho_\mathbf{k}}_{q,A}$. 
The latter quantity is essentially the  density perturbation in $\vec k$ space induced by an external perturbation with amplitude $A$ and wavenumber $\vec q$. 

From Eq.~(\ref{eq:rho_tot}), we see that $\braket{\hat\rho_\mathbf{k}}_{q,A}$ has non-zero components at multiples of the perturbing field wavenumber, i.e. at $k=\eta \vec q$ with $\eta$ being an integer number. We refer to $\braket{\hat\rho_{\eta \mathbf{q}}}_{q,A}$ at $\eta=1$, $\eta=2$, $\eta=3$ as density perturbations at the first, second and third harmonics, respectively.  Next, using the density response $\braket{\hat\rho_\mathbf{k}}_{q,A}$, we arrive at the following definitions of the \textit{density response functions}\cite{PhysRevResearch.3.033231, Mikhailov_Annalen, PhysRevLett.75.689}: 
\begin{eqnarray}\label{eq:rho1}
\braket{\hat\rho_\mathbf{q}}_{q,A} &=& \chi^{(1)}(q) A + \chi^{(1,\textnormal{cubic})}(q) A^3  {+\dots}\ ,\\
\label{eq:rho2}
\braket{\hat\rho_\mathbf{2q}}_{q,A} &=& \chi^{(2)}(q) A^2 {+\dots} \ , \\
\label{eq:rho3}
\braket{\hat\rho_\mathbf{3q}}_{q,A} &=& \chi^{(3)}(q) A^3 {+\dots} \ ,
\end{eqnarray}
where $\chi^{(1)}(q)$ is the \textit{linear response function}, $\chi^{(1,\textnormal{cubic})}(q)$ is the \textit{cubic response function at the first harmonic}, $\chi^{(2)}(q)$ is the \textit{quadratic response function}, and  $\chi^{(3)}(q)$ is the \textit{cubic response function at the third harmonic}. {Evidently, Eqs.~(\ref{eq:rho1})-(\ref{eq:rho3}) are given by expansions in terms of the perturbation amplitude $A$, and are accurate up to the third order. While the density response is only given by a single term at the wave number of the original perturbation within LRT, the consideration of nonlinear effects leads to a richer picture including the excitation of higher-order harmonics.}

In the ideal Fermi gas approximation, the  linear response function $\chi^{(1)}_0$ is given by the  {(temperature-dependent)} Lindhard function \cite{Lindhard}.
On the same level of description, Mikhailov expressed the ideal quadratic  response function $\chi^{(2)}_0(q)$ and ideal cubic   response function at the third harmonic $\chi^{(3)}_0(q)$ in terms of the Lindhard function ~\cite{Mikhailov_Annalen,Mikhailov_PRL}:
\begin{equation}\label{eq:Mikhailov2}
\chi^{(2)}_0(q) = \frac{2}{q^2}\left( \chi^{(1)}_0(2q)-\chi^{(1)}_0(q)\right),
\end{equation}
\begin{equation}\label{eq:Mikhailov3}
\chi_0^{(3)}(q)=\frac{3\chi_0^{(1)}(3q)-8\chi^{(1)}_0(2q)+5\chi_0^{(1)}(q)}{3q^4}.
\end{equation}

Next, on the mean field level, usually called RPA, the results for $\chi^{(2)}(q)$  and $\chi^{(3)}(q)$ can be obtained by taking into account screening on the level of the linear response and dropping quadratic or higher order corrections to screening \cite{PhysRevResearch.3.033231}:
\begin{equation}\label{eq:chi2_RPA}
    \chi^{(2)}_{\rm RPA}( q)= \frac{\chi^{(2)}_{0}( q)}{\left[1-v(q)\chi^{(1)}_{0}(q)\right]^{2} \left[1-v(2q)\chi^{(1)}_{0}( 2q)\right]}.
\end{equation}
and
\begin{equation}\label{eq:chi3_RPA}
    \chi^{(3)}_{\rm RPA}( q)= \frac{\chi^{(3)}_{0}( q)}{\left[1-v(q)\chi^{(1)}_{0}(q)\right]^{3} \left[1-v(3q)\chi^{(1)}_{0}( 3q)\right]}.
\end{equation}


Finally, some electronic correlation effects beyond the mean-field level can be taken into account using a local field correction (LFC) $G(k)$ in the denominator~\cite{PhysRevResearch.3.033231}:
 \begin{equation}
     \chi^{(2)}_{\rm LFC}( q) =  \chi^{(2)}_{0}( q) \left[1-v(q)\left[1-G(q)\right]\chi^{(1)}_{0}(q)\right]^{-2}\times \left[1-v(2q)\left[1-G(2q)\right]\chi^{(1)}_{0}( 2q)\right]^{-1}, \label{eq:chi2_LFC}
 \end{equation}
and 
 \begin{equation}
     \chi^{(3)}_{\rm LFC}( q) =  \chi^{(3)}_{0}( q) \left[1-v(q)\left[1-G(q)\right]\chi^{(1)}_{0}(q)\right]^{-3} \times \left[1-v(3q)\left[1-G(3q)\right]\chi^{(1)}_{0}( 3q)\right]^{-1}.\label{eq:chi3_LFC}
 \end{equation}
Similarly to the screened equations (\ref{eq:chi2_RPA}) and (\ref{eq:chi3_RPA}),  Eqs.~(\ref{eq:chi2_LFC}) and (\ref{eq:chi3_LFC}) take into account electronic exchange-correlations  on the basis of the linear response theory. However, contrary to the case of the LFC in the linear response function, the insertion of the LFC here  cannot give an exact result as further terms are missing. Equations (\ref{eq:chi2_LFC}) and   (\ref{eq:chi3_LFC}) are easy-to-compute solutions for the case of a harmonically perturbed electron gas since the static LFC at the parameters of interest  is readily available from a machine-learning (ML) representation which is based on QMC simulation results ~\cite{dornheim_ML,Dornheim_PRL_2020_ESA}.

Finally, we note that there are no satisfactory analytical results for the  ideal cubic response function at the first harmonic $\chi^{(1,\textnormal{cubic})}_0(q)$.
Nevertheless, there is a formal relation between  $\chi^{(1,\textnormal{cubic})}_0$, and  the cubic response function of correlated electrons which follows from the perturbative analysis based on the Green functions method \cite{PhysRevResearch.3.033231} :
\begin{equation}\label{eq:cubic_first_LFC}
    \chi^{(1,\textnormal{cubic})}_{\rm LFC}( q)= \frac{\chi^{(1,\textnormal{cubic})}_{0}( q)}{\left[1-v(q)\left[1-G(q)\right]\chi^{(1)}_{0}(q)\right]^{4}}.
\end{equation}

Note that the mean-field result for the cubic response function at the first harmonic follows from Eq.~(\ref{eq:cubic_first_LFC}) by setting $G=0$,
\begin{equation}\label{eq:cubic_first_RPA}
    \chi^{(1,\textnormal{cubic})}_{\rm RPA}( q)= \frac{\chi^{(1,\textnormal{cubic})}_{0}( q)}{\left[1-v(q)\chi^{(1)}_{0}(q)\right]^{4}}.
\end{equation}

 {In this work, we} use the KS-DFT method to compute  the set of density response functions defined by Eqs. (\ref{eq:rho1})-(\ref{eq:rho3}) and subsequently verify the quality of the KS-DFT  results 
by comparing with PIMC results at $\theta=1$. 
Then we use KS-DFT results to analyze the analytical approximations given by Eqs. (\ref{eq:Mikhailov2})-(\ref{eq:chi3_LFC}) in the wide range of parameters inaccessible for QMC methods. This allows us to unambiguously asses the importance of the neglected higher order (nonlinear) screening and LFC effects.


\section{Simulation details}\label{sec:simulation_details}

The computational workflow consists of four main steps: First, the thermal KS-DFT simulations \cite{Mermin_DFT_1965} of the free electron gas perturbed by an external field 
$V(\vec r)=2A\cos{\left(\vec q\cdot \vec r\right)}$  are performed for different $A$ and $\vec q$ values;
Second, the wave functions from KS-DFT simulations are used to compute the total density distribution along the direction of the wave vector $\vec q$;
Third, the density perturbation components in $k$ space $\braket{\hat\rho_\mathbf{k}}_{q,A}$ are computed using Eq. (\ref{eq:rho_tot});
Finally,  the density response functions are found by fitting data for $\braket{\hat\rho_\mathbf{k}}_{q,A}$ using Eqs. (\ref{eq:rho1})-(\ref{eq:rho3}).

To begin with, we consider a strongly correlated electron gas with $r_s=6$ at $\theta=1$ and  $r_s=5$ at $\theta=0.01$. At $r_s=6$, we compare the results with the available finite temperature PIMC data for the linear and non-linear density response functions \cite{PhysRevResearch.3.033231}. At $r_s=5$, we compare with the diffusion quantum Monte Carlo (DMC) results for the linear density response function computed by Moroni, Ceperley, and Senatore \cite{PhysRevLett.75.689}.
Furthermore, we investigate a metallic density with  $r_s=2$ at three different values of the degeneracy parameter, $\theta=1$, $\theta=0.5$, and $\theta=0.01$.
In this case, we also benchmark results against PIMC data at $\theta=1$ and compare  with  the linear density  response function from the DMC  simulations at $\theta=0.01$.

The  KS-DFT simulations of the free electron gas were performed using the GPAW code~\cite{GPAW1, GPAW2, ase-paper, ase-paper2}, which is a real-space implementation of the projector augmented-wave method. The number of particles in the main simulation box is varied in the range from $14$ to $66$. Accordingly, the main cubic cell size is defined by $r_s$ and $N$ as $L=r_s (\frac{4}{3}\pi N)^{1/3}$. The direction of the perturbation is set to be along the $z$ axis. Due to periodic boundary conditions, the value of the perturbation  wave number of the external harmonic field is defined by the reciprocal lattice vectors of the main simulation cell $q=\eta \times 4\pi/L$, with $\eta$ being a  positive integer number. 
We used a Monkhorst-Pack~\cite{PhysRevB.13.5188} sampling of the Brillouin zone with a \emph{k}-point grid of $N_k\times N_k\times N_k$, with $N_k=12$ at $r_s=2$, and $N_k=8$ at $r_s=6$ and $r_s=5$. The calculations were performed using a plane-wave basis where the cutoff energy has been converged to $800~{\rm eV}$ at  $\theta=1$ and $r_s=2$, and to  $440~{\rm eV}$ at the rest of the $r_s$ and $\theta$ values.
The number of orbitals is set to $N_b=500$ at $r_s=2$ and $\theta=1$ with the smallest occupation number $f_{\rm min}\lesssim 10^{-7}$.
We set $N_b=240$ at $r_s=6$ and $\theta=1$, and $N_b=140$ at $r_s=2$ and $\theta=0.5$ ($f_{\rm min}\lesssim 10^{-6}$). At $\theta=0.01$, we set $N_b=70$ for $N=66$ particles, and $N_b=2 N$ for $N=20$ and $N=14$ particles (with $f_{\rm min}= 0$).

At $r_s=2$, the perturbation amplitudes are set in the range from $A=0.01$ to $A=0.1$ with a step of $\Delta A=0.03$ (here $A$ is in Hartree atomic units).
At $r_s=6$ and $r_s=5$, the perturbation amplitudes are in the range from $A=0.002$ to $A=0.017$ with the step $\Delta A=0.005$.
These values of the perturbation amplitudes used for the calculation of the density response functions  were found empirically guided by the requirement $\Delta n/n_0\ll 1$, and by testing the validity of Eqs. (\ref{eq:rho_tot})-(\ref{eq:rho3}).  Examples of the dependence of the density perturbation $\braket{\hat\rho_\mathbf{k}}_{q,A}$ on $A$ as well as the application of Eqs. (\ref{eq:rho_tot})-(\ref{eq:rho3}) are illustrated in Appendix \ref{sec:app}.

The exchange-correlation (XC) functional in our KS-DFT simulations is the local density approximation (LDA) in the Perdew-Zunger parametrization~\cite{Perdew_LDA}. 
Recently, it was demonstrated for $\theta=1$ that commonly used GGA functionals such as PBE ~\cite{PBE}, PBEsol~\cite{PBEsol}, AM05~\cite{PhysRevB.72.085108}  and the meta-GGA functional SCAN~\cite{SCAN} are not able to provide  a superior description compared to LDA in the case of the  free electron gas  perturbed by an external field with fixed wave number $\vec q$ when $\Delta n/n_0<1$ \cite{moldabekov_jcp21, moldabekov2021benchmarking}. We do not aim to further study this problem in this work. Therefore, we do not consider other types of XC functionals beyond LDA.

In addition to the LDA based calculations, we performed simulations with zero XC functional (NullXC).
This allows us to obtain exact results for the density response on the mean-field level. The value of this type of KS-DFT calculations 
allows us to asses the accuracy of the corresponding 
theoretical mean-field expressions given in 
Eqs. (\ref{eq:chi2_RPA}) and (\ref{eq:chi3_RPA}). Furthermore, once the analytical results have been verified, the KS-DFT calculations on the mean-field level
can be combined with the LFC to compute a highly accurate  response function. We demonstrate that this is the case  for the quadratic response function and the cubic response function at the first harmonic using Eqs. (\ref{eq:chi2_LFC}) and (\ref{eq:cubic_first_LFC}).

\section{Results}\label{s:results}

\subsection{Strongly correlated hot electrons}

We start the discussion of our simulation results by considering the strongly correlated electron gas with the density parameter $r_s=6$ and at the reduced temperature $\theta=1$.
This corresponds to WDM generated in evaporation experiments \cite{Zastrau}. At these parameters, we can benchmark the KS-DFT calculations against previous PIMC calculations~\cite{PhysRevResearch.3.033231,Dornheim_PRL_2020}.

\subsubsection{Linear density response in WDM regime}\label{sss:chi1_rs6}

First, we verify that our KS-DFT calculations provide accurate data for the linear static density response function $\chi^{\rm (1)}(q)$,
which allows us to systematically analyze and exclude the possibility of finite size effects~\cite{Dornheim_JCP_2021}. We start this analysis by comparing the linear static density response function computed using KS-DFT with the exact analytical results on the mean-field level and with $\chi^{(1)}$ of the correlated electron gas computed using the exact data for the LFC~\cite{dornheim_ML}. 


\begin{figure}
\center
\includegraphics [width=0.45 \textwidth]{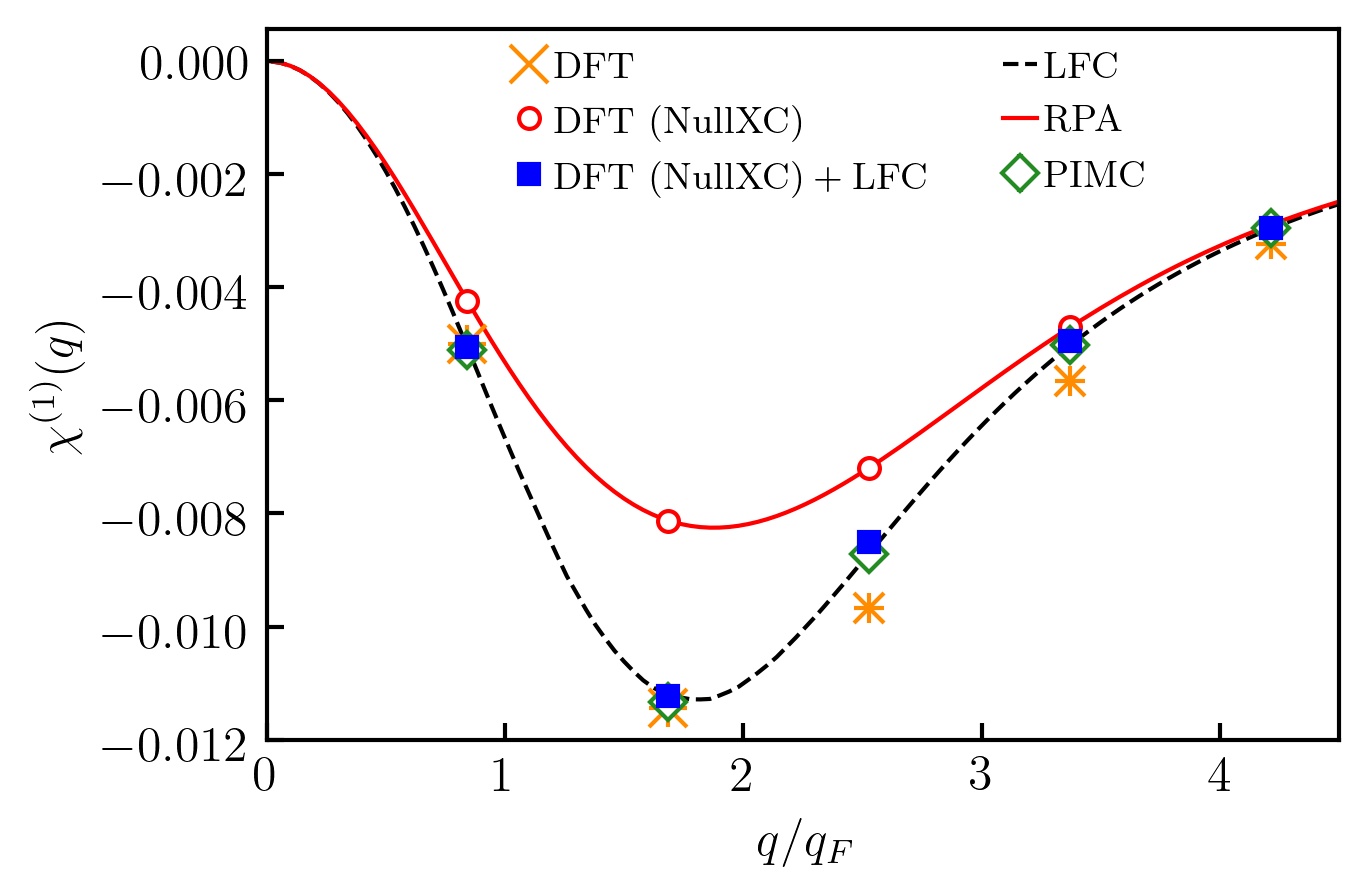}
\caption{\label{fig:chi1}
Linear static density response function at $r_s=6$ and $\theta=1$.}
\end{figure} 

In Fig.~\ref{fig:chi1}, we present the KS-DFT results computed using $N=14$ electrons.
In this case, the cell size is $L=12.335~{\rm \angstrom}$ and the accessible values of the wave numbers are multiples of  $q_{\rm min}\simeq 0.8427 q_F$.

From Fig.~\ref{fig:chi1}, first of all, we see that the linear density response function computed using KS-DFT with zero XC functional,  $\widetilde \chi^{(1)}_{\rm NullXC}$, accurately reproduces the exact random phase approximation (RPA) result for the static linear response function,
\begin{equation}\label{eq:chi1_RPA}
     \chi^{(1)}_{\rm RPA}(q)=\frac{\chi^{(1)}_0(q)}{1-v(q) \chi^{(1)}_0(q)},
\end{equation}
 where $v(q)=4\pi/q^2$ and $\chi^{(1)}_0(q)$ is the Lindhard function. This shows that finite size effects in our KS-DFT calculations with as few as $14$ electrons is negligible in the considered case.  {For completeness, we note that this is consistent with previous findings of PIMC simulations at similar conditions~\cite{dornheim_ML}.}

Secondly, we combine the density response function computed using the KS-DFT with zero XC functional, $\widetilde \chi^{(1)}_{\rm NullXC}$, with the LFC to find the linear density response function of correlated electrons:
\begin{equation}\label{eq:dft_chi1_LFC}
     \widetilde \chi_{\rm LFC}^{(1)}(q)=\frac{\widetilde \chi^{(1)}_{\rm NullXC}(q)}{1+v(q)G(q)\widetilde \chi^{(1)}_{\rm NullXC}(q)},
\end{equation}
where $G(q)$ is computed using the ML representation of the LFC ~\cite{dornheim_ML}.

From Fig.~\ref{fig:chi1}, we see that  $\widetilde \chi_{\rm LFC}^{(1)}(q)$ is in excellent agreement with the exact value computed using the Lindhard function, 
 \begin{equation}
 \chi_{\rm LFC}^{(1)}(q) = \frac{\chi^{(1)}_0(q)}{1-v(q)\left(1-G(q)\right) \chi^{(1)}_0(q)}
     = \frac{\chi^{(1)}_{\rm RPA}(q)}{1+v(q)G(q) \chi^{(1)}_{\rm RPA}(q)}\label{eq:chi1_LFC},
 \end{equation}
where after the second equality we used Eq.~(\ref{eq:chi1_RPA}) to express $\chi_{\rm LFC}^{(1)}(q)$ in terms of $\chi^{(1)}_{\rm RPA}(q)$.  

Note that Eq.~(\ref{eq:dft_chi1_LFC}) follows from Eq.~(\ref{eq:chi1_LFC}) after the substitution $\widetilde \chi^{(1)}_{\rm NullXC} \to \chi^{(1)}_{\rm RPA}$ and $\widetilde \chi_{\rm LFC}^{(1)} \to \chi_{\rm LFC}^{(1)}$.

Now, after verifying that our calculations are not affected by the finite size effect, we compare the LDA based KS-DFT calculations of the linear density response function, $\widetilde \chi_{\rm LDA}^{(1)} (q)$,  with the exact result $\chi_{\rm LFC}^{(1)}(q)$. From Fig.~\ref{fig:chi1}, we see that  $\widetilde \chi_{\rm LDA}^{(1)}  (q)$ is in good agreement with $\chi_{\rm LFC}^{(1)}(q)$ at $q<2q_F$ (with $q_F$ being the Fermi wave number) and exhibits significant disagreements at $q>2q_F$. 
To understand this finding, we recall that the LDA corresponds to the long wave length approximation of the LFC with $G_{\rm LDA}=\gamma k^2$, where $\gamma$ is defined by the 
compressibility sum rule \cite{PhysRevB.103.165102}. This approximation is applicable at $q\lesssim q_F$ and increasingly deviates from the exact result with the increase in the wave number beyond $2q_F$ \cite{moldabekov_jcp21}.  Note that all $\widetilde \chi_{\rm LDA}^{(1)} (q)$, $\widetilde \chi_{\rm LFC}^{(1)}(q)$,  and  $\chi_{\rm LFC}^{(1)}(q)$ tend to $ \chi^{(1)}_{\rm RPA}(q)$  in the limit of large wave numbers since the screening factor dominates over XC effects in this limit. This can be seen from Eqs.~(\ref{eq:chi1_LFC}) and (\ref{eq:dft_chi1_LFC}), where the LFC is suppressed by the factor $q^{-2}$.

The insight that we have gained considering $\widetilde \chi^{(1)}_{\rm NullXC}$, $\widetilde \chi^{(1)}(q)$, and $\widetilde \chi_{\rm LDA} (q)$  will help us to understand the KS-DFT results for the higher-order nonlinear density response functions discussed in the next subsection. 
Further, we use a tilde over a symbol to differentiate response functions calculated using KS-DFT from the theoretical definitions.



\begin{figure*}[!t]
\minipage{0.344\textwidth}
  \includegraphics[width=\linewidth]{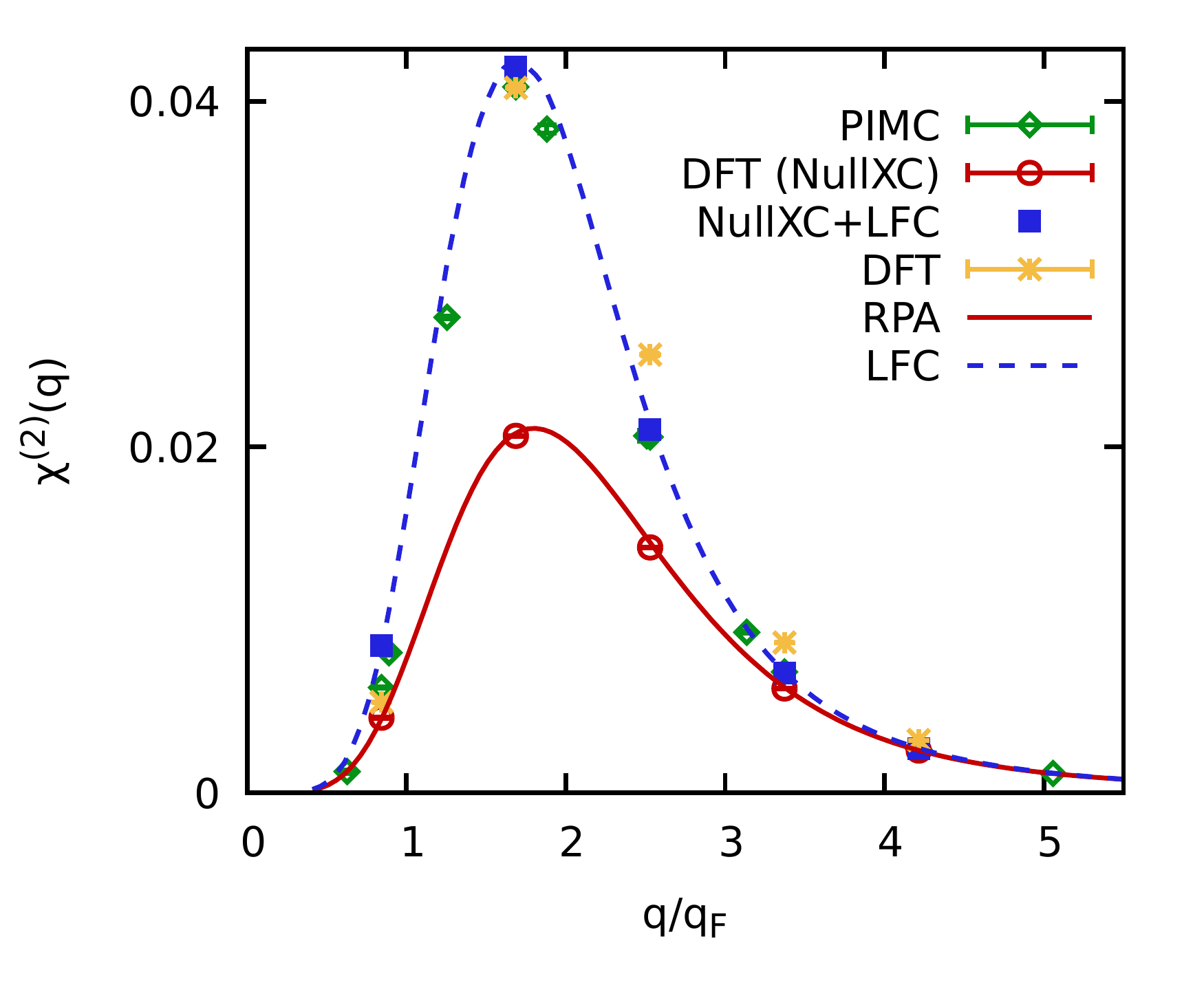}
\endminipage
\minipage{0.344\textwidth}
  \includegraphics[width=\linewidth]{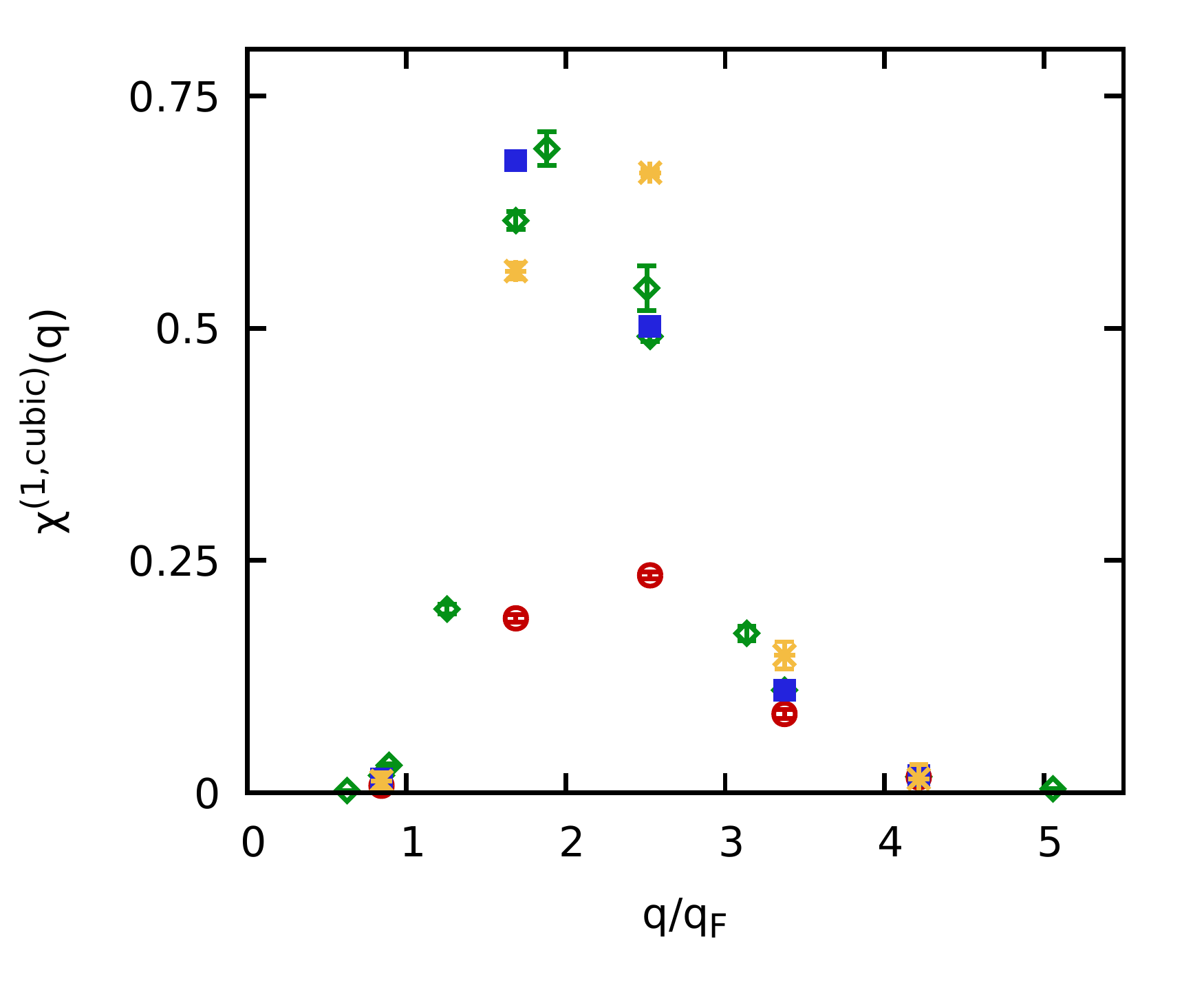}
\endminipage
\minipage{0.344\textwidth}%
  \includegraphics[width=\linewidth]{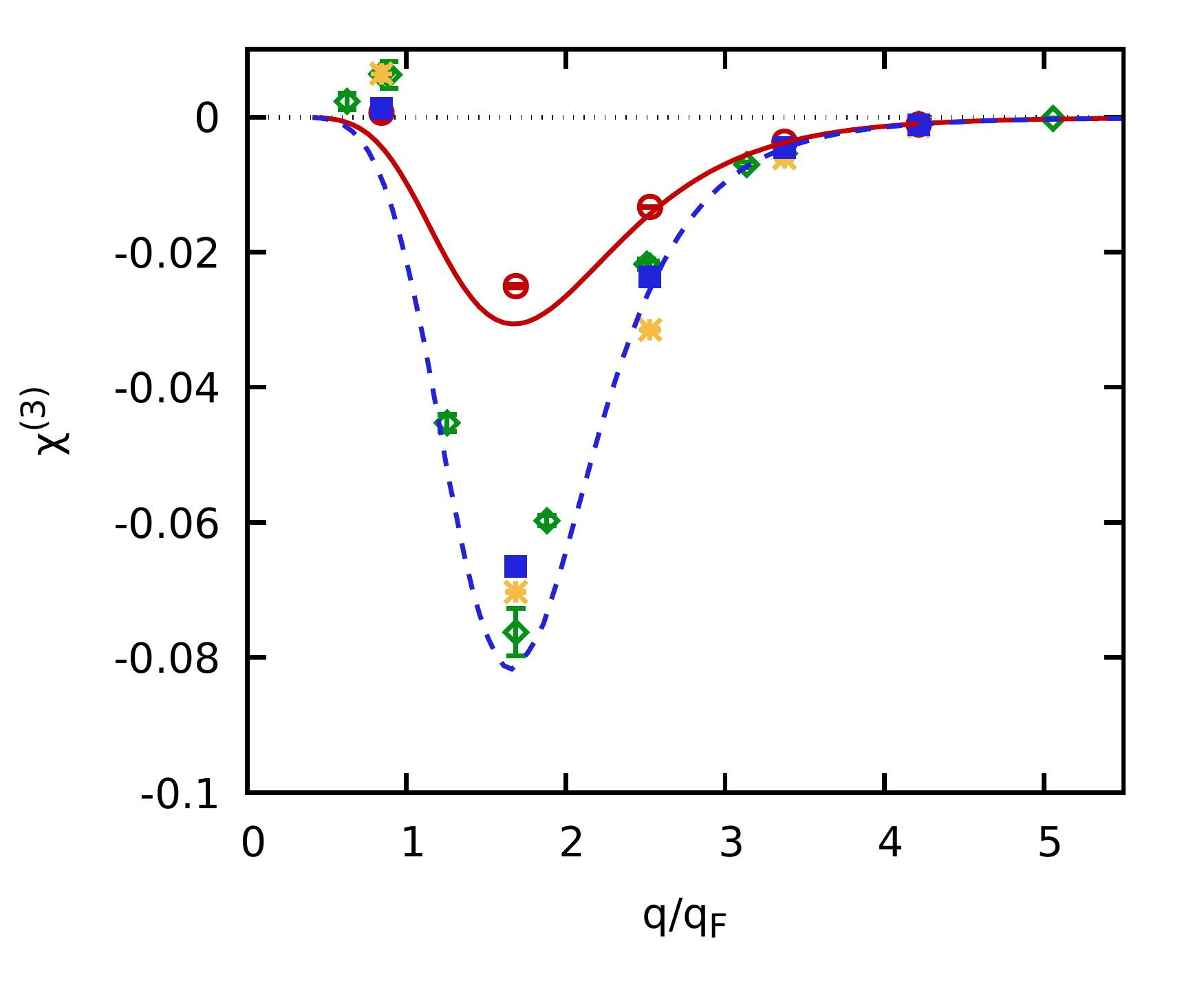}
\endminipage
  \caption{
  Non-linear static density response functions at $r_s=6$ and $\theta=1$.
  Left:  The quadratic response function at the second harmonic. 
  Middle: The cubic response function at the first harmonic.
  Right:  The cubic response function at the third harmonic. }
  \label{fig:NLR_rs6_theta1}
  \end{figure*}



\subsubsection{Non-linear density response in WDM regime}

 {Our new} results for the non-linear density response functions at $r_s=6$ and $\theta=1$ are presented in Fig. ~\ref{fig:NLR_rs6_theta1}.
In particular, the left panel shows the results for the quadratic response function (defined by Eq. (\ref{eq:rho2})), the middle panel presents data for the cubic response function at the first harmonic (the cubic term in Eq. (\ref{eq:rho1})), and the right panel shows the results for the cubic response function at the third harmonic (defined by Eq. (\ref{eq:rho2})).

First of all, we observe that the LDA XC functional based calculations are generally in good agreement with the 
PIMC results at $q<2q_F$ and overestimate  the considered non-linear density response functions at  $q>2q_F$.
The reason for this behavior of the LDA based calculations is the inaccuracy of the LFC incorporated in the LDA at $q>2q_F$ as it has been discussed in Sec.~\ref{sss:chi1_rs6} above. 

Next, from the left panel of Fig.~\ref{fig:NLR_rs6_theta1}, we see that KS-DFT calculations with XC set to zero (NullXC) are in excellent agreement with the theoretical RPA curve for the quadratic response function. This confirms the high quality of the analytical result Eq. (\ref{eq:chi2_RPA}) in the WDM regime. 
In the case of the cubic response function at the third harmonic as shown in the rightmost panel of Fig.~\ref{fig:NLR_rs6_theta1}, we observe that Eq.~(\ref{eq:chi3_RPA}) is accurate at $q>2q_F$, but overestimates the response at $q<2q_F$. Note that the LDA based KS-DFT results, the KS-DFT calculations without XC (NullXC), and the PIMC results all have positive sign at $q<q_F$, while the theoretical curves fail to capture the change in the sign of the cubic response function at the third harmonic with decrease in wave number. 


Let us next combine the KS-DFT data for the quadratic response computed with zero XC functional, $\widetilde \chi_{\rm NullXC}^{(2)}(q)$, with the LFC.
For that, we express $\chi^{(2)}_{\rm LFC}(q)$  via  $\chi_{\rm RPA}^{(2)}(q)$ using  Eqs. (\ref{eq:chi2_LFC}) and (\ref{eq:chi2_RPA}).
Then we perform substitutions  $\widetilde \chi^{(2)}_{\rm LFC}(q)  \to \chi^{(2)}_{\rm LFC}(q)$ and $ \widetilde \chi_{\rm NullXC}^{(2)}(q)\to  \chi_{\rm RPA}^{(2)}(q)$. As the result, we have the following relation:
 
\begin{equation}\label{eq:chi2_NullXC}
    \widetilde \chi^{(2)}_{\rm LFC}(q)= \widetilde \chi_{\rm NullXC}^{(2)}(q) \times \frac{\left[1-v(q)\widetilde \chi^{(1)}_{0}(q)\right]^{2} \left[1-v(2q)\widetilde \chi^{(1)}_{0}( 2q)\right]}{\left[1-v(q)\left[1-G(q)\right]\widetilde \chi^{(1)}_{0}(q)\right]^{2} \left[1-v(2q)\left[1-G(2q)\right]\widetilde  \chi^{(1)}_{0}( 2q)\right]^{}},
\end{equation}
 
 where $\widetilde \chi^{(1)}_{0}(q)$ can be extracted from  $\widetilde \chi_{\rm NullXC}^{(1)}(q)$ as
 \begin{equation}
     \widetilde \chi_{0}^{(1)}(q)=\frac{\widetilde \chi_{\rm NullXC}^{(1)}(q)}{1+v(q) \widetilde \chi_{\rm NullXC}^{(1)}(q)}.
 \end{equation}
 
Comparing the results calculated using Eq.~(\ref{eq:chi2_NullXC}) with the PIMC data, we conclude that the relationship (\ref{eq:chi2_LFC}) is fulfilled  with high accuracy.

 Similarly, we derive the connection between $\widetilde \chi^{(1,\textnormal{cubic})}_{\rm LFC}(q)$ and $\widetilde \chi^{(1,\textnormal{cubic})}_{\rm NullXC}(q)$ using Eqs.~(\ref{eq:cubic_first_LFC}) and (\ref{eq:cubic_first_RPA}), and replacing $\chi^{(1,\textnormal{cubic})}_{\rm LFC}(q)\to \widetilde \chi^{(1,\textnormal{cubic})}_{\rm LFC}(q)$ and $ \chi^{(1,\textnormal{cubic})}_{\rm RPA}(q)\to \widetilde \chi^{(1,\textnormal{cubic})}_{\rm NullXC}(q)$. As the result we find:
 \begin{equation}\label{eq:cubic_first_DFT}
    \widetilde \chi^{(1,\textnormal{cubic})}_{\rm LFC}(q)= \widetilde \chi^{(1,\textnormal{cubic})}_{\rm NullXC}(q) \times \frac{{\left[1-v(q)\widetilde \chi^{(1)}_{0}(q)\right]^{4}}}{{\left[1-v(q)\left[1-G(q)\right]\widetilde \chi^{(1)}_{0}(q)\right]^{4}}}.
\end{equation}
 
Using $\widetilde \chi^{(1,\textnormal{cubic})}_{\rm NullXC}(q)$ obtained from KS-DFT simulations with zero XC and the LFC computed using the ML representation~\cite{dornheim_ML}, we have found that
Eq.~(\ref{eq:cubic_first_DFT})  reproduces the PIMC results in the entire range of the wave numbers as it is can be seen from the middle panel of Fig.~\ref{fig:NLR_rs6_theta1}.  {It is only the availability of $\widetilde \chi^{(1,\textnormal{cubic})}_{\rm NullXC}(q)$ that allows us to estimate the cubic response at the first harmonic with PIMC accuracy as no analytical theory for $\chi_0^{(1,\textnormal{cubic})}$ currently exists. }

To further explore the combination of the KS-DFT calculations with zero XC functional and the ML representation of the LFC, we next analyze the
quality of the theoretical result Eq.~(\ref{eq:chi3_LFC}) for the cubic response at the third harmonic. Using Eqs.~(\ref{eq:chi3_RPA})  and (\ref{eq:chi3_LFC}), and 
replacing $\chi^{(3)}_{\rm LFC}(q)$ by  $\widetilde \chi^{(3)}_{\rm LFC}(q)$ and $\chi^{(3)}_{\rm RPA}(q)$ by $\widetilde \chi^{(3)}_{\rm NullXC}(q)$, we arrive a the following relation between $\chi^{(3)}_{\rm NullXC}(q)$ computed using the KS-DFT calculations with zero XC functional   and the LFC:
 
\begin{equation}\label{eq:dft_ch3_LFC}
    \widetilde \chi^{(3)}_{\rm LFC}(q)= \widetilde \chi_{\rm NullXC}^{(3)}(q) \times \frac{\left[1-v(q)\widetilde \chi^{(1)}_{0}(q)\right]^{3} \left[1-v(3q)\widetilde \chi^{(1)}_{0}( 3q)\right]}{\left[1-v(q)\left[1-G(q)\right]\widetilde \chi^{(1)}_{0}(q)\right]^{3} \left[1-v(3q)\left[1-G(3q)\right]\widetilde  \chi^{(1)}_{0}( 3q)\right]}.
\end{equation}

The comparison of $\widetilde \chi^{(3)}_{\rm LFC}(q)$ with the PIMC results is presented in the right panel of Fig.~\ref{fig:NLR_rs6_theta1}.
From this figure we see that  $\widetilde \chi^{(3)}_{\rm LFC}(q)$ significantly deviates from the PIMC data at $q<2q_F$.
This means that  the relation (\ref{eq:chi3_LFC}) does not provide an adequate description of the correlated electron gas.
This is expected since we have already demonstrated above that the RPA result Eq.~(\ref{eq:chi3_RPA}) is inadequate at  $q<2q_F$. 
Therefore, the description of the screening on the mean-field level must first be improved to describe the actual system.

\subsection{Strongly correlated and strongly degenerate electrons}

Next  we  investigate the strongly degenerate case with $r_s=5$ and $\theta=0.01$. 
In this regime, we are able to verify our KS-DFT calculations by comparing with  the accurate DMC calculations of the linear static density response function, $\chi^{(1)}$,  by Moroni, Ceperley, and Senatore \cite{PhysRevLett.75.689}.

Additionally, we further assess possible finite size effects at low temperature by comparing the simulation results for  $N=14$ particles to the results computed using  $N=20$, $N=38$, and $N=66$ particles.
In this case, the cell size is $L=10.28~{\rm \angstrom}$ (for $N=14$), $L=11.577~{\rm \angstrom}$ (for $N=20$), $L=14.34~{\rm \angstrom}$ (for $N=38$), and $L=17.236~{\rm \angstrom}$ (for $N=66$).
Correspondingly, the accessible values of the wave numbers are multiples of  $q_{\rm min}\simeq 0.8427 q_F$ (for $N=14$), $q_{\rm min}\simeq 0.74822 q_F$ (for $N=20$), $q_{\rm min}\simeq 0.604 q_F$ (for $N=38$), and  $q_{\rm min}\simeq 0.50 q_F$ (for $N=66$).



\begin{figure}
\center
\includegraphics[width=0.45 \textwidth]{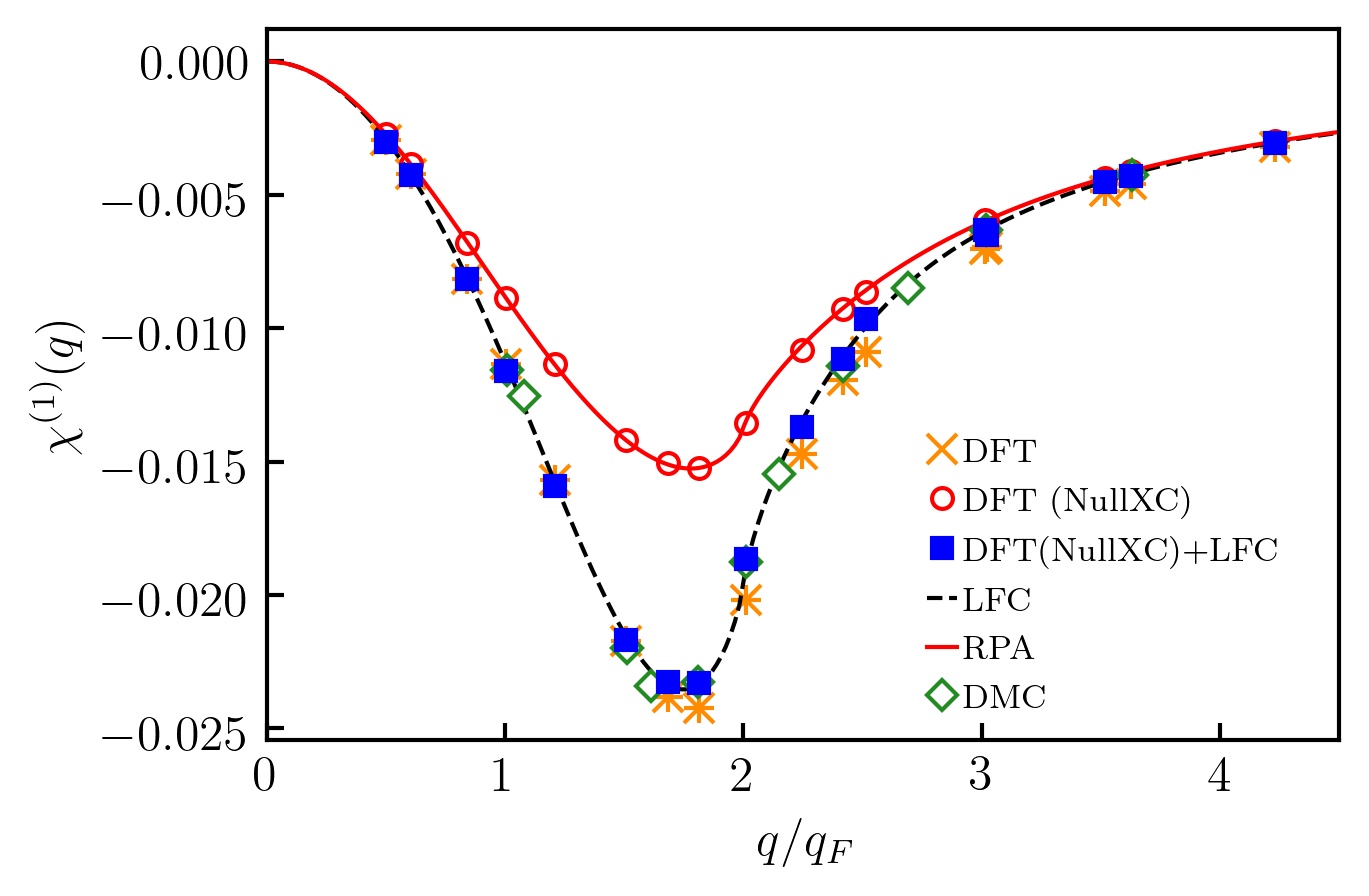}
\caption{\label{fig:chi1_rs5}
Comparison of the DMC data by Moroni, Ceperley, and Senatore \cite{PhysRevLett.75.689} with DFT results for the linear response function at $r_s=5$ and $\theta=0.01$.}
\end{figure}

\subsubsection{Linear density response in the limit of strong degeneracy}\label{sss:chi1_rs5}

In Fig.~\ref{fig:chi1_rs5}, we present results for the static linear density response function.
Evidently, the results for $\widetilde \chi^{(1)}_{\rm NullXC}$ computed using different numbers of particles accurately reproduces the exact mean-field level result $\chi^{(1)}_{\rm RPA}$, Eq.~(\ref{eq:chi1_RPA}). Furthermore, at all considered numbers of particles, the combination of $\widetilde \chi^{(1)}_{\rm NullXC}$ with the LFC by using Eq.~(\ref{eq:dft_chi1_LFC}) allows to reproduce the exact result given by Eq.~(\ref{eq:chi1_LFC}).
Therefore, the reduction of the number of particles from $N=66$ to $N=38$, then to $N=20$, and further to $N=14$ does not lead to a deterioration of the quality of the data for $\widetilde \chi^{(1)}_{\rm NullXC}$. This confirms the  {remarkable} convergence of the KS-DFT simulations for as few as  $N=14$ particles.  

To get a picture about the quality of the LDA based calculations, we compare $\widetilde \chi^{(1)}_{\rm LDA}$ with the DMC results by Moroni et al.~\cite{PhysRevLett.75.689}
and with $\chi^{(1)}_{\rm LFC}$ computed using the ML representation of the LFC by Dornheim et al.~\cite{dornheim_ML}. Despite the fact that the LDA is designed to describe only the long wave length limit of the LFC, we observe that in the strongly degenerate case, the LDA based KS-DFT calculations  provide high quality results for the linear density response function with a level of accuracy similar to the ground state quantum Monte-Carlo calculations.

\begin{figure*}[!t]
\minipage{0.332\textwidth}
  \includegraphics[width=\linewidth]{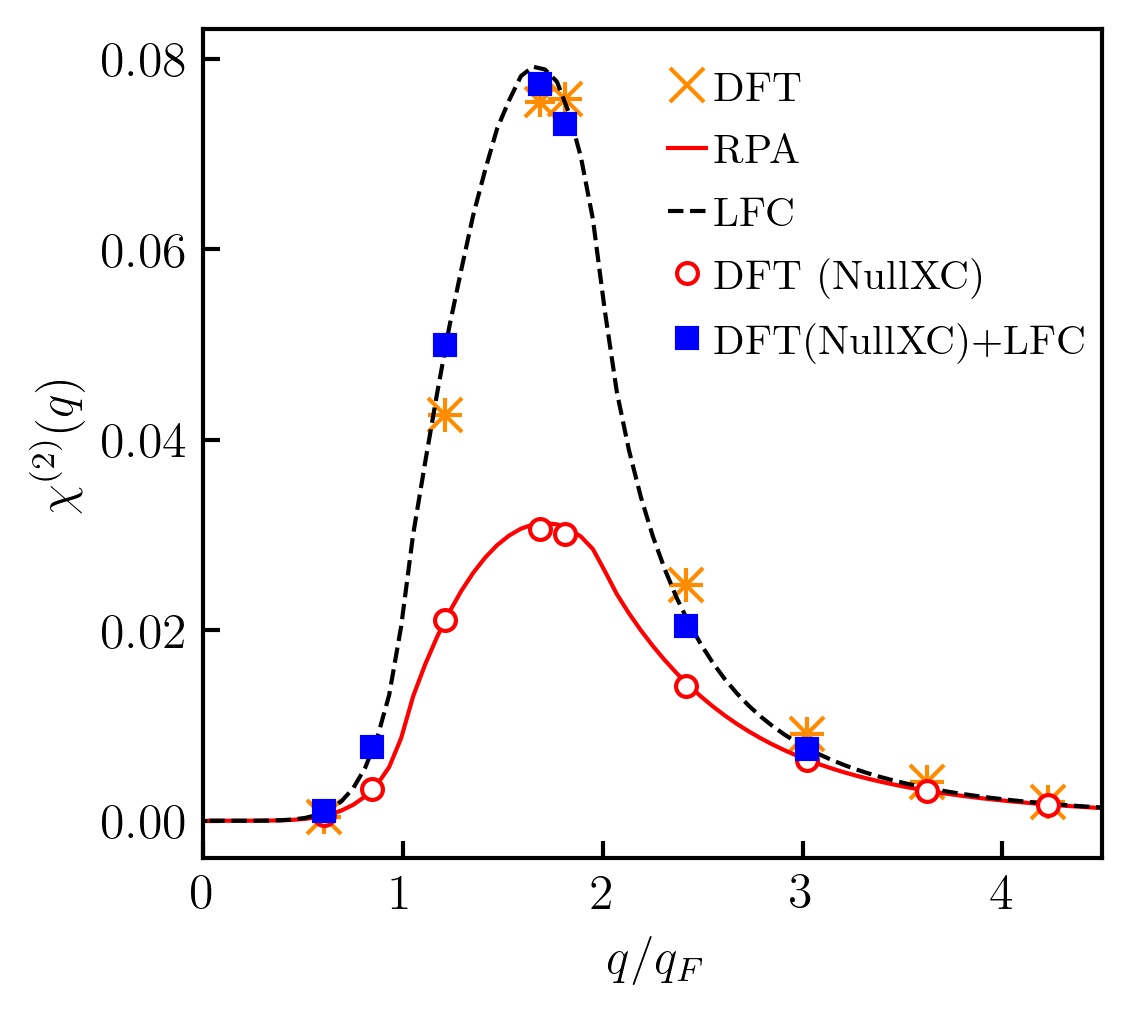}
\endminipage
\minipage{0.33\textwidth}
  \includegraphics[width=\linewidth]{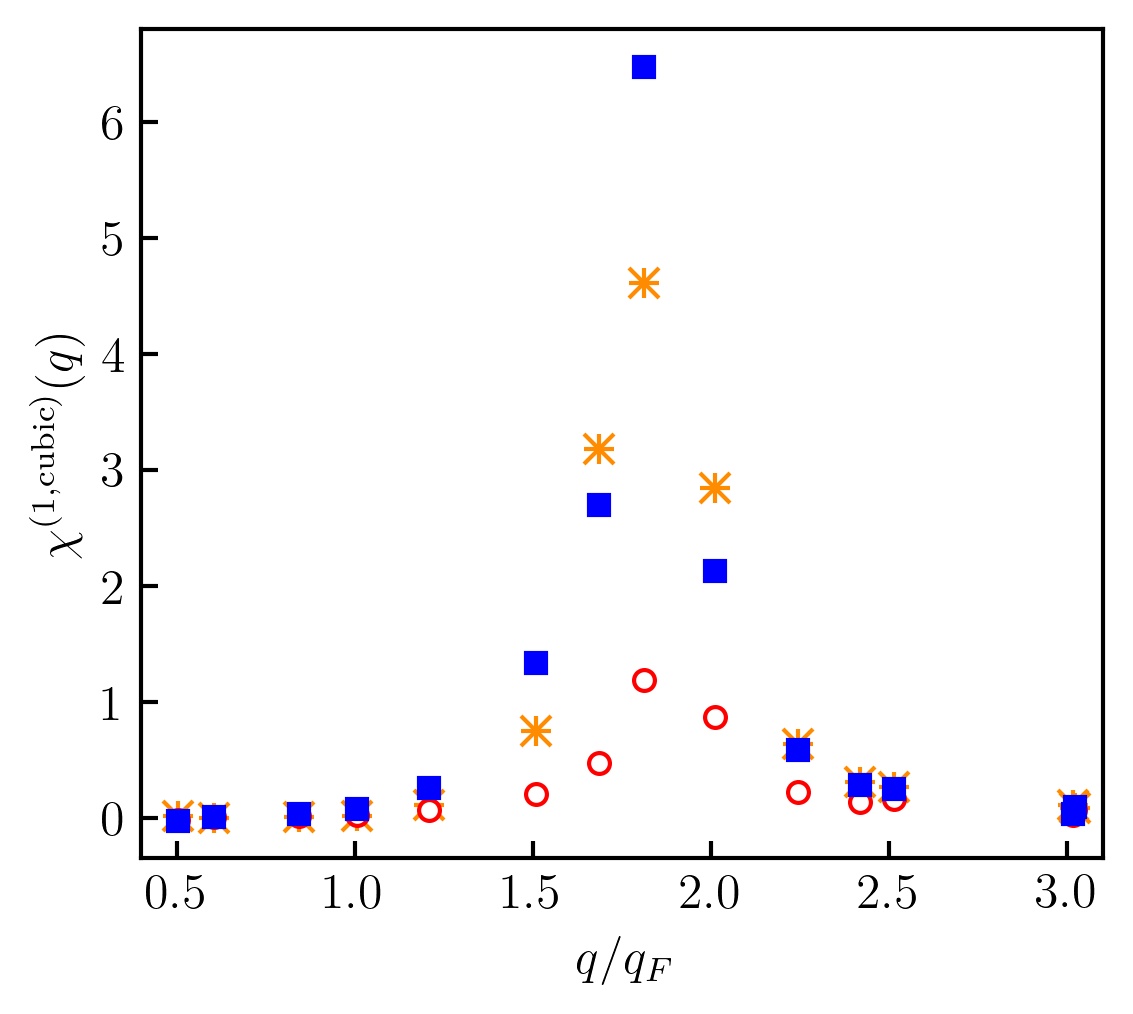}
\endminipage
\minipage{0.334\textwidth}%
  \includegraphics[width=\linewidth]{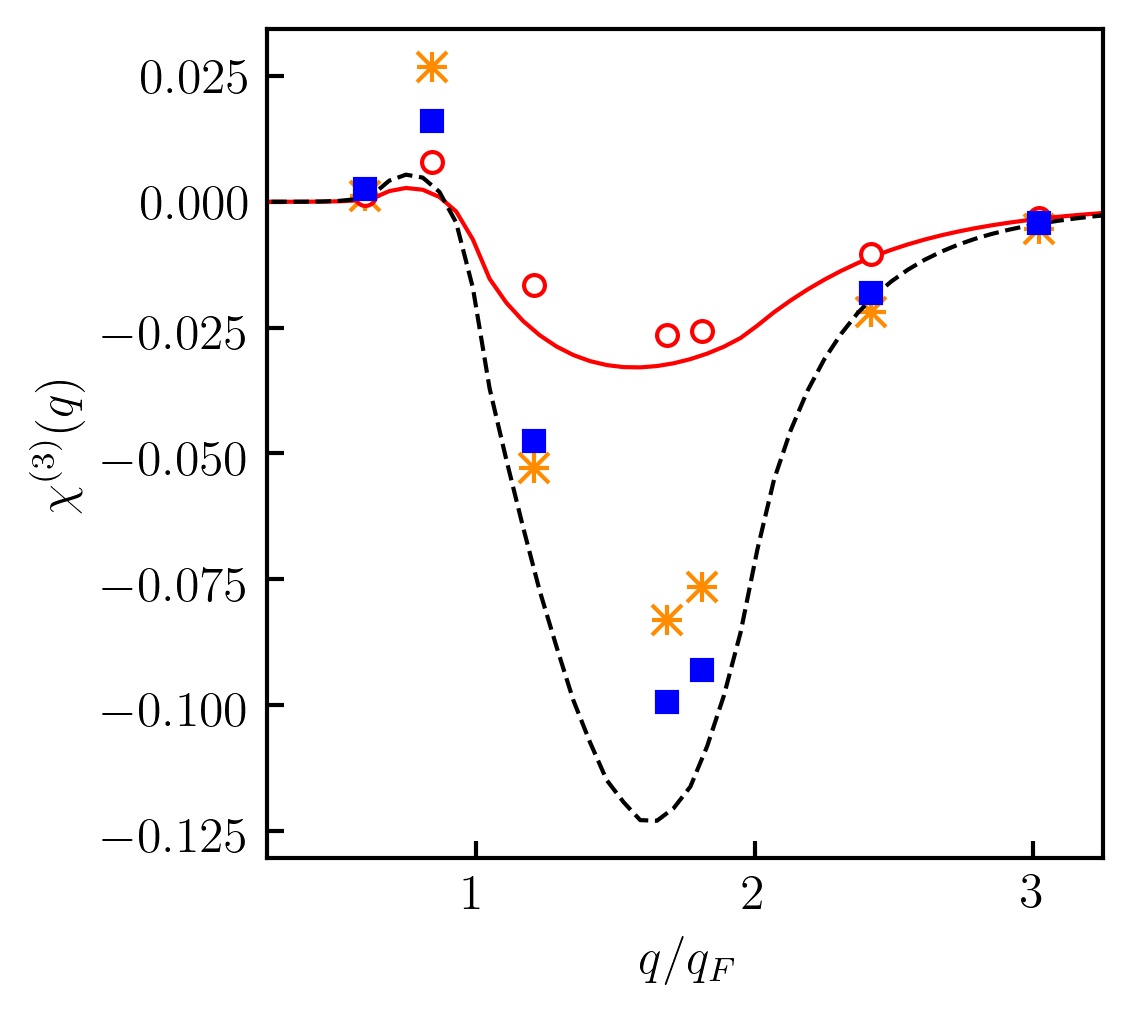}
\endminipage
  \caption{\label{fig:NLR_rs5}
  Non-linear response functions at $r_s=5$ and $\theta=0.01$.
  Left:  The quadratic response function at the second harmonic. 
  Middle: The cubic response function at the first harmonic.
  Right:  The cubic response function at the third harmonic. }
  \end{figure*}



\subsubsection{Non-linear density response in the limit of strong degeneracy}

After successfully testing the accuracy of our KS-DFT simulations on the linear density response function,
we analyze results for the higher order density response functions presented in Fig.~\ref{fig:NLR_rs5}. 
In the left panel, we see that $\widetilde \chi^{(2)}_{\rm NullXC}$ is in excellent agreement  with  $\chi^{(2)}_{\rm RPA}$ and that $\widetilde \chi^{(2)}_{\rm LFC}$ is also reproduces  $\chi^{(2)}_{\rm LFC}$. This confirms the correctness of the the analytical results for the quadratic response function given by Eqs.~(\ref{eq:chi2_RPA}) and (\ref{eq:chi2_LFC}) in the limit of strong degeneracy.  The LDA based data $\widetilde \chi^{(2)}_{\rm LDA}$  provides an adequate description of 
the quadratic response function and captures the effect of the stronger response when XC effects are included compared to the mean-field level results $\chi^{(2)}_{\rm RPA}$ and $\widetilde \chi^{(2)}_{\rm NullXC}$. Certain quantitative disagreements between   $\widetilde \chi^{(2)}_{\rm LDA}$  and $\chi^{(2)}_{\rm LFC}$ can be understood by noting that accurate data for LFC (beyond LDA) is needed to correctly describe the quadratic response.  {In fact, Dornheim \textit{et al.}~\cite{JPSP21} have recently pointed out that the quadratic response is directly related to three-body correlations, which explains this sensitivity to XC-effects.}

The middle panel of Fig. \ref{fig:NLR_rs5} presents data for the cubic response at the first harmonic.
Compared to the partially degenerate case with $\theta=1$, the results exhibit a much sharper peak and much stronger response at $1.5q_F<q<2q_F$.
At these wave numbers, the difference between $\chi^{(1,cubic)}_{\rm LFC}$ and $\widetilde \chi^{(1,cubic)}_{\rm LDA}$ 
 {are most likely a direct consequence of the} 
fact that 
 the  cubic response depends on the fourth power of the LFC, meaning that any deviations from the correct response in the first order gets amplified when a higher order response is considered. 

The right panel of  Fig.~\ref{fig:NLR_rs5}  shows the cubic response at the third harmonic. 
In this case, the theoretical result for the response at the mean-field level given by Eq.~(\ref{eq:chi3_RPA}), $\chi^{(3)}_{\rm RPA}$, fails to reproduce the exact data  $\widetilde \chi^{(3)}_{\rm NullXC}$.
This extends the conclusion that Eq.~(\ref{eq:chi3_RPA}) fails to correctly describe the screened response from the WDM regime considered earlier to the case of strong degeneracy. 
As a consequence, being built upon Eq.~(\ref{eq:chi3_RPA}), the LFC result $\chi^{(3)}_{\rm LFC}$ defined by  Eq.~(\ref{eq:chi3_LFC}) also does not provide the correct description of the cubic response of the correlated electron gas at the third harmonic. 
This makes the analysis based on the comparison of  $\widetilde \chi^{(3)}_{\rm LFC}$ and  $\widetilde \chi^{(3)}_{\rm LDA}$ less meaningful.
Since the LDA is an approximation to the true XC effects, $\widetilde \chi^{(3)}_{\rm LDA}$  cannot be considered to be the exact result. Nevertheless, it provides the correct quantitative outcome. Particularly, we see  from the right panel of  Fig.~\ref{fig:NLR_rs5} that XC effects lead to a stronger response of the system compared to $\chi^{(3)}_{\rm NullXC}$.
We stress that $\widetilde \chi^{(3)}_{\rm NullXC}$ is still the exact ab initio result for the cubic response on the mean-field level. 
Therefore, it can be used to verify theoretical derivations. Once a more accurate theoretical result for $\chi^{(3)}_{\rm RPA}$  {that includes nonlinear screening effects that are neglected in Eq.~(\ref{eq:chi3_RPA})} is derived, the correct way to include the LFC should  
 {directly follow.}

\subsection{Free electron gas at metallic density}
As a particularly important regime from the point of view of applications, we next consider $r_s=2$, which is a characteristic metallic density. 
In this case, we investigate three different values of the degeneracy parameter, namely $\theta=1$, $\theta=0.5$, and $\theta=0.01$.
For $\theta=1$, we have performed series of calculations with $N=14$, $N=20$, and $N=34$.
At $\theta=0.5$, we have considered $N=14$ and $N=34$ particles. At $\theta=0.01$, we have performed simulations with $N=14$, $N=20$, $N=38$, and $N=66$ particles.
In agreement with the calculations in the strongly correlated case,  there is no noticeable finite size effect for $r_s=2$ at these numbers of particles.



\begin{figure*}[!t]
\minipage{0.33\textwidth}
  \includegraphics[width=\linewidth]{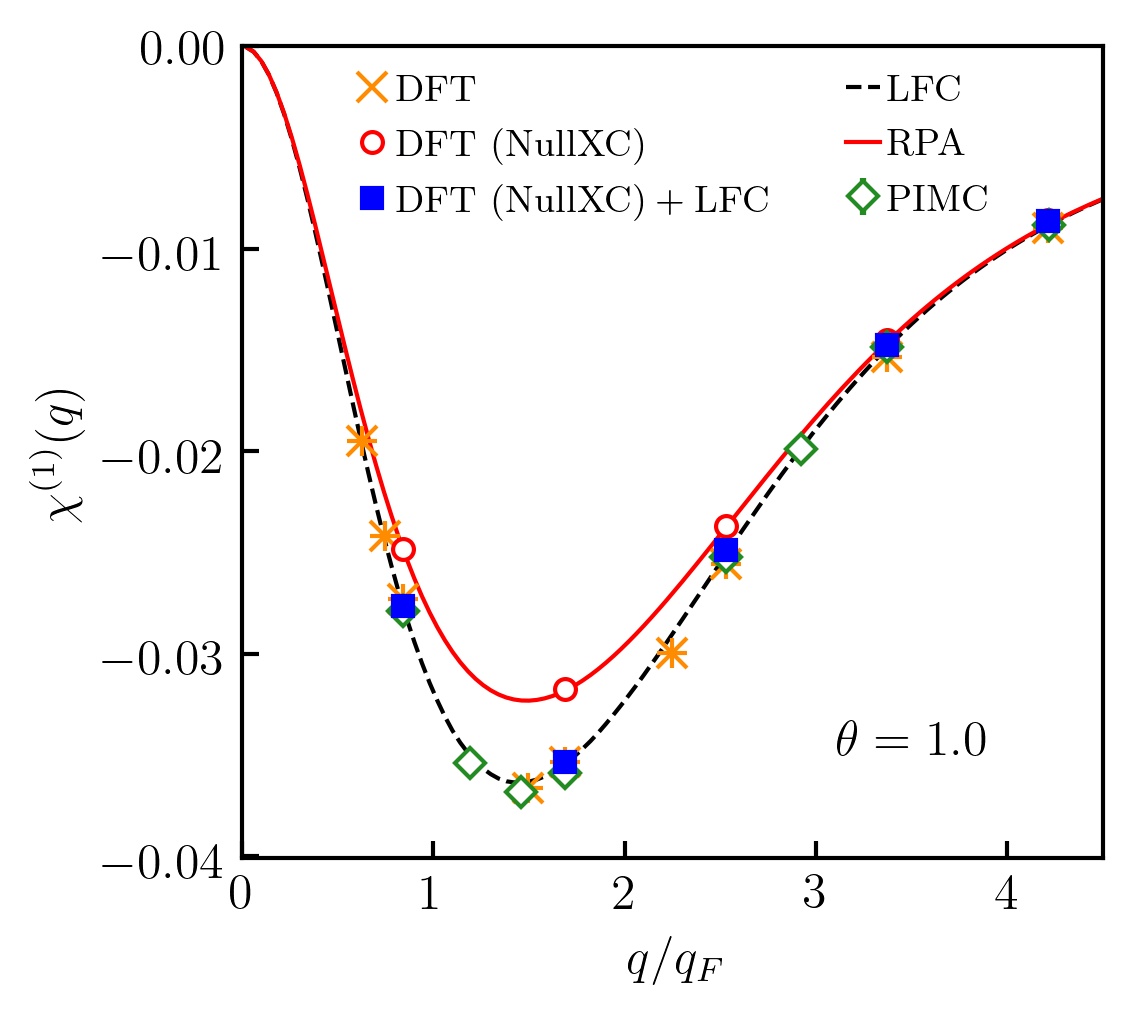}
\endminipage
\minipage{0.33\textwidth}
  \includegraphics[width=\linewidth]{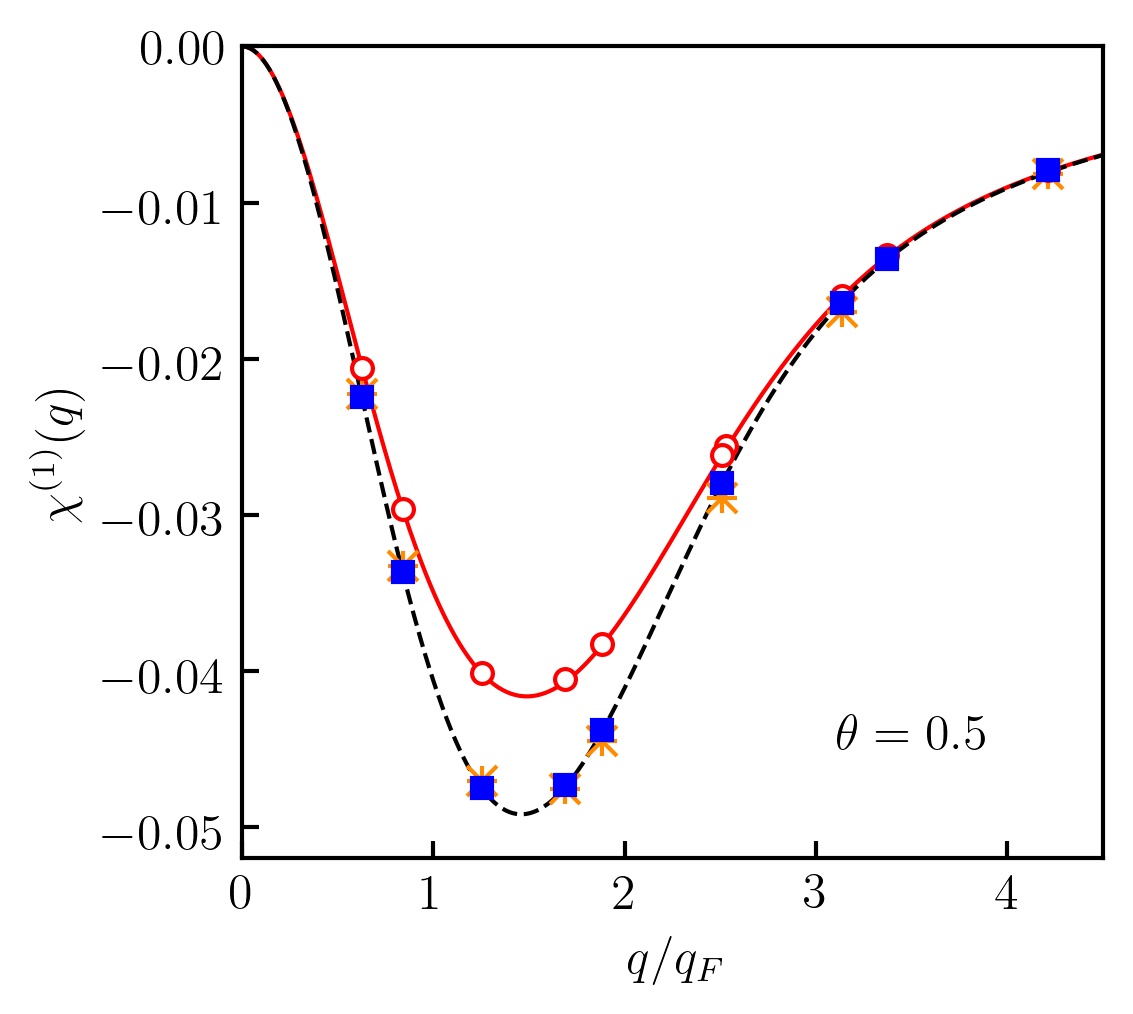}
\endminipage
\minipage{0.33\textwidth}%
  \includegraphics[width=\linewidth]{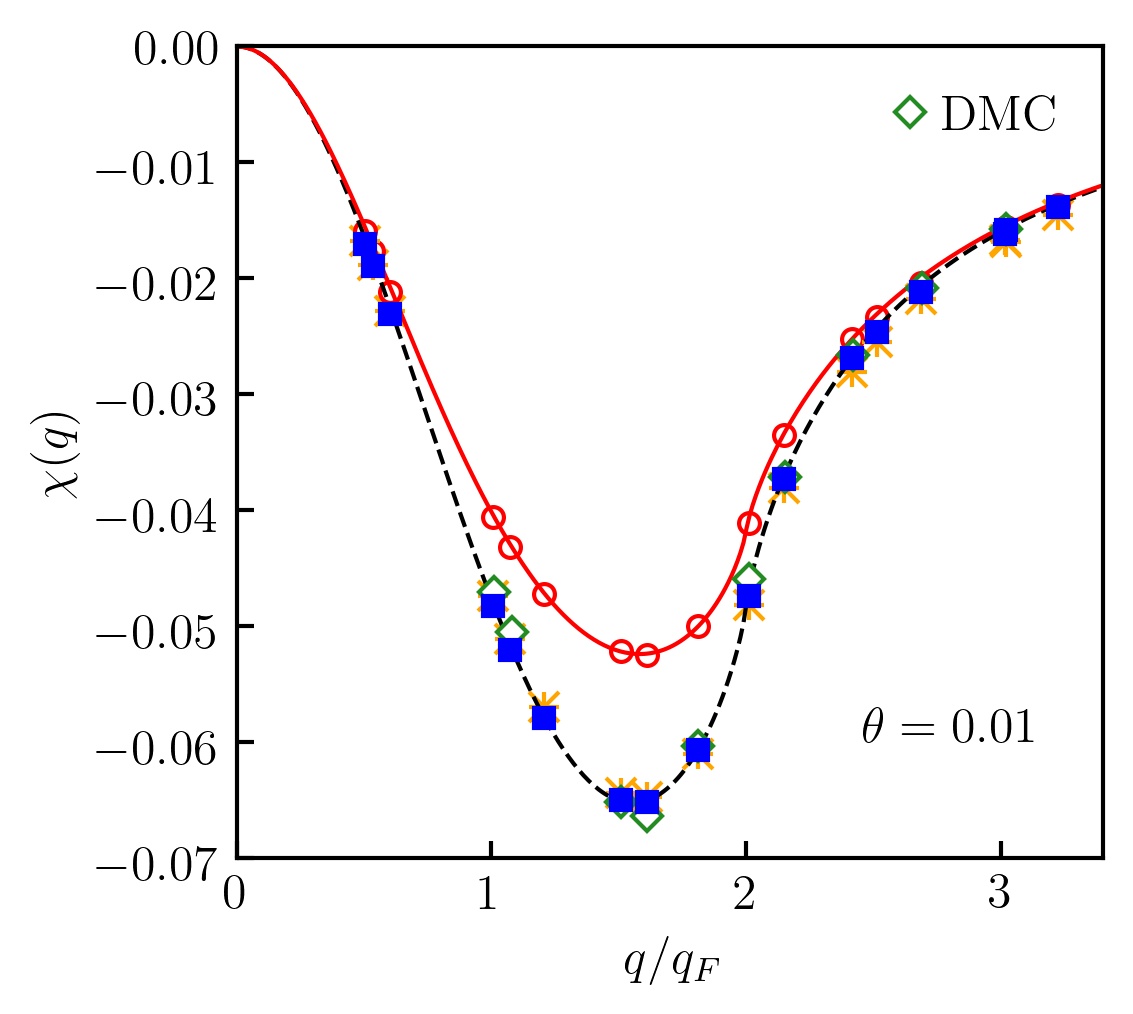}
\endminipage
\caption{\label{fig:chi1_rs2}
Linear static density response function for $r_s=2$  at $\theta=1$ (the left panel), at $\theta=0.5$ (the middle panel), and at $\theta=0.01$ (the right panel). }\label{fig:awesome_image3}
\end{figure*}


\subsubsection{Linear density response}\label{sss:chi1_rs2}

In Fig.~\ref{fig:chi1_rs2}, we present results for the linear density response function at $r_s=2$ and compare the KS-DFT data with PIMC results and with  $\chi^{(1)}_{\rm LFC}$ at $\theta=1$ in the left panel.
We observe that the LDA based results $\widetilde \chi^{(1)}_{\rm LDA}$ are in good agreement with both the PIMC data and $\chi^{(1)}_{\rm LFC}$. 
At $\theta=0.5$, too, we find good agreement of $\widetilde \chi^{(1)}_{\rm LDA}$ with  $\chi^{(1)}_{\rm LFC}$ as it can be seen from the middle panel of Fig.~\ref{fig:chi1_rs2}. In the limit of $\theta\to0$, we compare  $\widetilde \chi^{(1)}_{\rm LDA}$ with the DMC data by Moroni \textit{et al.}~\cite{PhysRevLett.75.689} as well as with $\chi^{(1)}_{\rm LFC}$. Evidently, the LDA based KS-DFT simulations provide an accurate description of the linear density response function in the strongly degenerate case too.  {We note that KS-DFT is more accurate in particular for $q>2q_\textnormal{F}$ compared to the previously considered case of $r_s=6$ due to the reduced impact of electronic XC-effects at the higher density.}

The comparison of  $\widetilde \chi^{(1)}_{\rm NullXC}$ with $\chi^{(1)}_{\rm RPA}$ that is also presented in  Fig.~\ref{fig:chi1_rs2} confirms the high accuracy of the KS-DFT results for the description of the linear response function on the mean-field level across temperature regimes. As a consequence,  $\widetilde \chi^{(1)}_{\rm LFC}$ is in an excellent agreement with $\chi^{(1)}_{\rm LFC}$  {over the entire wavenumber range}.




\begin{figure*}[!t]
\minipage{0.37\textwidth}
  \includegraphics[width=\linewidth]{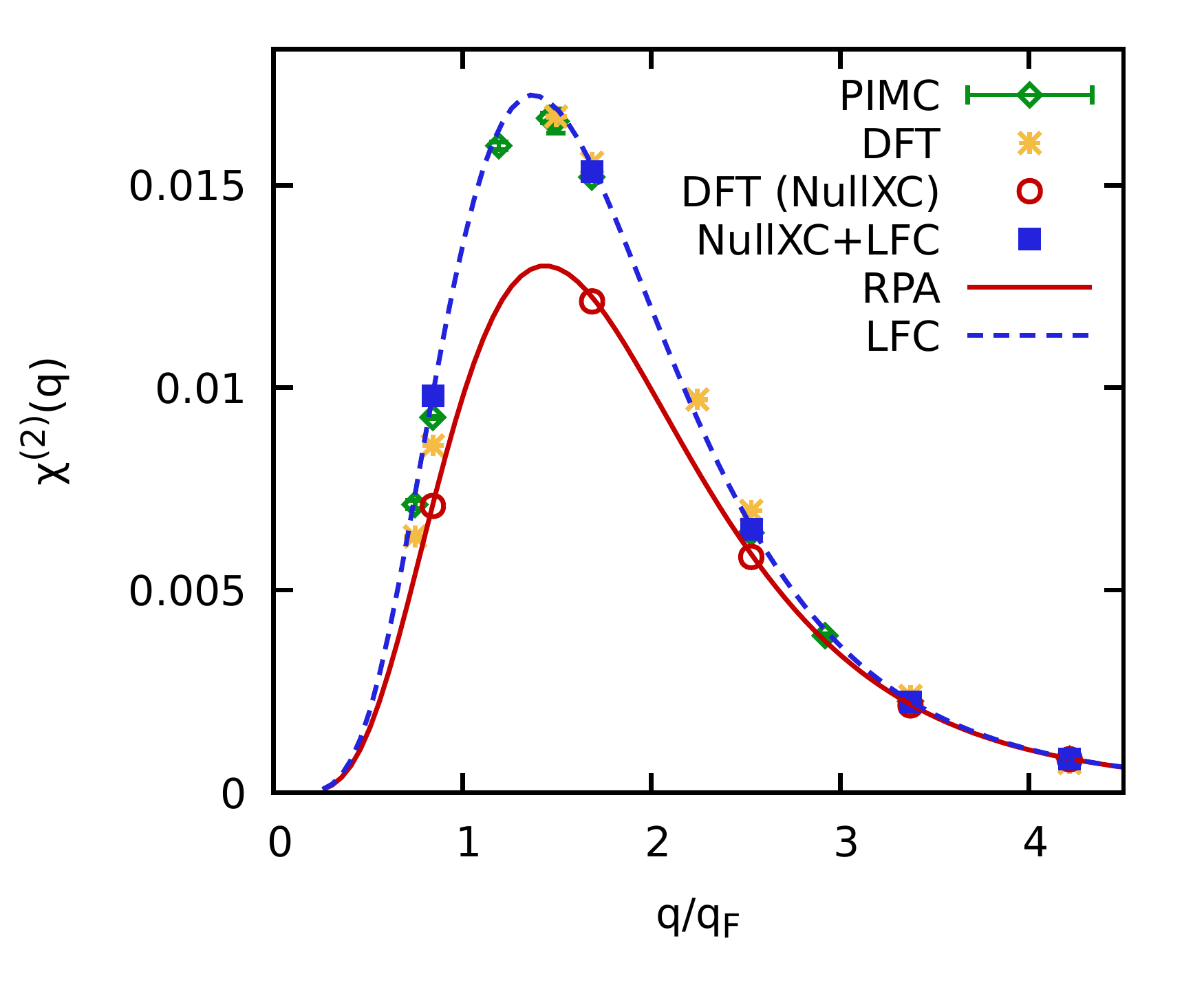}
\endminipage
\minipage{0.32\textwidth}
  \includegraphics[width=\linewidth]{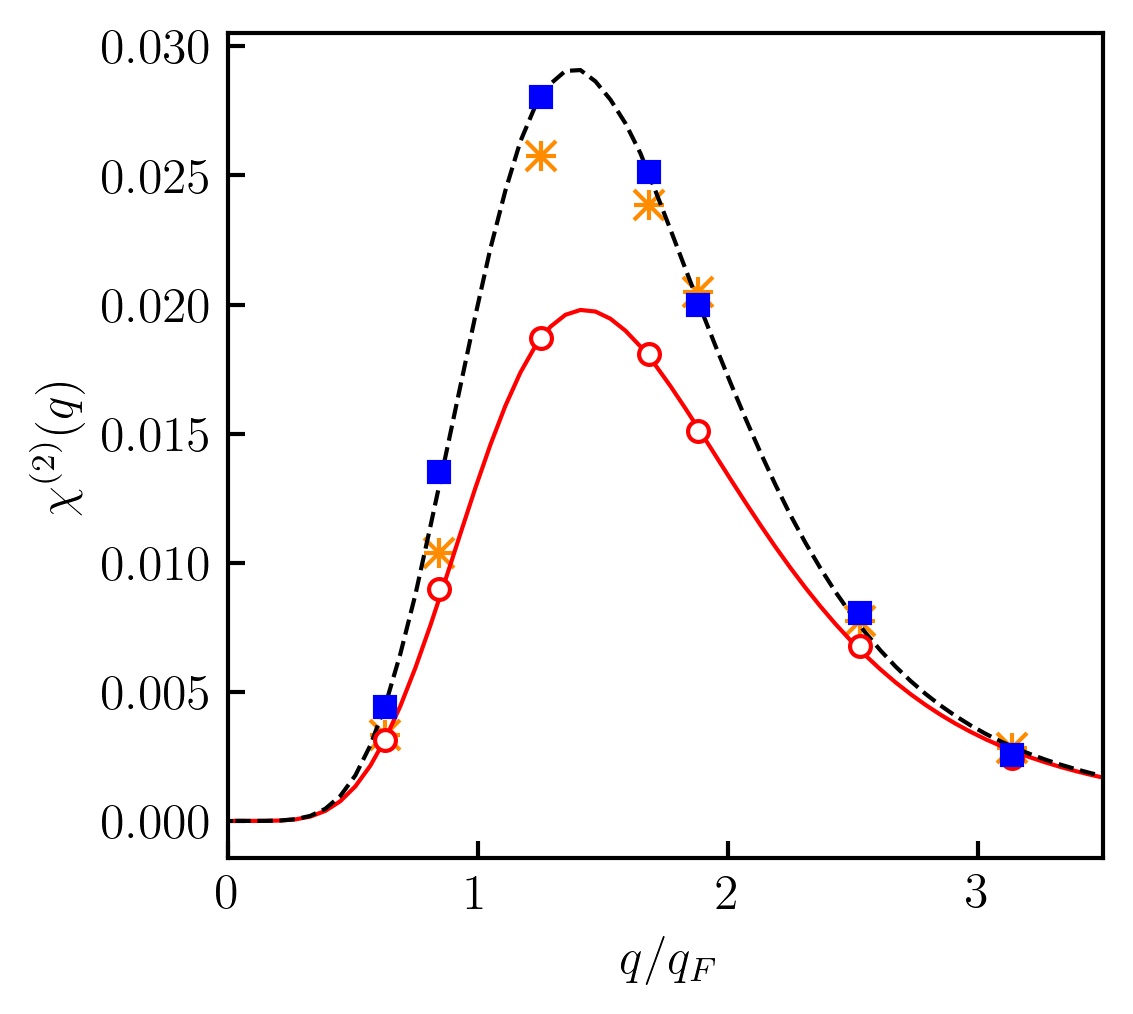}
\endminipage
\minipage{0.32\textwidth}%
  \includegraphics[width=\linewidth]{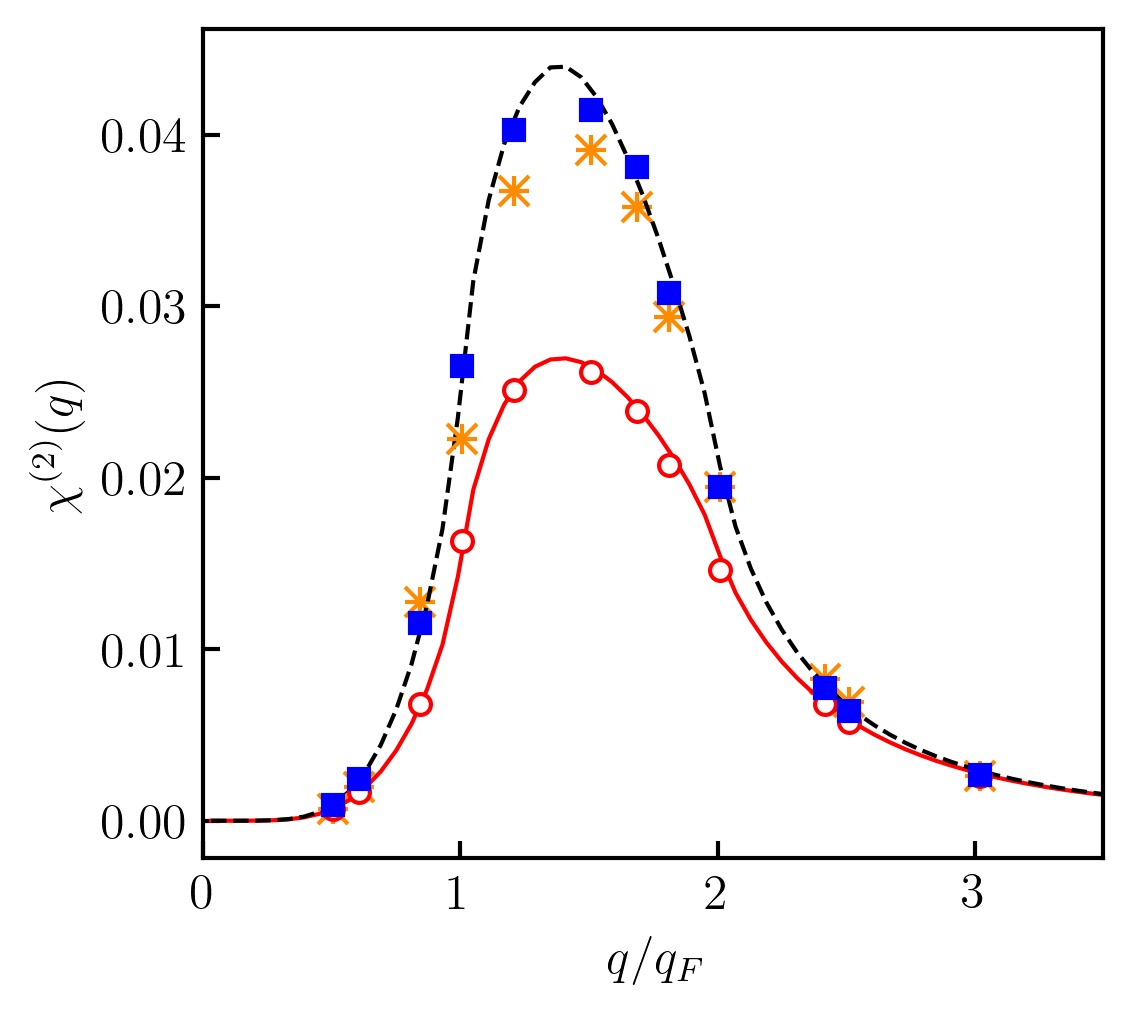}
\endminipage
\caption{Quadratic static density response function for $r_s=2$  at $\theta=1$ (the left panel), at $\theta=0.5$ (the middle panel), and at $\theta=0.01$ (the right panel).}\label{fig:chi2_rs2}
\end{figure*}



\subsubsection{Quadratic density response}\label{sss:chi2_rs2}

The results for the quadratic response function at $r_s=2$ are shown in Fig.~\ref{fig:chi2_rs2}.
The quadratic response $\widetilde \chi^{(2)}_{\rm LDA}$  closely reproduces both the PIMC data and $\chi^{(2)}_{\rm LFC}$ at $\theta=1$ as it is demonstrated in the left panel of  Fig.~\ref{fig:chi2_rs2}. The agreement between  $\widetilde \chi^{(2)}_{\rm LDA}$ and $\chi^{(2)}_{\rm LFC}$  somewhat deteriorates with the decrease in the temperature from $\theta=1$ to $\theta=0.5$ and further to $\theta=0.01$.  This is shown in the middle and right panels of   Fig.~\ref{fig:chi2_rs2}. Nevertheless, the LDA, which is an XC functional purely based on the uniform electron gas model, provides an overall impressively accurate description of the quadratic response function at all considered wave numbers of the perturbation. 

Similarly to the discussed cases of the strongly coupled electrons,  $\widetilde \chi^{(2)}_{\rm NullXC}$ and $\chi^{(2)}_{\rm RPA}$ are in close agreement with each other 
at $\theta=1$, $\theta=0.5$, and $\theta=0.01$. This is a clear illustration of the high accuracy of the theoretical result Eq.~(\ref{eq:chi2_RPA}) for the mean-field description.
Consequently, we find almost the same result using $\widetilde \chi^{(2)}_{\rm LFC}$ and $\chi^{(2)}_{\rm LFC}$.

From comparing amplitudes of the quadratic response function in  Fig.~\ref{fig:chi2_rs2} at different temperatures, one can see that the response of the system at the second harmonic becomes stronger upon decreasing the temperature of the electrons. For example, the decrease of the temperature from the partially degenerate case ($\theta=1$) to the strongly degenerate case (here represented by $\theta=0.01$) leads to an increase of the maximum value of the quadratic response function by a factor of 2.5.  For comparison, the amplitude of the linear response function increases about two times with the decrease of the temperature from $\theta=1$ to $\theta=0.01$ at the same conditions.



\begin{figure*}[!t]
\minipage{0.37\textwidth}
  \includegraphics[width=\linewidth]{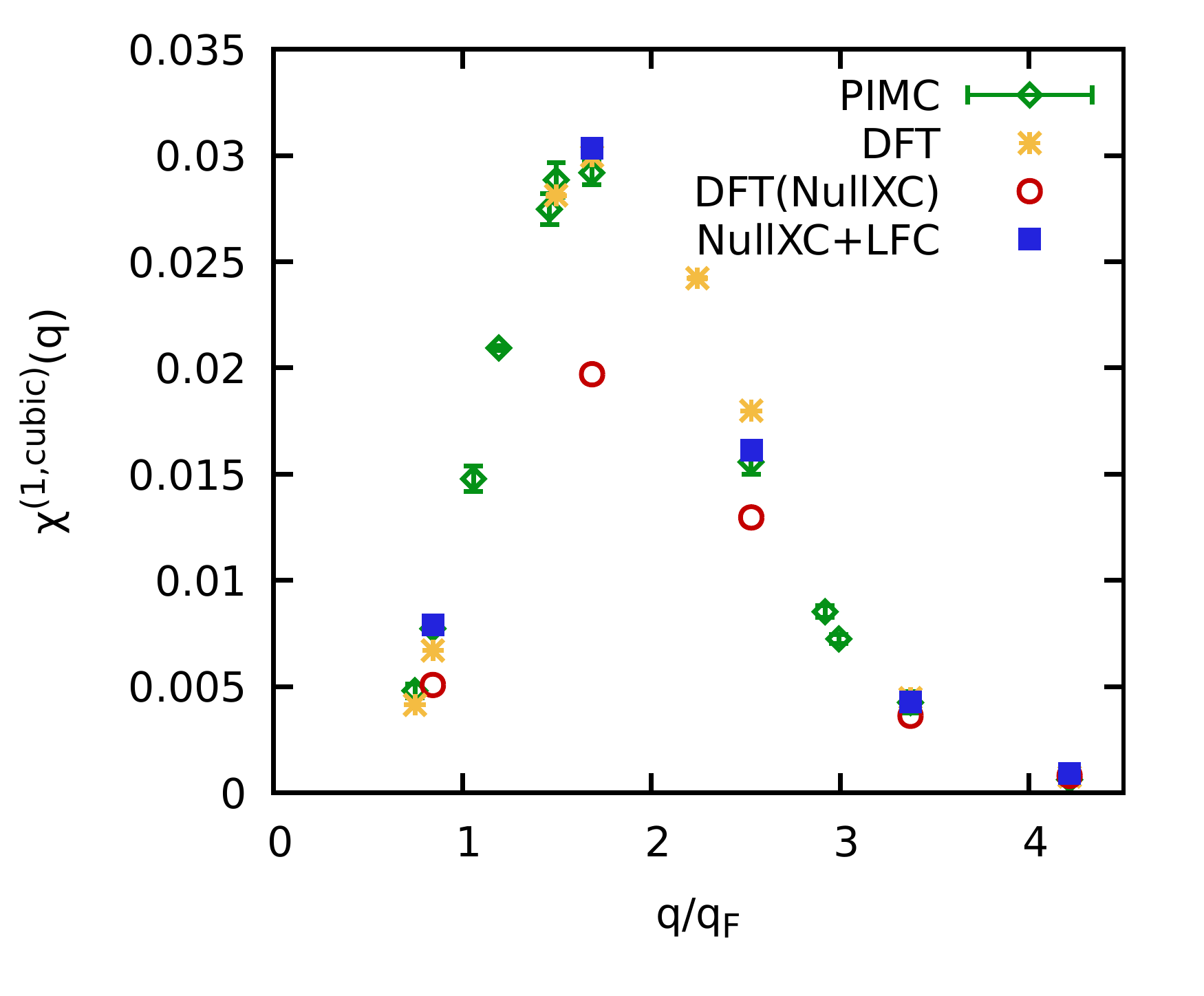}
\endminipage
\minipage{0.32\textwidth}
  \includegraphics[width=\linewidth]{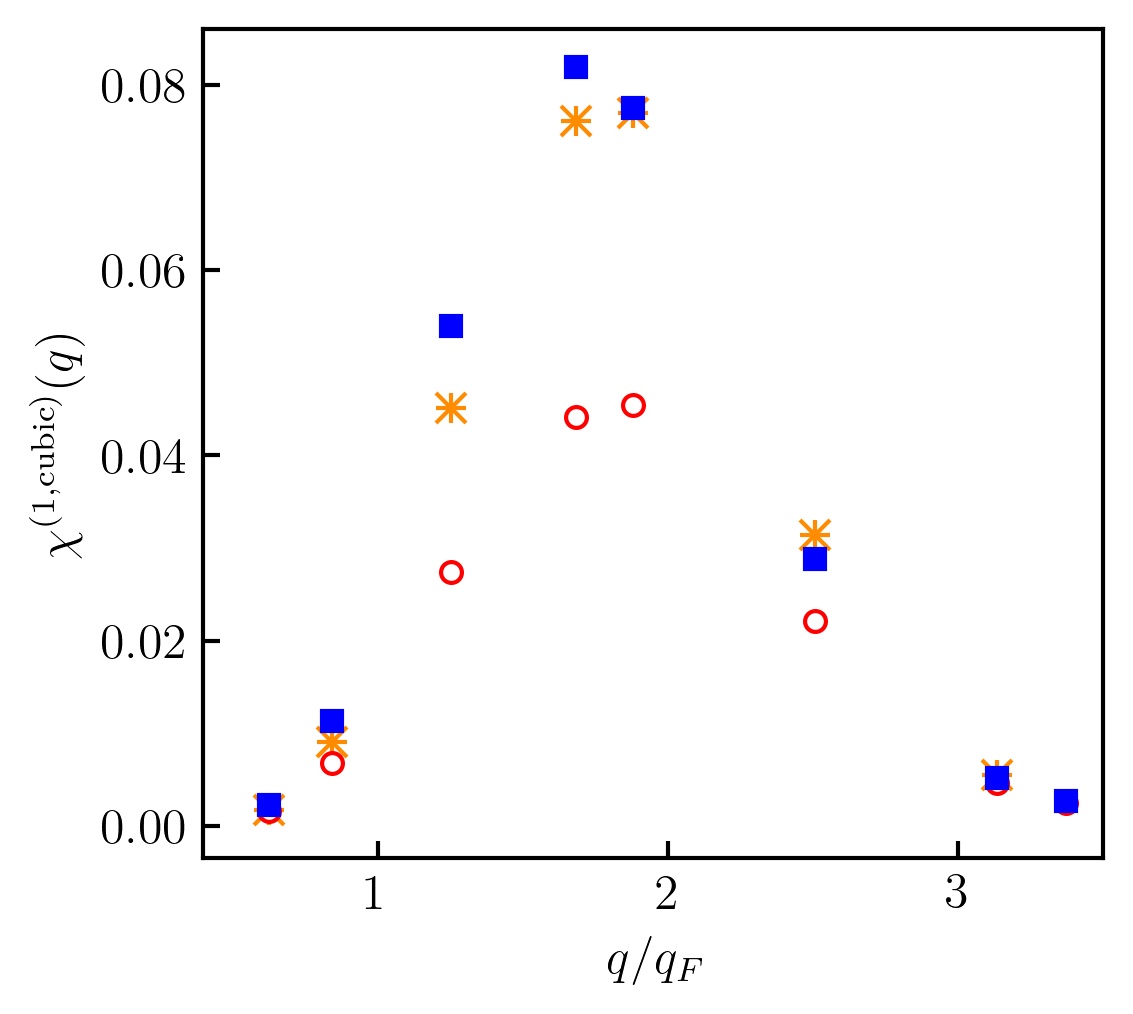}
\endminipage
\minipage{0.32\textwidth}%
  \includegraphics[width=\linewidth]{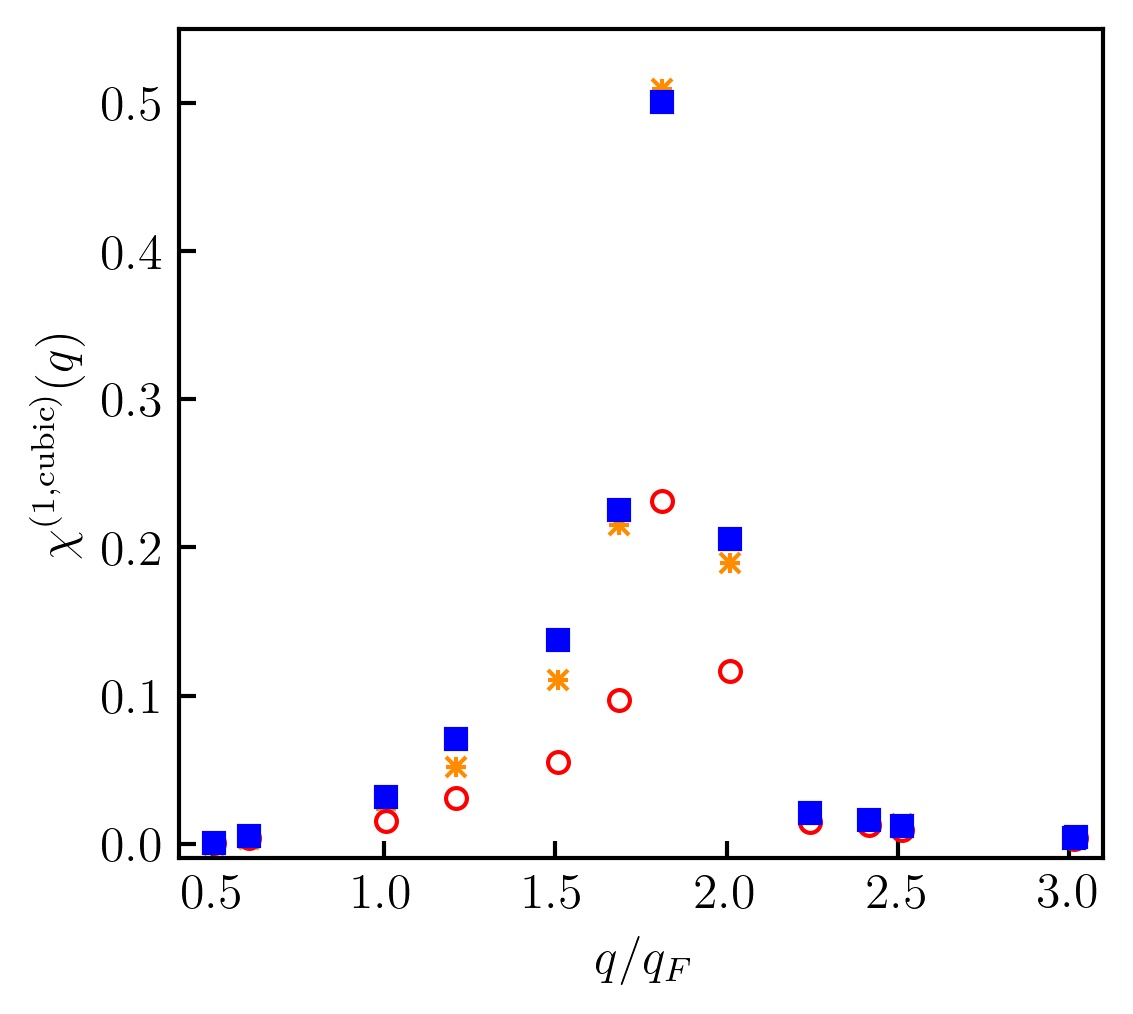}
\endminipage
\caption{Cubic static density response function at the first harmonic for $r_s=2$  at $\theta=1$ (the left panel), at $\theta=0.5$ (the middle panel), and at $\theta=0.01$ (the right panel).}\label{fig:chi13_rs2}
\end{figure*}



\subsubsection{Cubic density response at the first harmonic}\label{sss:chi13_rs2}

Let us next investigate the cubic response function at the first harmonic, which is shown in Fig.~\ref{fig:chi13_rs2}.
The comparison with the PIMC data at $\theta=1$ is provided in the left panel of Fig.~\ref{fig:chi13_rs2}.
From this comparison it is evident that the LDA based KS-DFT calculations are able to provide a very accurate description of the cubic response at $r_s=2$ and $\theta=1$, 
as they are in good agreement to both the PIMC data and $\widetilde \chi^{(1,\textnormal{cubic})}_{\rm LFC}$. This is an indication that the relation Eq.~(\ref{eq:cubic_first_LFC}) holds, since $\widetilde \chi^{(1,\textnormal{cubic})}_{\rm LFC}$ is computed using the KS-DFT data $\widetilde \chi^{(1,\textnormal{cubic})}_{\rm NullXC}$ and the LFC according to Eq.~(\ref{eq:cubic_first_DFT}).

Decreasing the electronic temperature leads to a substantial increase in the amplitude of the cubic response function at the first harmonic.
This is visible from the comparison of the amplitudes of the cubic response at $\theta=1$ (left), at $\theta=0.5$ (middle), and at $\theta=0.01$ (right) in Fig.~\ref{fig:chi13_rs2}. For example, the maximum of the cubic response function at the first harmonic at $\theta=0.01$ is about 16 times larger than at $\theta=1$.
The consequences of this remarkable behavior are discussed in more detail in Sec. 5.



\begin{figure*}[!t]
\minipage{0.37\textwidth}
  \includegraphics[width=\linewidth]{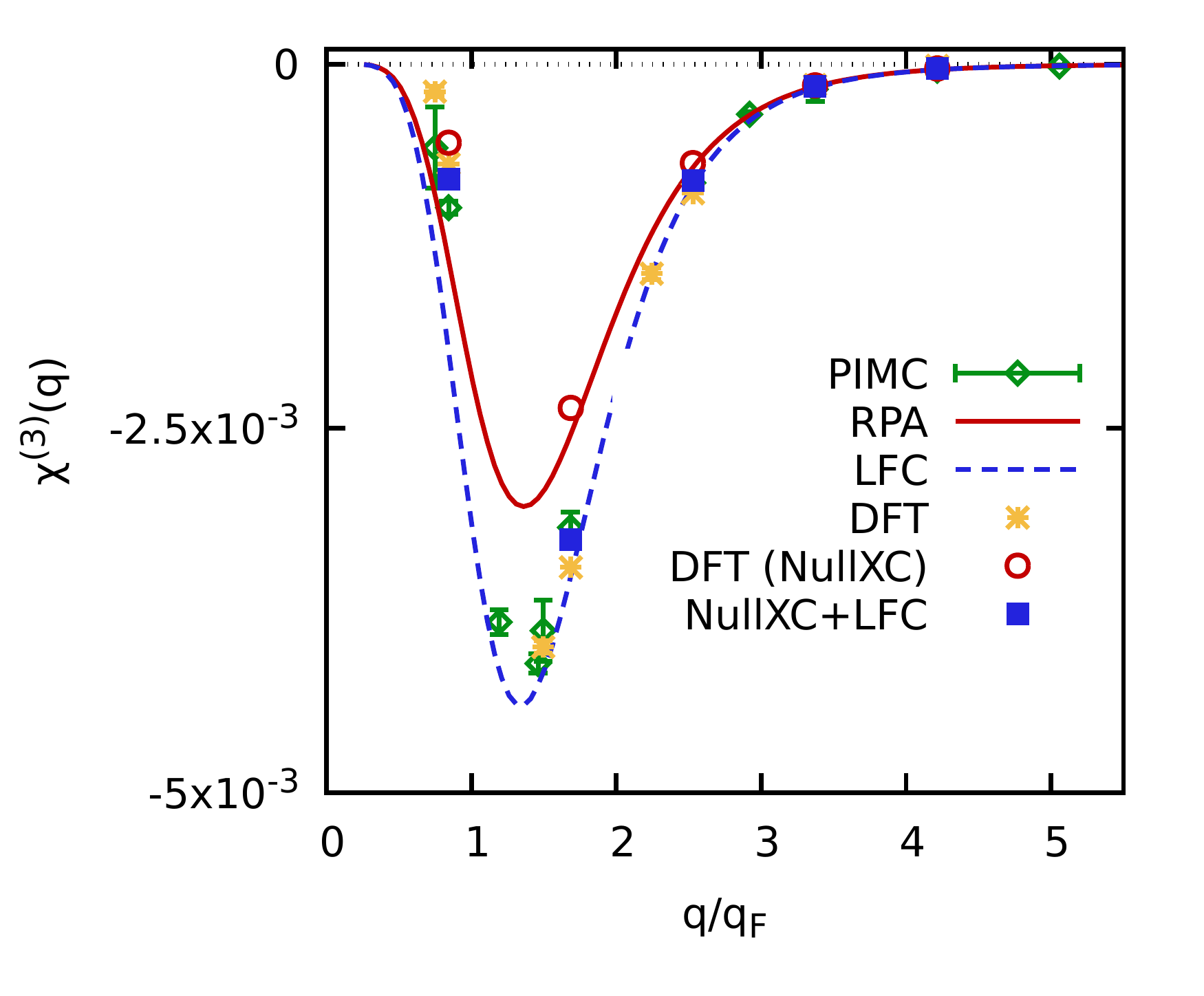}
\endminipage
\minipage{0.32\textwidth}
  \includegraphics[width=\linewidth]{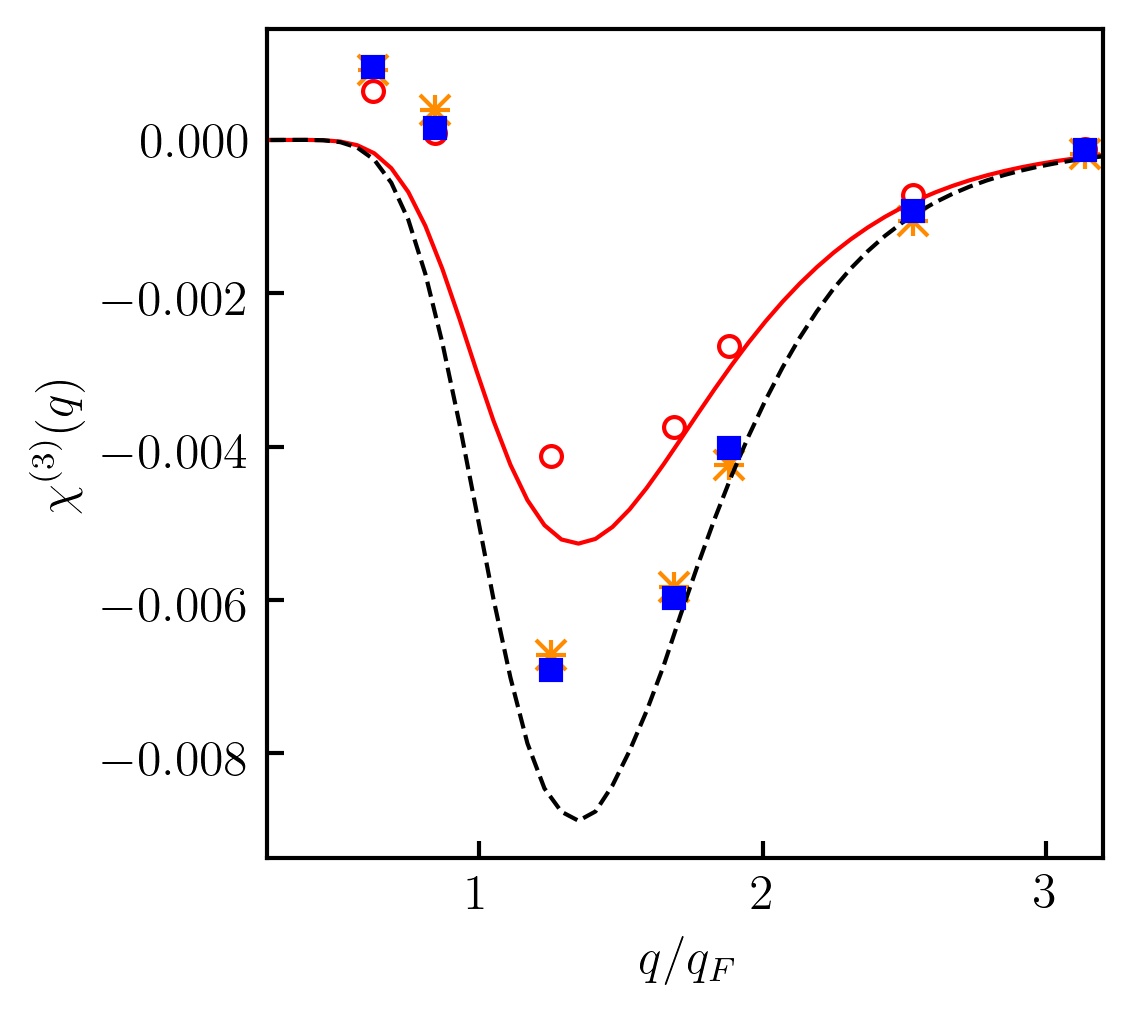}
\endminipage
\minipage{0.32\textwidth}%
  \includegraphics[width=\linewidth]{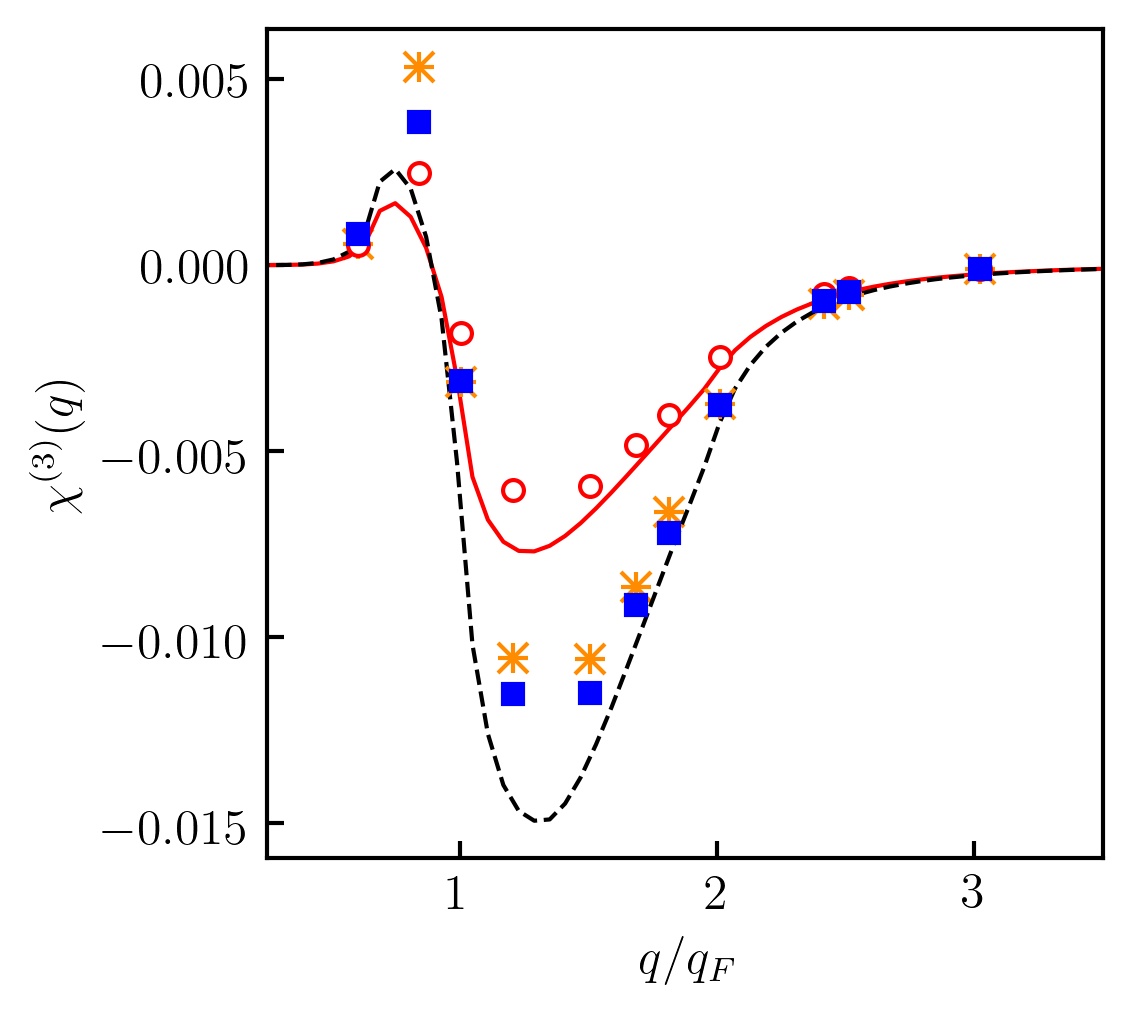}
\endminipage
\caption{Cubic static density response function at the third harmonic for $r_s=2$  at $\theta=1$ (the left panel), at $\theta=0.5$ (the middle panel), and at $\theta=0.01$ (the right panel).}\label{fig:chi33_rs2}
\end{figure*}



\subsubsection{Cubic density response at the third harmonic}\label{sss:chi33_rs2}

Finally, we present the results for the cubic response at the third harmonic in Fig.~\ref{fig:chi33_rs2}, and the left panel depicts the comparison with PIMC data at $\theta=1$.
Evidently, the LDA based calculations $\widetilde \chi^{(3)}_{\rm LDA}$ are in good agreement with the latter.
Next, $\chi^{(3)}_{\rm RPA}$ overestimates the strength of response compare  to $\widetilde \chi^{(3)}_{\rm NullXC}$ at $q<q_F$.
In contrast to the response at stronger coupling ($r_s=6$ and $\theta=1$), at $r_s=2$ we do not see the change in the sign of the cubic response at the third harmonic upon increasing the wave number from small $q<q_F$ to large $q>q_F$ values. However, this behavior is restored with a decrease in the temperature to $\theta=0.5$ as it is clearly visible in the middle panel of Fig.~\ref{fig:chi33_rs2}. Importantly, such behavior at $\theta=0.5$ is not captured by the analytical results given in Eqs.~(\ref{eq:chi3_RPA}) and (\ref{eq:chi3_LFC}). At $q<1.5q_F$,  this leads to a significant disagreement between the analytical result in the mean field-level approximation, Eq.~(\ref{eq:chi3_RPA}),  and the exact result computed using null XC in the KS-DFT simulations. 
However, in the limit of strong degeneracy depicted in the right panel of Fig.~\ref{fig:chi33_rs2}, Eq.~(\ref{eq:chi3_RPA}) again provides the qualitatively correct result by reproducing the change in the sign of the corresponding response function
, but remains in quantitative disagreement with $\widetilde \chi^{(3)}_{\rm NullXC}$ in the vicinity of the positive maximum and the negative minimum.

Interestingly, despite the poor performance of Eq.~(\ref{eq:chi3_RPA}), the agreement between $\widetilde \chi^{(3)}_{\rm LDA}$  and $\widetilde \chi^{(3)}_{\rm LFC}$ is rather good at all considered temperature regimes. This is explained by the fact that XC effects are less pronounced compared to the above considered strongly correlated cases.

\section{Conclusions and Outlook}\label{s:end}

In this work we have  {explored a new methodology for the study of the nonlinear electronic density response based on KS-DFT. As a particular example, we have investigated the free electron gas model and} demonstrated that the KS-DFT method is an effective and valuable tool for the investigation of various
nonlinear electronic density response functions across temperature regimes. This conclusion is important for parameters where quantum Monte-Carlo methods experience significant difficulties or fail to converge at all due to the fermion sign problem. This approximately corresponds to $\theta<1$ and $r_s\gtrsim 2$ \cite{dornheim_physrep_18, JCP_Lee, Yilmaz}. A particularly effective method to gauge and guide the 
 {development of new theoretical approaches}
is given by the KS-DFT simulation with zero XC functional.
This is due to the fact that theoretical models of correlated electrons are built upon mean-field approximations, in combination with the electronic LFC. 
 Therefore, a disqualification of the analytical results for the  non-linear density response functions in the mean-field approximation automatically rules out the applicability of the results obtained using LFC.

As a demonstration of the KS-DFT method based analysis of the theoretical results, we have considered the quadratic and cubic response functions at different values of the density and degeneracy parameters. First of all, we have confirmed the validity of the analytical results for the quadratic response function [Eqs.~(\ref{eq:chi2_RPA}) and (\ref{eq:chi2_LFC})] for partially to strongly degenerate electrons. This confirms and complements the earlier PIMC based analysis  at $\theta\geq 1$ \cite{PhysRevResearch.3.033231}.  Secondly, it has been shown that the analytical results for the cubic response function at the third harmonic [Eqs.~(\ref{eq:chi3_RPA}) and (\ref{eq:chi3_LFC})] are quantitatively inaccurate at $q\lesssim 1.5~ q_F$ for all considered values of $\theta$. Moreover, Eqs.~(\ref{eq:chi3_RPA}) and (\ref{eq:chi3_LFC}) are qualitatively inadequate at $\theta=0.5$.

The application of the KS-DFT method to study the cubic response at the first harmonic [as defined in Eq.~(\ref{eq:rho1})] has allowed us to observe a change of its characteristics with the decrease of the temperature to $\theta<1$. We have revealed that the decrease of the temperature from the partially degenerate regime with $\theta\sim 1$ to the strongly degenerate regime ($\theta\ll1$) leads to a significant increase in the maximum of the cubic response at the first harmonic.
Let us demonstrate the implication of this finding for electrons at a metallic density, $r_s=2$.
Using Eqs.~(\ref{eq:rho1})-(\ref{eq:rho3}) and the data presented in Figs. \ref{fig:chi1_rs2}-\ref{fig:chi33_rs2},
one can deduce that, at $r_s=2$, both the quadratic response and the cubic response at the third harmonic remain inferior to the linear density response function if the perturbation amplitude $A<1$,  at all considered $\theta$ values and wave numbers of the perturbation. In contrast, the cubic response function at the first harmonic becomes dominant over the linear response function if $A>0.8$ at $\theta=0.5$ and if $A>0.36$ at $\theta=0.01$. The applicability of the perturbative analysis requires the smallness of the higher order correction compared to the first order term in Eq.~(\ref{eq:rho1}).
Therefore, at $\theta\ll1$, not the quadratic response, but the cubic response at the first harmonic leads to the strongest restriction on the applicability of the non-linear density response theory of free electrons with respect to the perturbation amplitude. 

Another important finding is that the LDA XC functional provides a  {remarkably} accurate description of the linear and nonlinear density response at metallic density, $r_s=2$, across the entire considered temperature range. 
 {This strongly indicates that our new KS-DFT based approach constitutes a reliable tool for the investigation of the nonlinear density response both at ambient conditions and in the WDM regime.}
At stronger coupling parameters, $r_s=6$ and $r_s=5$, the LDA XC functional based KS-DFT calculations of the correlated electron gas provide qualitatively correct behavior but ought to be considered with caution from a quantitative point of view.  

We conclude this study by pointing out that  {the present methodology is very general and can be directly applied to arbitrary materials; the only approximation is given the choice of the XC-functional. In particular,} simulations including ions are much more problematic and computationally expensive for the quantum Monte-Carlo methods compared to the considered case of a free electron gas model. Therefore, the KS-DFT method is particularly valuable for multi-component systems.
This proof of concept of the capability of KS-DFT for the estimation of the NLRT of an electron gas is thus a pivotal first step before extending our considerations to real materials.


\section*{Acknowledgments}

This work was funded by the Center for Advanced Systems Understanding (CASUS) which is financed by the German Federal Ministry of Education and Research (BMBF) and by the Saxon Ministry for Science, Art, and Tourism (SMWK) with tax funds on the basis of the budget approved by the Saxon State Parliament. We gratefully acknowledge computation time at the Norddeutscher Verbund f\"ur Hoch- und H\"ochstleistungsrechnen (HLRN) under grant shp00026, and on the Bull Cluster at the Center for Information Services and High Performance Computing (ZIH) at Technische Universit\"at Dresden.

\section*{Author Declarations}
\subsection{Conflict of Interest}
The authors have no conflicts to disclose.


\section*{Appendix: Illustration of the extraction of response functions from the density perturbation}\label{sec:app}

Here we illustrate the extraction of the density response functions from the  density perturbation components in Fourier space $\braket{\hat\rho_\mathbf{k}}_{q,A}$. As an example, we consider $r_s=2$ and $\theta=1$. The results are computed by performing KS-DFT simulations of $N=14$ electrons in the main cell using the LDA XC functional. 

\begin{figure}
\center
\includegraphics [width=0.4 \textwidth]{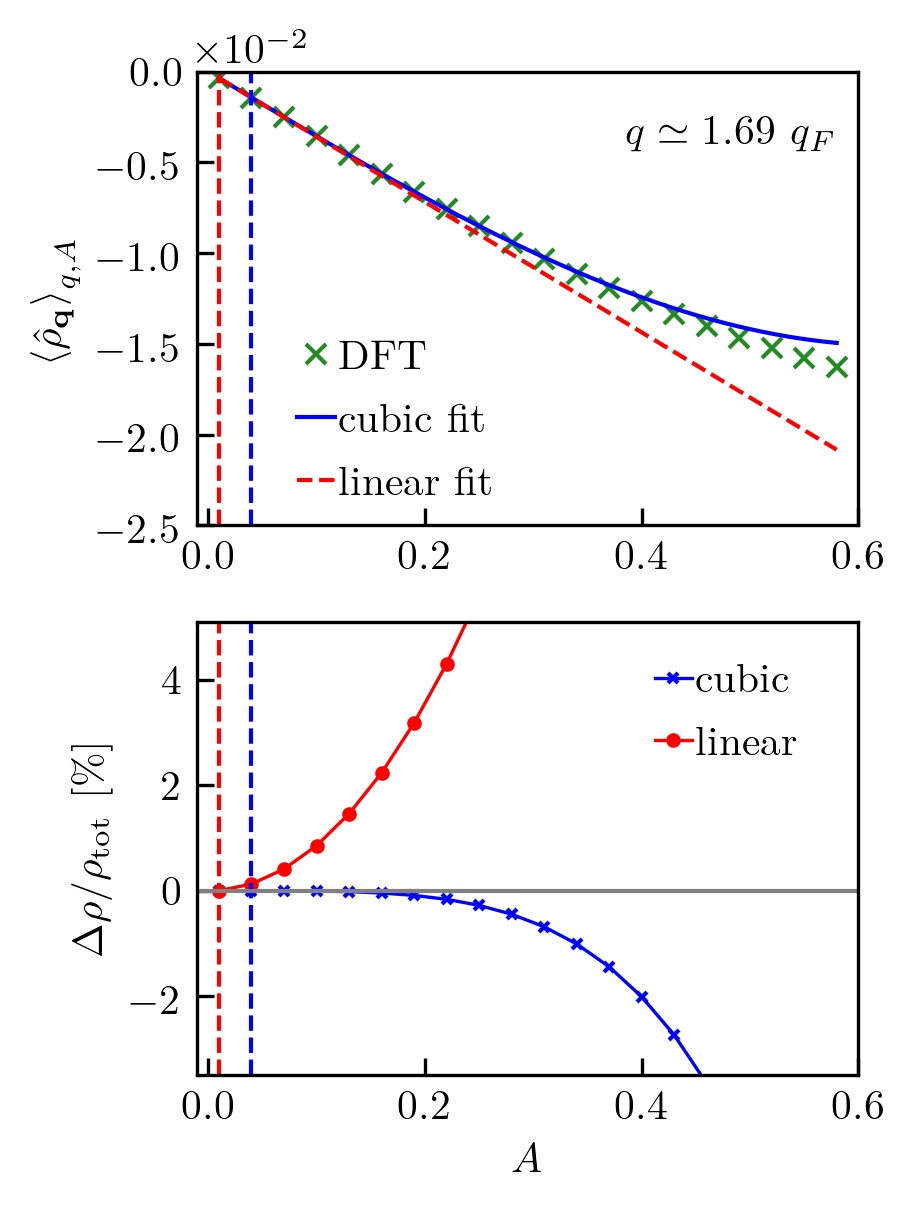}
\caption{\label{fig:rho1}
The density perturbation component  $\braket{\hat\rho_\mathbf{k}}_{q,A}$ at $k=q$ is used to find the linear density response and the cubic density response at the first harmonic at $r_s=2$ and $\theta=1$.}
\end{figure}

In the top panel of Fig.~\ref{fig:rho1}, we show the density perturbation at the first harmonic ($k=q$) with the perturbation wave number $q\simeq 1.69 q_F$.
 The results for 
Eq.~(\ref{eq:rho1}) are given by the solid line, where $\chi^{(1)}$ and $\chi^{(1,\textnormal{cubic})}$ in Eq.~(\ref{eq:rho1}) are found using the first two points (as indicated by the vertical lines).  
This corresponds to a cubic fit which takes into account both the linear density response and the cubic density response at the first harmonic.
In this way, we extract data for these response functions. For comparison, we also  show the linear response result neglecting the cubic contribution (dashed line). We see that the inclusion of the cubic contribution allows us to extend the description of the density response beyond $A=0.16$ up to $A=0.4$ (with an error less than $2~\%$). 
The corresponding values of the relative difference between the simulation results and the data computed using  Eq.~(\ref{eq:rho1}) are depicted in the bottom panel of Fig.~\ref{fig:rho1}, where the difference is divided by the value of the total density perturbation $\rho_{\rm tot}$. The latter allows us to estimate the resulting error in the description of the total density.

\begin{figure}
\center
\includegraphics [width=0.4 \textwidth]{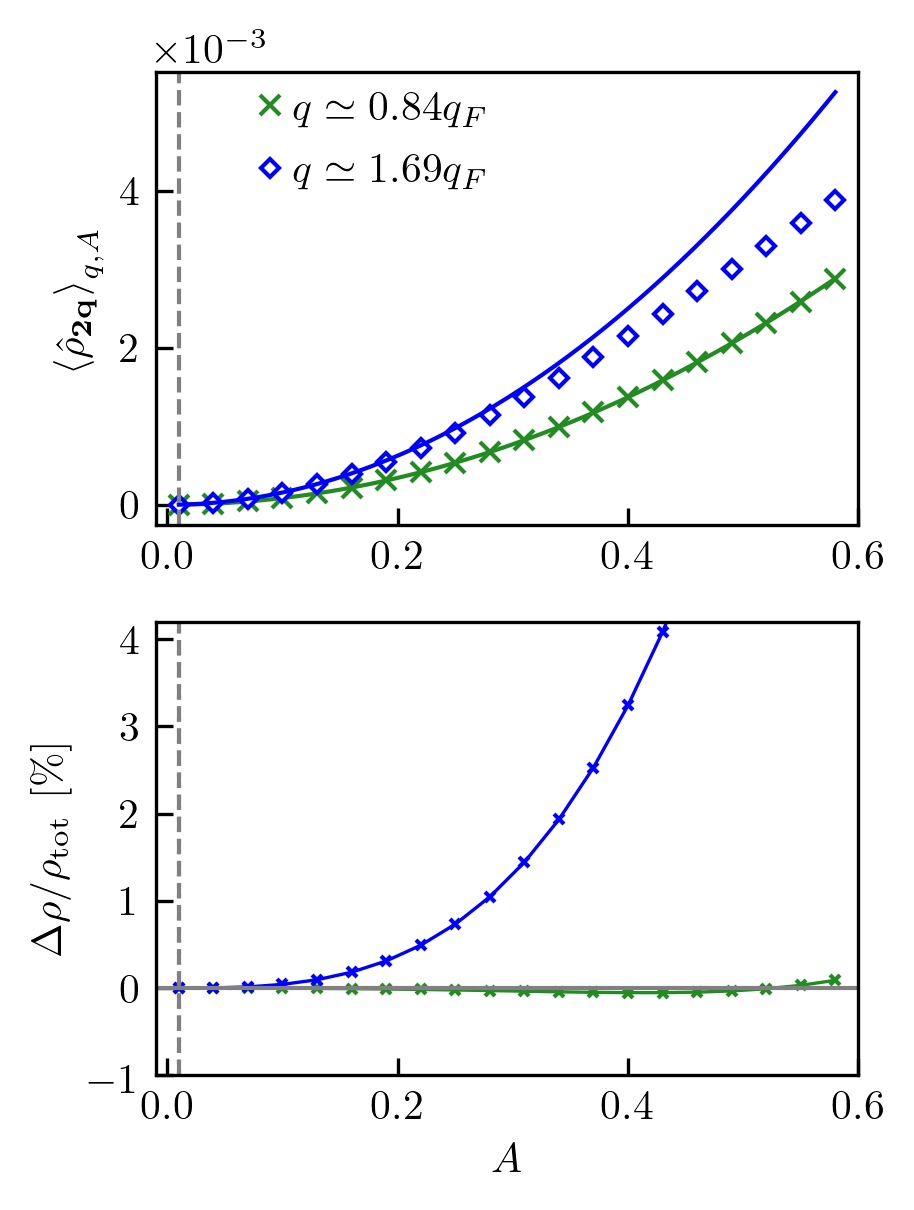}
\caption{\label{fig:rho2}
The density perturbation component  $\braket{\hat\rho_\mathbf{k}}_{q,A}$ at $k=2q$ is used to find the quadratic density response at $r_s=2$ and $\theta=1$.}
\end{figure} 

 {A similar analysis is presented for the density response at the second harmonic of the original perturbation ($k=2q$) in Fig.~\ref{fig:rho2}, and the top panel depicts}
 {results for the} perturbation wave numbers $q\simeq 1.69 q_F$ and $q\simeq 0.84 q_F$. 
The KS-DFT results are computed using the perturbation amplitudes in the range  $0.01\leq A\leq0.6$. The solid lines show the fit according to Eq.~(\ref{eq:rho2}) based only on the first data point at $A=0.01$ (indicated by the vertical line). This allows one to find the value of the quadratic response. The corresponding values of the relative difference between the KS-DFT data and the result of the fit based on  Eq.~(\ref{eq:rho2}) are given  in the bottom panel of Fig.~\ref{fig:rho2}.  

\begin{figure}
\center
\includegraphics [width=0.4 \textwidth]{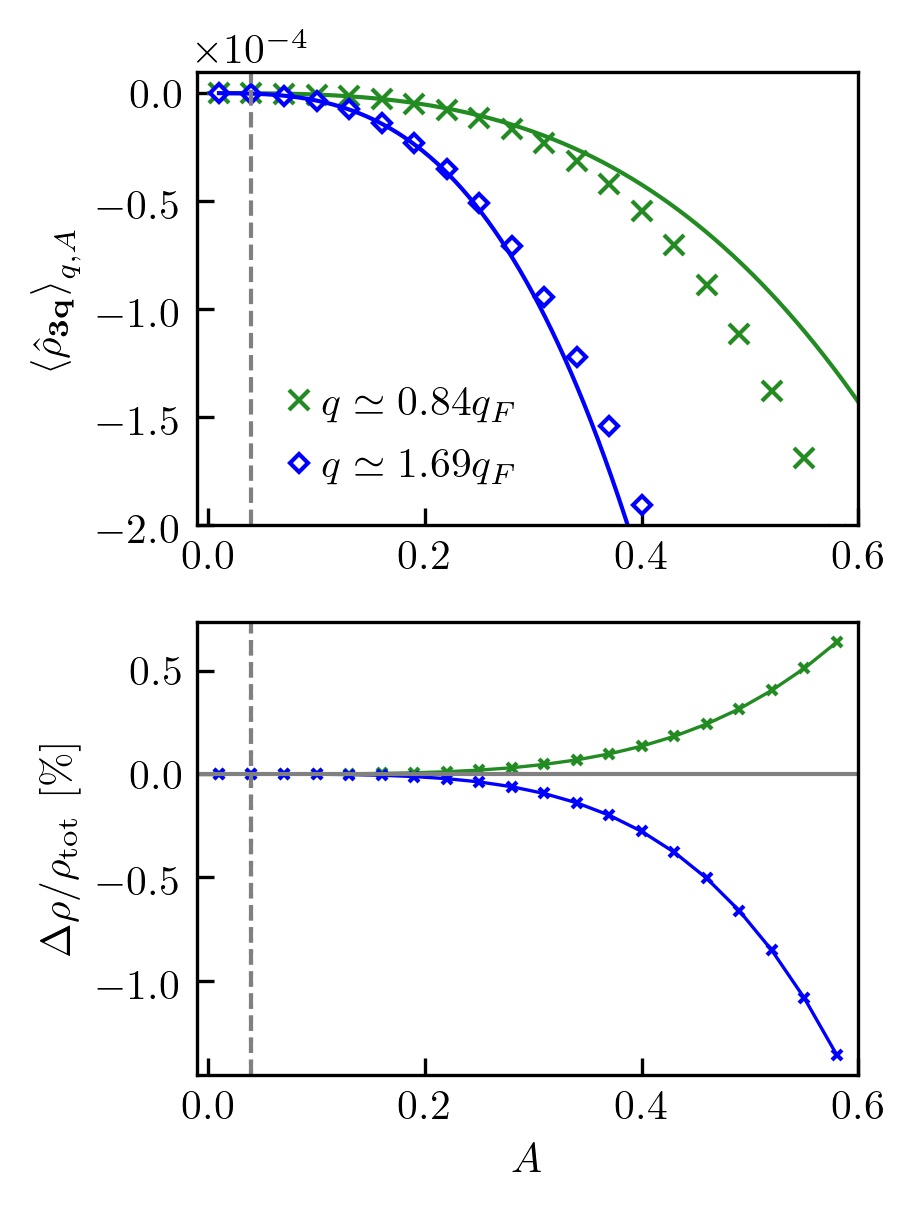}
\caption{\label{fig:rho3}
The density perturbation component  $\braket{\hat\rho_\mathbf{k}}_{q,A}$ at $k=3q$ is used to find the cubic density response at $r_s=2$ and $\theta=1$.}
\end{figure}

Finally, we show the results for the density response at the third harmonic ($k=3q$) in the top panel of Fig.~\ref{fig:rho3}.
 ~In this case, the KS-DFT simulations are performed for  $0.01\leq A\leq0.6$. The cubic response function at the third harmonic is found by using the data for $\braket{\hat\rho_\mathbf{3q}}_{q,A}$ at $A=0.04$. The values of the relative difference between the KS-DFT data and the results computed using  Eq.~(\ref{eq:rho3}) are provided in the bottom panel of Fig.~\ref{fig:rho3}.


All KS-DFT based results for the linear and nonlinear density response that have been presented in this work for different wave numbers, densities, and temperatures have been obtained by performing a similar analysis to those shown in Figs.~\ref{fig:rho1}-\ref{fig:rho3}, but involving a reduced number of $A$ values. The specific ranges of the considered $A$ values are given in Sec.~\ref{sec:simulation_details}.

\section*{Data Availability}
The data that support the findings of this study are available from the corresponding author upon reasonable request.

\bibliography{ref}

\providecommand{\latin}[1]{#1}
\makeatletter
\providecommand{\doi}
  {\begingroup\let\do\@makeother\dospecials
  \catcode`\{=1 \catcode`\}=2 \doi@aux}
\providecommand{\doi@aux}[1]{\endgroup\texttt{#1}}
\makeatother
\providecommand*\mcitethebibliography{\thebibliography}
\csname @ifundefined\endcsname{endmcitethebibliography}
  {\let\endmcitethebibliography\endthebibliography}{}
\begin{mcitethebibliography}{83}
\providecommand*\natexlab[1]{#1}
\providecommand*\mciteSetBstSublistMode[1]{}
\providecommand*\mciteSetBstMaxWidthForm[2]{}
\providecommand*\mciteBstWouldAddEndPuncttrue
  {\def\EndOfBibitem{\unskip.}}
\providecommand*\mciteBstWouldAddEndPunctfalse
  {\let\EndOfBibitem\relax}
\providecommand*\mciteSetBstMidEndSepPunct[3]{}
\providecommand*\mciteSetBstSublistLabelBeginEnd[3]{}
\providecommand*\EndOfBibitem{}
\mciteSetBstSublistMode{f}
\mciteSetBstMaxWidthForm{subitem}{(\alph{mcitesubitemcount})}
\mciteSetBstSublistLabelBeginEnd
  {\mcitemaxwidthsubitemform\space}
  {\relax}
  {\relax}

\bibitem[Nolting and Brewer(2009)Nolting, and Brewer]{nolting2009fundamentals}
Nolting,~W.; Brewer,~W. \emph{Fundamentals of Many-body Physics: Principles and
  Methods}; Springer Berlin Heidelberg, 2009\relax
\mciteBstWouldAddEndPuncttrue
\mciteSetBstMidEndSepPunct{\mcitedefaultmidpunct}
{\mcitedefaultendpunct}{\mcitedefaultseppunct}\relax
\EndOfBibitem
\bibitem[Panoiu \latin{et~al.}(2018)Panoiu, Sha, Lei, and Li]{Panoiu_2018}
Panoiu,~N.~C.; Sha,~W. E.~I.; Lei,~D.~Y.; Li,~G.-C. Nonlinear optics in
  plasmonic nanostructures. \emph{Journal of Optics} \textbf{2018}, \emph{20},
  083001\relax
\mciteBstWouldAddEndPuncttrue
\mciteSetBstMidEndSepPunct{\mcitedefaultmidpunct}
{\mcitedefaultendpunct}{\mcitedefaultseppunct}\relax
\EndOfBibitem
\bibitem[Lee \latin{et~al.}(2014)Lee, Tymchenko, Argyropoulos, Chen, Lu,
  Demmerle, Boehm, Amann, Al{\`u}, and Belkin]{Lee2014}
Lee,~J.; Tymchenko,~M.; Argyropoulos,~C.; Chen,~P.-Y.; Lu,~F.; Demmerle,~F.;
  Boehm,~G.; Amann,~M.-C.; Al{\`u},~A.; Belkin,~M.~A. Giant nonlinear response
  from plasmonic metasurfaces coupled to intersubband transitions.
  \emph{Nature} \textbf{2014}, \emph{511}, 65--69\relax
\mciteBstWouldAddEndPuncttrue
\mciteSetBstMidEndSepPunct{\mcitedefaultmidpunct}
{\mcitedefaultendpunct}{\mcitedefaultseppunct}\relax
\EndOfBibitem
\bibitem[Dalstein \latin{et~al.}(2018)Dalstein, Revel, Humbert, and
  Busson]{doi:10.1063/1.5021553}
Dalstein,~L.; Revel,~A.; Humbert,~C.; Busson,~B. Nonlinear optical response of
  a gold surface in the visible range: A study by two-color sum-frequency
  generation spectroscopy. I. Experimental determination. \emph{The Journal of
  Chemical Physics} \textbf{2018}, \emph{148}, 134701\relax
\mciteBstWouldAddEndPuncttrue
\mciteSetBstMidEndSepPunct{\mcitedefaultmidpunct}
{\mcitedefaultendpunct}{\mcitedefaultseppunct}\relax
\EndOfBibitem
\bibitem[Ventura \latin{et~al.}(2019)Ventura, Passos, Lopes, and dos
  Santos]{Ventura_2019}
Ventura,~G.~B.; Passos,~D.; Lopes,~J. M. V.~P.; dos Santos,~J. M. B.~L. A study
  of the nonlinear optical response of the plain graphene and gapped graphene
  monolayers beyond the Dirac approximation. \emph{Journal of Physics:
  Condensed Matter} \textbf{2019}, \relax
\mciteBstWouldAddEndPunctfalse
\mciteSetBstMidEndSepPunct{\mcitedefaultmidpunct}
{}{\mcitedefaultseppunct}\relax
\EndOfBibitem
\bibitem[Dornheim \latin{et~al.}(2020)Dornheim, Vorberger, and
  Bonitz]{Dornheim_PRL_2020}
Dornheim,~T.; Vorberger,~J.; Bonitz,~M. Nonlinear Electronic Density Response
  in Warm Dense Matter. \emph{Phys. Rev. Lett.} \textbf{2020}, \emph{125},
  085001\relax
\mciteBstWouldAddEndPuncttrue
\mciteSetBstMidEndSepPunct{\mcitedefaultmidpunct}
{\mcitedefaultendpunct}{\mcitedefaultseppunct}\relax
\EndOfBibitem
\bibitem[Fuchs \latin{et~al.}(2015)Fuchs, Trigo, Chen, Ghimire, Shwartz,
  Kozina, Jiang, Henighan, Bray, Ndabashimiye, Bucksbaum, Feng, Herrmann,
  Carini, Pines, Hart, Kenney, Guillet, Boutet, Williams, Messerschmidt,
  Seibert, Moeller, Hastings, and Reis]{Fuchs2015}
Fuchs,~M.; Trigo,~M.; Chen,~J.; Ghimire,~S.; Shwartz,~S.; Kozina,~M.;
  Jiang,~M.; Henighan,~T.; Bray,~C.; Ndabashimiye,~G.; Bucksbaum,~P.~H.;
  Feng,~Y.; Herrmann,~S.; Carini,~G.~A.; Pines,~J.; Hart,~P.; Kenney,~C.;
  Guillet,~S.; Boutet,~S.; Williams,~G.~J.; Messerschmidt,~M.; Seibert,~M.~M.;
  Moeller,~S.; Hastings,~J.~B.; Reis,~D.~A. Anomalous nonlinear X-ray Compton
  scattering. \emph{Nature Physics} \textbf{2015}, \emph{11}, 964--970\relax
\mciteBstWouldAddEndPuncttrue
\mciteSetBstMidEndSepPunct{\mcitedefaultmidpunct}
{\mcitedefaultendpunct}{\mcitedefaultseppunct}\relax
\EndOfBibitem
\bibitem[Benuzzi-Mounaix \latin{et~al.}(2014)Benuzzi-Mounaix, Mazevet, Ravasio,
  Vinci, Denoeud, Koenig, Amadou, Brambrink, Festa, Levy, Harmand, Brygoo,
  Huser, Recoules, Bouchet, Morard, Guyot, de~Resseguier, Myanishi, Ozaki,
  Dorchies, Gaudin, Leguay, Peyrusse, Henry, Raffestin, Pape, Smith, and
  Musella]{Benuzzi_Mounaix_2014}
Benuzzi-Mounaix,~A.; Mazevet,~S.; Ravasio,~A.; Vinci,~T.; Denoeud,~A.;
  Koenig,~M.; Amadou,~N.; Brambrink,~E.; Festa,~F.; Levy,~A.; Harmand,~M.;
  Brygoo,~S.; Huser,~G.; Recoules,~V.; Bouchet,~J.; Morard,~G.; Guyot,~F.;
  de~Resseguier,~T.; Myanishi,~K.; Ozaki,~N.; Dorchies,~F.; Gaudin,~J.;
  Leguay,~P.~M.; Peyrusse,~O.; Henry,~O.; Raffestin,~D.; Pape,~S.~L.;
  Smith,~R.; Musella,~R. Progress in warm dense matter study with applications
  to planetology. \emph{Physica Scripta} \textbf{2014}, \emph{T161},
  014060\relax
\mciteBstWouldAddEndPuncttrue
\mciteSetBstMidEndSepPunct{\mcitedefaultmidpunct}
{\mcitedefaultendpunct}{\mcitedefaultseppunct}\relax
\EndOfBibitem
\bibitem[Saumon \latin{et~al.}(1992)Saumon, Hubbard, Chabrier, and van
  Horn]{saumon1}
Saumon,~D.; Hubbard,~W.~B.; Chabrier,~G.; van Horn,~H.~M. The role of the
  molecular-metallic transition of hydrogen in the evolution of Jupiter,
  Saturn, and brown dwarfs. \emph{Astrophys. J} \textbf{1992}, \emph{391},
  827--831\relax
\mciteBstWouldAddEndPuncttrue
\mciteSetBstMidEndSepPunct{\mcitedefaultmidpunct}
{\mcitedefaultendpunct}{\mcitedefaultseppunct}\relax
\EndOfBibitem
\bibitem[Hu \latin{et~al.}(2010)Hu, Militzer, Goncharov, and Skupsky]{Hu2010}
Hu,~S.~X.; Militzer,~B.; Goncharov,~V.~N.; Skupsky,~S. Strong Coupling and
  Degeneracy Effects in Inertial Confinement Fusion Implosions. \emph{Phys.
  Rev. Lett.} \textbf{2010}, \emph{104}, 235003\relax
\mciteBstWouldAddEndPuncttrue
\mciteSetBstMidEndSepPunct{\mcitedefaultmidpunct}
{\mcitedefaultendpunct}{\mcitedefaultseppunct}\relax
\EndOfBibitem
\bibitem[Lazicki \latin{et~al.}(2021)Lazicki, McGonegle, Rygg, Braun, Swift,
  Gorman, Smith, Heighway, Higginbotham, Suggit, Fratanduono, Coppari,
  Wehrenberg, Kraus, Erskine, Bernier, McNaney, Rudd, Collins, Eggert, and
  Wark]{Lazicki2021}
Lazicki,~A.; McGonegle,~D.; Rygg,~J.~R.; Braun,~D.~G.; Swift,~D.~C.;
  Gorman,~M.~G.; Smith,~R.~F.; Heighway,~P.~G.; Higginbotham,~A.;
  Suggit,~M.~J.; Fratanduono,~D.~E.; Coppari,~F.; Wehrenberg,~C.~E.;
  Kraus,~R.~G.; Erskine,~D.; Bernier,~J.~V.; McNaney,~J.~M.; Rudd,~R.~E.;
  Collins,~G.~W.; Eggert,~J.~H.; Wark,~J.~S. Metastability of diamond
  ramp-compressed to 2 terapascals. \emph{Nature} \textbf{2021}, \emph{589},
  532--535\relax
\mciteBstWouldAddEndPuncttrue
\mciteSetBstMidEndSepPunct{\mcitedefaultmidpunct}
{\mcitedefaultendpunct}{\mcitedefaultseppunct}\relax
\EndOfBibitem
\bibitem[Dattelbaum \latin{et~al.}(2021)Dattelbaum, Watkins, Firestone, Huber,
  Gustavsen, Ringstrand, Coe, Podlesak, Gleason, Lee, Galtier, and
  Sandberg]{Dattelbaum2021}
Dattelbaum,~D.~M.; Watkins,~E.~B.; Firestone,~M.~A.; Huber,~R.~C.;
  Gustavsen,~R.~L.; Ringstrand,~B.~S.; Coe,~J.~D.; Podlesak,~D.;
  Gleason,~A.~E.; Lee,~H.~J.; Galtier,~E.; Sandberg,~R.~L. Carbon clusters
  formed from shocked benzene. \emph{Nature Communications} \textbf{2021},
  \emph{12}, 5202\relax
\mciteBstWouldAddEndPuncttrue
\mciteSetBstMidEndSepPunct{\mcitedefaultmidpunct}
{\mcitedefaultendpunct}{\mcitedefaultseppunct}\relax
\EndOfBibitem
\bibitem[L{\"u}tgert \latin{et~al.}(2021)L{\"u}tgert, Vorberger, Hartley,
  Voigt, R{\"o}del, Schuster, Benuzzi-Mounaix, Brown, Cowan, Cunningham,
  D{\"o}ppner, Falcone, Fletcher, Galtier, Glenzer, Laso~Garcia, Gericke,
  Heimann, Lee, McBride, Pelka, Prencipe, Saunders, Sch{\"o}lmerich,
  Sch{\"o}rner, Sun, Vinci, Ravasio, and Kraus]{Luetgert2021}
L{\"u}tgert,~J.; Vorberger,~J.; Hartley,~N.~J.; Voigt,~K.; R{\"o}del,~M.;
  Schuster,~A.~K.; Benuzzi-Mounaix,~A.; Brown,~S.; Cowan,~T.~E.;
  Cunningham,~E.; D{\"o}ppner,~T.; Falcone,~R.~W.; Fletcher,~L.~B.;
  Galtier,~E.; Glenzer,~S.~H.; Laso~Garcia,~A.; Gericke,~D.~O.; Heimann,~P.~A.;
  Lee,~H.~J.; McBride,~E.~E.; Pelka,~A.; Prencipe,~I.; Saunders,~A.~M.;
  Sch{\"o}lmerich,~M.; Sch{\"o}rner,~M.; Sun,~P.; Vinci,~T.; Ravasio,~A.;
  Kraus,~D. Measuring the structure and equation of state of polyethylene
  terephthalate at megabar pressures. \emph{Scientific Reports} \textbf{2021},
  \emph{11}, 12883\relax
\mciteBstWouldAddEndPuncttrue
\mciteSetBstMidEndSepPunct{\mcitedefaultmidpunct}
{\mcitedefaultendpunct}{\mcitedefaultseppunct}\relax
\EndOfBibitem
\bibitem[Dornheim \latin{et~al.}(2018)Dornheim, Groth, and Bonitz]{review}
Dornheim,~T.; Groth,~S.; Bonitz,~M. The uniform electron gas at warm dense
  matter conditions. \emph{Phys. Reports} \textbf{2018}, \emph{744},
  1--86\relax
\mciteBstWouldAddEndPuncttrue
\mciteSetBstMidEndSepPunct{\mcitedefaultmidpunct}
{\mcitedefaultendpunct}{\mcitedefaultseppunct}\relax
\EndOfBibitem
\bibitem[Loos and Gill(2016)Loos, and Gill]{loos}
Loos,~P.-F.; Gill,~P. M.~W. The uniform electron gas. \emph{Comput. Mol. Sci}
  \textbf{2016}, \emph{6}, 410--429\relax
\mciteBstWouldAddEndPuncttrue
\mciteSetBstMidEndSepPunct{\mcitedefaultmidpunct}
{\mcitedefaultendpunct}{\mcitedefaultseppunct}\relax
\EndOfBibitem
\bibitem[Dornheim \latin{et~al.}(2021)Dornheim, B\"ohme, Moldabekov, Vorberger,
  and Bonitz]{PhysRevResearch.3.033231}
Dornheim,~T.; B\"ohme,~M.; Moldabekov,~Z.~A.; Vorberger,~J.; Bonitz,~M. Density
  response of the warm dense electron gas beyond linear response theory:
  Excitation of harmonics. \emph{Phys. Rev. Research} \textbf{2021}, \emph{3},
  033231\relax
\mciteBstWouldAddEndPuncttrue
\mciteSetBstMidEndSepPunct{\mcitedefaultmidpunct}
{\mcitedefaultendpunct}{\mcitedefaultseppunct}\relax
\EndOfBibitem
\bibitem[Dornheim \latin{et~al.}(2021)Dornheim, Moldabekov, and
  Vorberger]{JCP21_nonlin}
Dornheim,~T.; Moldabekov,~Z.~A.; Vorberger,~J. Nonlinear density response from
  imaginary-time correlation functions: Ab initio path integral Monte Carlo
  simulations of the warm dense electron gas. \emph{The Journal of Chemical
  Physics} \textbf{2021}, \emph{155}, 054110\relax
\mciteBstWouldAddEndPuncttrue
\mciteSetBstMidEndSepPunct{\mcitedefaultmidpunct}
{\mcitedefaultendpunct}{\mcitedefaultseppunct}\relax
\EndOfBibitem
\bibitem[Dornheim(2019)]{PhysRevE.100.023307}
Dornheim,~T. Fermion sign problem in path integral Monte Carlo simulations:
  Quantum dots, ultracold atoms, and warm dense matter. \emph{Phys. Rev. E}
  \textbf{2019}, \emph{100}, 023307\relax
\mciteBstWouldAddEndPuncttrue
\mciteSetBstMidEndSepPunct{\mcitedefaultmidpunct}
{\mcitedefaultendpunct}{\mcitedefaultseppunct}\relax
\EndOfBibitem
\bibitem[Troyer and Wiese(2005)Troyer, and Wiese]{troyer}
Troyer,~M.; Wiese,~U.~J. Computational Complexity and Fundamental Limitations
  to Fermionic Quantum {M}onte {C}arlo Simulations. \emph{Phys. Rev. Lett}
  \textbf{2005}, \emph{94}, 170201\relax
\mciteBstWouldAddEndPuncttrue
\mciteSetBstMidEndSepPunct{\mcitedefaultmidpunct}
{\mcitedefaultendpunct}{\mcitedefaultseppunct}\relax
\EndOfBibitem
\bibitem[Mermin(1965)]{Mermin_DFT_1965}
Mermin,~N.~D. Thermal Properties of the Inhomogeneous Electron Gas. \emph{Phys.
  Rev.} \textbf{1965}, \emph{137}, A1441--A1443\relax
\mciteBstWouldAddEndPuncttrue
\mciteSetBstMidEndSepPunct{\mcitedefaultmidpunct}
{\mcitedefaultendpunct}{\mcitedefaultseppunct}\relax
\EndOfBibitem
\bibitem[Gross and Kohn(1985)Gross, and Kohn]{Gross_Kohn_PRL_1985}
Gross,~E. K.~U.; Kohn,~W. Local density-functional theory of
  frequency-dependent linear response. \emph{Phys. Rev. Lett.} \textbf{1985},
  \emph{55}, 2850--2852\relax
\mciteBstWouldAddEndPuncttrue
\mciteSetBstMidEndSepPunct{\mcitedefaultmidpunct}
{\mcitedefaultendpunct}{\mcitedefaultseppunct}\relax
\EndOfBibitem
\bibitem[Martin \latin{et~al.}(2004)Martin, Martin, and
  Press]{martin2004electronic}
Martin,~R.; Martin,~R.; Press,~C.~U. \emph{Electronic Structure: Basic Theory
  and Practical Methods}; Cambridge University Press, 2004\relax
\mciteBstWouldAddEndPuncttrue
\mciteSetBstMidEndSepPunct{\mcitedefaultmidpunct}
{\mcitedefaultendpunct}{\mcitedefaultseppunct}\relax
\EndOfBibitem
\bibitem[Zhang \latin{et~al.}(2016)Zhang, Wang, Kang, Zhang, and
  He]{Zhang_POP_2016}
Zhang,~S.; Wang,~H.; Kang,~W.; Zhang,~P.; He,~X.~T. Extended application of
  Kohn-Sham first-principles molecular dynamics method with plane wave
  approximation at high energy—From cold materials to hot dense plasmas.
  \emph{Physics of Plasmas} \textbf{2016}, \emph{23}, 042707\relax
\mciteBstWouldAddEndPuncttrue
\mciteSetBstMidEndSepPunct{\mcitedefaultmidpunct}
{\mcitedefaultendpunct}{\mcitedefaultseppunct}\relax
\EndOfBibitem
\bibitem[Bethkenhagen \latin{et~al.}(2021)Bethkenhagen, Sharma, Suryanarayana,
  Pask, Sadigh, and Hamel]{bethkenhagen2021thermodynamic}
Bethkenhagen,~M.; Sharma,~A.; Suryanarayana,~P.; Pask,~J.~E.; Sadigh,~B.;
  Hamel,~S. Thermodynamic, structural, and transport properties of dense carbon
  up to 10 million Kelvin from Kohn-Sham density functional theory
  calculations. 2021\relax
\mciteBstWouldAddEndPuncttrue
\mciteSetBstMidEndSepPunct{\mcitedefaultmidpunct}
{\mcitedefaultendpunct}{\mcitedefaultseppunct}\relax
\EndOfBibitem
\bibitem[Wesolowski and Wang(2013)Wesolowski, and Wang]{wesolowski2013recent}
Wesolowski,~T.; Wang,~Y. \emph{Recent Progress in Orbital-free Density
  Functional Theory}; Recent advances in computational chemistry; World
  Scientific, 2013\relax
\mciteBstWouldAddEndPuncttrue
\mciteSetBstMidEndSepPunct{\mcitedefaultmidpunct}
{\mcitedefaultendpunct}{\mcitedefaultseppunct}\relax
\EndOfBibitem
\bibitem[Bonitz \latin{et~al.}(2020)Bonitz, Dornheim, Moldabekov, Zhang,
  Hamann, Kählert, Filinov, Ramakrishna, and Vorberger]{new_POP}
Bonitz,~M.; Dornheim,~T.; Moldabekov,~Z.~A.; Zhang,~S.; Hamann,~P.;
  Kählert,~H.; Filinov,~A.; Ramakrishna,~K.; Vorberger,~J. Ab initio
  simulation of warm dense matter. \emph{Physics of Plasmas} \textbf{2020},
  \emph{27}, 042710\relax
\mciteBstWouldAddEndPuncttrue
\mciteSetBstMidEndSepPunct{\mcitedefaultmidpunct}
{\mcitedefaultendpunct}{\mcitedefaultseppunct}\relax
\EndOfBibitem
\bibitem[Kritcher \latin{et~al.}(2020)Kritcher, Swift, D{\"o}ppner, Bachmann,
  Benedict, Collins, DuBois, Elsner, Fontaine, Gaffney, Hamel, Lazicki,
  Johnson, Kostinski, Kraus, MacDonald, Maddox, Martin, Neumayer, Nikroo,
  Nilsen, Remington, Saumon, Sterne, Sweet, Correa, Whitley, Falcone, and
  Glenzer]{Kritcher2020}
Kritcher,~A.~L.; Swift,~D.~C.; D{\"o}ppner,~T.; Bachmann,~B.; Benedict,~L.~X.;
  Collins,~G.~W.; DuBois,~J.~L.; Elsner,~F.; Fontaine,~G.; Gaffney,~J.~A.;
  Hamel,~S.; Lazicki,~A.; Johnson,~W.~R.; Kostinski,~N.; Kraus,~D.;
  MacDonald,~M.~J.; Maddox,~B.; Martin,~M.~E.; Neumayer,~P.; Nikroo,~A.;
  Nilsen,~J.; Remington,~B.~A.; Saumon,~D.; Sterne,~P.~A.; Sweet,~W.;
  Correa,~A.~A.; Whitley,~H.~D.; Falcone,~R.~W.; Glenzer,~S.~H. A measurement
  of the equation of state of carbon envelopes of white dwarfs. \emph{Nature}
  \textbf{2020}, \emph{584}, 51--54\relax
\mciteBstWouldAddEndPuncttrue
\mciteSetBstMidEndSepPunct{\mcitedefaultmidpunct}
{\mcitedefaultendpunct}{\mcitedefaultseppunct}\relax
\EndOfBibitem
\bibitem[Booth \latin{et~al.}(2015)Booth, Robinson, Hakel, Clarke, Dance,
  Doria, Gizzi, Gregori, Koester, Labate, Levato, Li, Makita, Mancini, Pasley,
  Rajeev, Riley, Wagenaars, Waugh, and Woolsey]{Booth2015}
Booth,~N.; Robinson,~A. P.~L.; Hakel,~P.; Clarke,~R.~J.; Dance,~R.~J.;
  Doria,~D.; Gizzi,~L.~A.; Gregori,~G.; Koester,~P.; Labate,~L.; Levato,~T.;
  Li,~B.; Makita,~M.; Mancini,~R.~C.; Pasley,~J.; Rajeev,~P.~P.; Riley,~D.;
  Wagenaars,~E.; Waugh,~J.~N.; Woolsey,~N.~C. Laboratory measurements of
  resistivity in warm dense plasmas relevant to the microphysics of brown
  dwarfs. \emph{Nature Communications} \textbf{2015}, \emph{6}, 8742\relax
\mciteBstWouldAddEndPuncttrue
\mciteSetBstMidEndSepPunct{\mcitedefaultmidpunct}
{\mcitedefaultendpunct}{\mcitedefaultseppunct}\relax
\EndOfBibitem
\bibitem[Schuster \latin{et~al.}(2020)Schuster, Hartley, Vorberger, D\"oppner,
  van Driel, Falcone, Fletcher, Frydrych, Galtier, Gamboa, Gericke, Glenzer,
  Granados, MacDonald, MacKinnon, McBride, Nam, Neumayer, Pak, Prencipe, Voigt,
  Saunders, Sun, and Kraus]{PhysRevB.101.054301}
Schuster,~A.~K.; Hartley,~N.~J.; Vorberger,~J.; D\"oppner,~T.; van Driel,~T.;
  Falcone,~R.~W.; Fletcher,~L.~B.; Frydrych,~S.; Galtier,~E.; Gamboa,~E.~J.;
  Gericke,~D.~O.; Glenzer,~S.~H.; Granados,~E.; MacDonald,~M.~J.;
  MacKinnon,~A.~J.; McBride,~E.~E.; Nam,~I.; Neumayer,~P.; Pak,~A.;
  Prencipe,~I.; Voigt,~K.; Saunders,~A.~M.; Sun,~P.; Kraus,~D. Measurement of
  diamond nucleation rates from hydrocarbons at conditions comparable to the
  interiors of icy giant planets. \emph{Phys. Rev. B} \textbf{2020},
  \emph{101}, 054301\relax
\mciteBstWouldAddEndPuncttrue
\mciteSetBstMidEndSepPunct{\mcitedefaultmidpunct}
{\mcitedefaultendpunct}{\mcitedefaultseppunct}\relax
\EndOfBibitem
\bibitem[Dornheim \latin{et~al.}(2021)Dornheim, Vorberger, Moldabekov, and
  Bonitz]{dornheim2021nonlinear}
Dornheim,~T.; Vorberger,~J.; Moldabekov,~Z.; Bonitz,~M. Nonlinear electronic
  density response of the warm dense electron gas: multiple perturbations and
  mode coupling. 2021\relax
\mciteBstWouldAddEndPuncttrue
\mciteSetBstMidEndSepPunct{\mcitedefaultmidpunct}
{\mcitedefaultendpunct}{\mcitedefaultseppunct}\relax
\EndOfBibitem
\bibitem[Dornheim \latin{et~al.}(2019)Dornheim, Groth, Filinov, and
  Bonitz]{Dornheim_permutation_cycles}
Dornheim,~T.; Groth,~S.; Filinov,~A.~V.; Bonitz,~M. Path integral Monte Carlo
  simulation of degenerate electrons: Permutation-cycle properties. \emph{The
  Journal of Chemical Physics} \textbf{2019}, \emph{151}, 014108\relax
\mciteBstWouldAddEndPuncttrue
\mciteSetBstMidEndSepPunct{\mcitedefaultmidpunct}
{\mcitedefaultendpunct}{\mcitedefaultseppunct}\relax
\EndOfBibitem
\bibitem[Groth \latin{et~al.}(2017)Groth, Dornheim, and Bonitz]{JCP_Simon17}
Groth,~S.; Dornheim,~T.; Bonitz,~M. Configuration path integral Monte Carlo
  approach to the static density response of the warm dense electron gas.
  \emph{The Journal of Chemical Physics} \textbf{2017}, \emph{147},
  164108\relax
\mciteBstWouldAddEndPuncttrue
\mciteSetBstMidEndSepPunct{\mcitedefaultmidpunct}
{\mcitedefaultendpunct}{\mcitedefaultseppunct}\relax
\EndOfBibitem
\bibitem[Yilmaz \latin{et~al.}(2020)Yilmaz, Hunger, Dornheim, Groth, and
  Bonitz]{Yilmaz}
Yilmaz,~A.; Hunger,~K.; Dornheim,~T.; Groth,~S.; Bonitz,~M. Restricted
  configuration path integral Monte Carlo. \emph{The Journal of Chemical
  Physics} \textbf{2020}, \emph{153}, 124114\relax
\mciteBstWouldAddEndPuncttrue
\mciteSetBstMidEndSepPunct{\mcitedefaultmidpunct}
{\mcitedefaultendpunct}{\mcitedefaultseppunct}\relax
\EndOfBibitem
\bibitem[Dornheim \latin{et~al.}(2015)Dornheim, Groth, Filinov, and
  Bonitz]{Dornheim_2015}
Dornheim,~T.; Groth,~S.; Filinov,~A.; Bonitz,~M. Permutation blocking path
  integral Monte Carlo: a highly efficient approach to the simulation of
  strongly degenerate non-ideal fermions. \emph{New Journal of Physics}
  \textbf{2015}, \emph{17}, 073017\relax
\mciteBstWouldAddEndPuncttrue
\mciteSetBstMidEndSepPunct{\mcitedefaultmidpunct}
{\mcitedefaultendpunct}{\mcitedefaultseppunct}\relax
\EndOfBibitem
\bibitem[Dornheim \latin{et~al.}(2015)Dornheim, Schoof, Groth, Filinov, and
  Bonitz]{JCP_tobias_15}
Dornheim,~T.; Schoof,~T.; Groth,~S.; Filinov,~A.; Bonitz,~M. Permutation
  blocking path integral Monte Carlo approach to the uniform electron gas at
  finite temperature. \emph{The Journal of Chemical Physics} \textbf{2015},
  \emph{143}, 204101\relax
\mciteBstWouldAddEndPuncttrue
\mciteSetBstMidEndSepPunct{\mcitedefaultmidpunct}
{\mcitedefaultendpunct}{\mcitedefaultseppunct}\relax
\EndOfBibitem
\bibitem[Lee \latin{et~al.}(2021)Lee, Morales, and Malone]{JCP_Lee}
Lee,~J.; Morales,~M.~A.; Malone,~F.~D. A phaseless auxiliary-field quantum
  Monte Carlo perspective on the uniform electron gas at finite temperatures:
  Issues, observations, and benchmark study. \emph{The Journal of Chemical
  Physics} \textbf{2021}, \emph{154}, 064109\relax
\mciteBstWouldAddEndPuncttrue
\mciteSetBstMidEndSepPunct{\mcitedefaultmidpunct}
{\mcitedefaultendpunct}{\mcitedefaultseppunct}\relax
\EndOfBibitem
\bibitem[Dornheim \latin{et~al.}(2018)Dornheim, Groth, and
  Bonitz]{dornheim_physrep_18}
Dornheim,~T.; Groth,~S.; Bonitz,~M. The uniform electron gas at warm dense
  matter conditions. \emph{Phys. Rep.} \textbf{2018}, \emph{744}, 1 -- 86\relax
\mciteBstWouldAddEndPuncttrue
\mciteSetBstMidEndSepPunct{\mcitedefaultmidpunct}
{\mcitedefaultendpunct}{\mcitedefaultseppunct}\relax
\EndOfBibitem
\bibitem[Witt and Carter(2019)Witt, and Carter]{PhysRevB.100.125106}
Witt,~W.~C.; Carter,~E.~A. Kinetic energy density of nearly free electrons. I.
  Response functionals of the external potential. \emph{Phys. Rev. B}
  \textbf{2019}, \emph{100}, 125106\relax
\mciteBstWouldAddEndPuncttrue
\mciteSetBstMidEndSepPunct{\mcitedefaultmidpunct}
{\mcitedefaultendpunct}{\mcitedefaultseppunct}\relax
\EndOfBibitem
\bibitem[Witt and Carter(2019)Witt, and Carter]{PhysRevB.100.125107}
Witt,~W.~C.; Carter,~E.~A. Kinetic energy density of nearly free electrons. II.
  Response functionals of the electron density. \emph{Phys. Rev. B}
  \textbf{2019}, \emph{100}, 125107\relax
\mciteBstWouldAddEndPuncttrue
\mciteSetBstMidEndSepPunct{\mcitedefaultmidpunct}
{\mcitedefaultendpunct}{\mcitedefaultseppunct}\relax
\EndOfBibitem
\bibitem[Shao \latin{et~al.}(2021)Shao, Mi, and Pavanello]{PhysRevB.104.045118}
Shao,~X.; Mi,~W.; Pavanello,~M. Revised Huang-Carter nonlocal kinetic energy
  functional for semiconductors and their surfaces. \emph{Phys. Rev. B}
  \textbf{2021}, \emph{104}, 045118\relax
\mciteBstWouldAddEndPuncttrue
\mciteSetBstMidEndSepPunct{\mcitedefaultmidpunct}
{\mcitedefaultendpunct}{\mcitedefaultseppunct}\relax
\EndOfBibitem
\bibitem[Sjostrom and Daligault(2013)Sjostrom, and
  Daligault]{PhysRevB.88.195103}
Sjostrom,~T.; Daligault,~J. Nonlocal orbital-free noninteracting free-energy
  functional for warm dense matter. \emph{Phys. Rev. B} \textbf{2013},
  \emph{88}, 195103\relax
\mciteBstWouldAddEndPuncttrue
\mciteSetBstMidEndSepPunct{\mcitedefaultmidpunct}
{\mcitedefaultendpunct}{\mcitedefaultseppunct}\relax
\EndOfBibitem
\bibitem[Sjostrom and Crockett(2015)Sjostrom, and Crockett]{PhysRevB.92.115104}
Sjostrom,~T.; Crockett,~S. Orbital-free extension to Kohn-Sham density
  functional theory equation of state calculations: Application to silicon
  dioxide. \emph{Phys. Rev. B} \textbf{2015}, \emph{92}, 115104\relax
\mciteBstWouldAddEndPuncttrue
\mciteSetBstMidEndSepPunct{\mcitedefaultmidpunct}
{\mcitedefaultendpunct}{\mcitedefaultseppunct}\relax
\EndOfBibitem
\bibitem[Moldabekov \latin{et~al.}(2018)Moldabekov, Bonitz, and
  Ramazanov]{zhandos_pop18}
Moldabekov,~Z.~A.; Bonitz,~M.; Ramazanov,~T.~S. Theoretical foundations of
  quantum hydrodynamics for plasmas. \emph{Phys. Plasmas} \textbf{2018},
  \emph{25}, 031903\relax
\mciteBstWouldAddEndPuncttrue
\mciteSetBstMidEndSepPunct{\mcitedefaultmidpunct}
{\mcitedefaultendpunct}{\mcitedefaultseppunct}\relax
\EndOfBibitem
\bibitem[Baghramyan \latin{et~al.}(2021)Baghramyan, Della~Sala, and
  Cirac\`{\i}]{PhysRevX.11.011049}
Baghramyan,~H.~M.; Della~Sala,~F.; Cirac\`{\i},~C. Laplacian-Level Quantum
  Hydrodynamic Theory for Plasmonics. \emph{Phys. Rev. X} \textbf{2021},
  \emph{11}, 011049\relax
\mciteBstWouldAddEndPuncttrue
\mciteSetBstMidEndSepPunct{\mcitedefaultmidpunct}
{\mcitedefaultendpunct}{\mcitedefaultseppunct}\relax
\EndOfBibitem
\bibitem[Graziani \latin{et~al.}(2021)Graziani, Moldabekov, Olson, and
  Bonitz]{graziani2021shock}
Graziani,~F.; Moldabekov,~Z.; Olson,~B.; Bonitz,~M. Shock Physics in Warm Dense
  Matter--a quantum hydrodynamics perspective. 2021\relax
\mciteBstWouldAddEndPuncttrue
\mciteSetBstMidEndSepPunct{\mcitedefaultmidpunct}
{\mcitedefaultendpunct}{\mcitedefaultseppunct}\relax
\EndOfBibitem
\bibitem[Manfredi \latin{et~al.}(2021)Manfredi, Hervieux, and
  Hurst]{Manfredi2021}
Manfredi,~G.; Hervieux,~P.-A.; Hurst,~J. Fluid descriptions of quantum plasmas.
  \emph{Reviews of Modern Plasma Physics} \textbf{2021}, \emph{5}, 7\relax
\mciteBstWouldAddEndPuncttrue
\mciteSetBstMidEndSepPunct{\mcitedefaultmidpunct}
{\mcitedefaultendpunct}{\mcitedefaultseppunct}\relax
\EndOfBibitem
\bibitem[Jiang \latin{et~al.}(2021)Jiang, Shao, and
  Pavanello]{PhysRevB.104.235110}
Jiang,~K.; Shao,~X.; Pavanello,~M. Nonlocal and nonadiabatic Pauli potential
  for time-dependent orbital-free density functional theory. \emph{Phys. Rev.
  B} \textbf{2021}, \emph{104}, 235110\relax
\mciteBstWouldAddEndPuncttrue
\mciteSetBstMidEndSepPunct{\mcitedefaultmidpunct}
{\mcitedefaultendpunct}{\mcitedefaultseppunct}\relax
\EndOfBibitem
\bibitem[Moldabekov \latin{et~al.}(2017)Moldabekov, Bonitz, and
  Ramazanov]{zhandos_cpp17_1d}
Moldabekov,~Z.; Bonitz,~M.; Ramazanov,~T. {Gradient correction and Bohm
  potential for two‐ and one‐dimensional electron gases at a finite
  temperature}. \emph{Contrib. Plasma Phys.} \textbf{2017}, \emph{57},
  499--505\relax
\mciteBstWouldAddEndPuncttrue
\mciteSetBstMidEndSepPunct{\mcitedefaultmidpunct}
{\mcitedefaultendpunct}{\mcitedefaultseppunct}\relax
\EndOfBibitem
\bibitem[Moldabekov \latin{et~al.}()Moldabekov, Dornheim, and Bonitz]{cpp21}
Moldabekov,~Z.~A.; Dornheim,~T.; Bonitz,~M. Screening of a test charge in a
  free-electron gas at warm dense matter and dense non-ideal plasma conditions.
  \emph{Contributions to Plasma Physics} e202000176\relax
\mciteBstWouldAddEndPuncttrue
\mciteSetBstMidEndSepPunct{\mcitedefaultmidpunct}
{\mcitedefaultendpunct}{\mcitedefaultseppunct}\relax
\EndOfBibitem
\bibitem[Moldabekov \latin{et~al.}(2018)Moldabekov, Groth, Dornheim, K\"ahlert,
  Bonitz, and Ramazanov]{PhysRevE.98.023207}
Moldabekov,~Z.~A.; Groth,~S.; Dornheim,~T.; K\"ahlert,~H.; Bonitz,~M.;
  Ramazanov,~T.~S. Structural characteristics of strongly coupled ions in a
  dense quantum plasma. \emph{Phys. Rev. E} \textbf{2018}, \emph{98},
  023207\relax
\mciteBstWouldAddEndPuncttrue
\mciteSetBstMidEndSepPunct{\mcitedefaultmidpunct}
{\mcitedefaultendpunct}{\mcitedefaultseppunct}\relax
\EndOfBibitem
\bibitem[Moldabekov \latin{et~al.}(2019)Moldabekov, K\"ahlert, Dornheim, Groth,
  Bonitz, and Ramazanov]{PhysRevE.99.053203}
Moldabekov,~Z.~A.; K\"ahlert,~H.; Dornheim,~T.; Groth,~S.; Bonitz,~M.;
  Ramazanov,~T.~S. Dynamical structure factor of strongly coupled ions in a
  dense quantum plasma. \emph{Phys. Rev. E} \textbf{2019}, \emph{99},
  053203\relax
\mciteBstWouldAddEndPuncttrue
\mciteSetBstMidEndSepPunct{\mcitedefaultmidpunct}
{\mcitedefaultendpunct}{\mcitedefaultseppunct}\relax
\EndOfBibitem
\bibitem[Moldabekov \latin{et~al.}(2015)Moldabekov, Schoof, Ludwig, Bonitz, and
  Ramazanov]{pop15}
Moldabekov,~Z.; Schoof,~T.; Ludwig,~P.; Bonitz,~M.; Ramazanov,~T. Statically
  screened ion potential and {B}ohm potential in a quantum plasma.
  \emph{Physics of Plasmas} \textbf{2015}, \emph{22}, 102104\relax
\mciteBstWouldAddEndPuncttrue
\mciteSetBstMidEndSepPunct{\mcitedefaultmidpunct}
{\mcitedefaultendpunct}{\mcitedefaultseppunct}\relax
\EndOfBibitem
\bibitem[Senatore \latin{et~al.}(1996)Senatore, Moroni, and
  Ceperley]{SENATORE1996851}
Senatore,~G.; Moroni,~S.; Ceperley,~D. Local field factor and effective
  potentials in liquid metals. \emph{Journal of Non-Crystalline Solids}
  \textbf{1996}, \emph{205-207}, 851--854\relax
\mciteBstWouldAddEndPuncttrue
\mciteSetBstMidEndSepPunct{\mcitedefaultmidpunct}
{\mcitedefaultendpunct}{\mcitedefaultseppunct}\relax
\EndOfBibitem
\bibitem[Porter \latin{et~al.}(2010)Porter, Ashcroft, and
  Chester]{PhysRevB.81.224113}
Porter,~J.~A.; Ashcroft,~N.~W.; Chester,~G.~V. Pair potentials for simple
  metallic systems: Beyond linear response. \emph{Phys. Rev. B} \textbf{2010},
  \emph{81}, 224113\relax
\mciteBstWouldAddEndPuncttrue
\mciteSetBstMidEndSepPunct{\mcitedefaultmidpunct}
{\mcitedefaultendpunct}{\mcitedefaultseppunct}\relax
\EndOfBibitem
\bibitem[Bonev and Ashcroft(2001)Bonev, and Ashcroft]{PhysRevB.64.224112}
Bonev,~S.~A.; Ashcroft,~N.~W. Hydrogen in jellium: First-principles pair
  interactions. \emph{Phys. Rev. B} \textbf{2001}, \emph{64}, 224112\relax
\mciteBstWouldAddEndPuncttrue
\mciteSetBstMidEndSepPunct{\mcitedefaultmidpunct}
{\mcitedefaultendpunct}{\mcitedefaultseppunct}\relax
\EndOfBibitem
\bibitem[Moldabekov \latin{et~al.}(2021)Moldabekov, Dornheim, and
  Cangi]{moldabekov2021thermal}
Moldabekov,~Z.~A.; Dornheim,~T.; Cangi,~A. Thermal Signals from Collective
  Electronic Excitations in Inhomogeneous Warm Dense Matter. 2021\relax
\mciteBstWouldAddEndPuncttrue
\mciteSetBstMidEndSepPunct{\mcitedefaultmidpunct}
{\mcitedefaultendpunct}{\mcitedefaultseppunct}\relax
\EndOfBibitem
\bibitem[Dornheim \latin{et~al.}(2021)Dornheim, Vorberger, and
  Moldabekov]{JPSP21}
Dornheim,~T.; Vorberger,~J.; Moldabekov,~Z.~A. Nonlinear Density Response and
  Higher Order Correlation Functions in Warm Dense Matter. \emph{Journal of the
  Physical Society of Japan} \textbf{2021}, \emph{90}, 104002\relax
\mciteBstWouldAddEndPuncttrue
\mciteSetBstMidEndSepPunct{\mcitedefaultmidpunct}
{\mcitedefaultendpunct}{\mcitedefaultseppunct}\relax
\EndOfBibitem
\bibitem[Glenzer and Redmer(2009)Glenzer, and Redmer]{RevModPhys.81.1625}
Glenzer,~S.~H.; Redmer,~R. X-ray Thomson scattering in high energy density
  plasmas. \emph{Rev. Mod. Phys.} \textbf{2009}, \emph{81}, 1625--1663\relax
\mciteBstWouldAddEndPuncttrue
\mciteSetBstMidEndSepPunct{\mcitedefaultmidpunct}
{\mcitedefaultendpunct}{\mcitedefaultseppunct}\relax
\EndOfBibitem
\bibitem[Bergara \latin{et~al.}(1999)Bergara, Pitarke, and
  Echenique]{PhysRevB.59.10145}
Bergara,~A.; Pitarke,~J.~M.; Echenique,~P.~M. Quadratic electronic response of
  a two-dimensional electron gas. \emph{Phys. Rev. B} \textbf{1999}, \emph{59},
  10145--10151\relax
\mciteBstWouldAddEndPuncttrue
\mciteSetBstMidEndSepPunct{\mcitedefaultmidpunct}
{\mcitedefaultendpunct}{\mcitedefaultseppunct}\relax
\EndOfBibitem
\bibitem[Hu and Zaremba(1988)Hu, and Zaremba]{PhysRevB.37.9268}
Hu,~C.~D.; Zaremba,~E. ${Z}^{3}$ correction to the stopping power of ions in an
  electron gas. \emph{Phys. Rev. B} \textbf{1988}, \emph{37}, 9268--9277\relax
\mciteBstWouldAddEndPuncttrue
\mciteSetBstMidEndSepPunct{\mcitedefaultmidpunct}
{\mcitedefaultendpunct}{\mcitedefaultseppunct}\relax
\EndOfBibitem
\bibitem[Rommel and Kalman(1996)Rommel, and Kalman]{PhysRevE.54.3518}
Rommel,~J.~M.; Kalman,~G. Analytical properties of the quadratic density
  response and quadratic dynamical structure functions: Conservation sum rules
  and frequency moments. \emph{Phys. Rev. E} \textbf{1996}, \emph{54},
  3518--3530\relax
\mciteBstWouldAddEndPuncttrue
\mciteSetBstMidEndSepPunct{\mcitedefaultmidpunct}
{\mcitedefaultendpunct}{\mcitedefaultseppunct}\relax
\EndOfBibitem
\bibitem[Dornheim \latin{et~al.}(2019)Dornheim, Vorberger, Groth, Hoffmann,
  Moldabekov, and Bonitz]{dornheim_ML}
Dornheim,~T.; Vorberger,~J.; Groth,~S.; Hoffmann,~N.; Moldabekov,~Z.;
  Bonitz,~M. The Static Local Field Correction of the Warm Dense Electron Gas:
  An ab Initio Path Integral {M}onte {C}arlo Study and Machine Learning
  Representation. \emph{J. Chem. Phys} \textbf{2019}, \emph{151}, 194104\relax
\mciteBstWouldAddEndPuncttrue
\mciteSetBstMidEndSepPunct{\mcitedefaultmidpunct}
{\mcitedefaultendpunct}{\mcitedefaultseppunct}\relax
\EndOfBibitem
\bibitem[Dornheim \latin{et~al.}(2020)Dornheim, Cangi, Ramakrishna, B\"ohme,
  Tanaka, and Vorberger]{Dornheim_PRL_2020_ESA}
Dornheim,~T.; Cangi,~A.; Ramakrishna,~K.; B\"ohme,~M.; Tanaka,~S.;
  Vorberger,~J. Effective Static Approximation: A Fast and Reliable Tool for
  Warm-Dense Matter Theory. \emph{Phys. Rev. Lett.} \textbf{2020}, \emph{125},
  235001\relax
\mciteBstWouldAddEndPuncttrue
\mciteSetBstMidEndSepPunct{\mcitedefaultmidpunct}
{\mcitedefaultendpunct}{\mcitedefaultseppunct}\relax
\EndOfBibitem
\bibitem[Moroni \latin{et~al.}(1995)Moroni, Ceperley, and
  Senatore]{PhysRevLett.75.689}
Moroni,~S.; Ceperley,~D.~M.; Senatore,~G. Static Response and Local Field
  Factor of the Electron Gas. \emph{Phys. Rev. Lett.} \textbf{1995}, \emph{75},
  689--692\relax
\mciteBstWouldAddEndPuncttrue
\mciteSetBstMidEndSepPunct{\mcitedefaultmidpunct}
{\mcitedefaultendpunct}{\mcitedefaultseppunct}\relax
\EndOfBibitem
\bibitem[Mikhailov(2012)]{Mikhailov_Annalen}
Mikhailov,~S. Second-order response of a uniform three-dimensional electron gas
  to a longitudinal electric field. \emph{Annalen der Physik} \textbf{2012},
  \emph{524}, 182--187\relax
\mciteBstWouldAddEndPuncttrue
\mciteSetBstMidEndSepPunct{\mcitedefaultmidpunct}
{\mcitedefaultendpunct}{\mcitedefaultseppunct}\relax
\EndOfBibitem
\bibitem[Lindhard(1954)]{Lindhard}
Lindhard,~J. On the properties of a gas of charged particles. \emph{Kgl. Danske
  Videnskab. Selskab Mat.-fys. Medd.} \textbf{1954}, \emph{28}\relax
\mciteBstWouldAddEndPuncttrue
\mciteSetBstMidEndSepPunct{\mcitedefaultmidpunct}
{\mcitedefaultendpunct}{\mcitedefaultseppunct}\relax
\EndOfBibitem
\bibitem[Mikhailov(2014)]{Mikhailov_PRL}
Mikhailov,~S.~A. Nonlinear Electromagnetic Response of a Uniform Electron Gas.
  \emph{Phys. Rev. Lett.} \textbf{2014}, \emph{113}, 027405\relax
\mciteBstWouldAddEndPuncttrue
\mciteSetBstMidEndSepPunct{\mcitedefaultmidpunct}
{\mcitedefaultendpunct}{\mcitedefaultseppunct}\relax
\EndOfBibitem
\bibitem[Mortensen \latin{et~al.}(2005)Mortensen, Hansen, and Jacobsen]{GPAW1}
Mortensen,~J.~J.; Hansen,~L.~B.; Jacobsen,~K.~W. Real-space grid implementation
  of the projector augmented wave method. \emph{Physical Review B}
  \textbf{2005}, \emph{71}, 035109\relax
\mciteBstWouldAddEndPuncttrue
\mciteSetBstMidEndSepPunct{\mcitedefaultmidpunct}
{\mcitedefaultendpunct}{\mcitedefaultseppunct}\relax
\EndOfBibitem
\bibitem[Enkovaara \latin{et~al.}(2010)Enkovaara, Rostgaard, Mortensen, Chen,
  Dułak, Ferrighi, Gavnholt, Glinsvad, Haikola, Hansen, Kristoffersen, Kuisma,
  Larsen, Lehtovaara, Ljungberg, Lopez-Acevedo, Moses, Ojanen, Olsen, Petzold,
  Romero, Stausholm-Møller, Strange, Tritsaris, Vanin, Walter, Hammer,
  Häkkinen, Madsen, Nieminen, Nørskov, Puska, Rantala, Schiøtz, Thygesen,
  and Jacobsen]{GPAW2}
Enkovaara,~J.; Rostgaard,~C.; Mortensen,~J.~J.; Chen,~J.; Dułak,~M.;
  Ferrighi,~L.; Gavnholt,~J.; Glinsvad,~C.; Haikola,~V.; Hansen,~H.~A.;
  Kristoffersen,~H.~H.; Kuisma,~M.; Larsen,~A.~H.; Lehtovaara,~L.;
  Ljungberg,~M.; Lopez-Acevedo,~O.; Moses,~P.~G.; Ojanen,~J.; Olsen,~T.;
  Petzold,~V.; Romero,~N.~A.; Stausholm-Møller,~J.; Strange,~M.;
  Tritsaris,~G.~A.; Vanin,~M.; Walter,~M.; Hammer,~B.; Häkkinen,~H.;
  Madsen,~G. K.~H.; Nieminen,~R.~M.; Nørskov,~J.~K.; Puska,~M.;
  Rantala,~T.~T.; Schiøtz,~J.; Thygesen,~K.~S.; Jacobsen,~K.~W. Electronic
  structure calculations with {GPAW}: a real-space implementation of the
  projector augmented-wave method. \emph{Journal of Physics: Condensed Matter}
  \textbf{2010}, \emph{22}, 253202\relax
\mciteBstWouldAddEndPuncttrue
\mciteSetBstMidEndSepPunct{\mcitedefaultmidpunct}
{\mcitedefaultendpunct}{\mcitedefaultseppunct}\relax
\EndOfBibitem
\bibitem[Larsen \latin{et~al.}(2017)Larsen, Mortensen, Blomqvist, Castelli,
  Christensen, Dułak, Friis, Groves, Hammer, Hargus, Hermes, Jennings, Jensen,
  Kermode, Kitchin, Kolsbjerg, Kubal, Kaasbjerg, Lysgaard, Maronsson, Maxson,
  Olsen, Pastewka, Peterson, Rostgaard, Schiøtz, Schütt, Strange, Thygesen,
  Vegge, Vilhelmsen, Walter, Zeng, and Jacobsen]{ase-paper}
Larsen,~A.~H.; Mortensen,~J.~J.; Blomqvist,~J.; Castelli,~I.~E.;
  Christensen,~R.; Dułak,~M.; Friis,~J.; Groves,~M.~N.; Hammer,~B.;
  Hargus,~C.; Hermes,~E.~D.; Jennings,~P.~C.; Jensen,~P.~B.; Kermode,~J.;
  Kitchin,~J.~R.; Kolsbjerg,~E.~L.; Kubal,~J.; Kaasbjerg,~K.; Lysgaard,~S.;
  Maronsson,~J.~B.; Maxson,~T.; Olsen,~T.; Pastewka,~L.; Peterson,~A.;
  Rostgaard,~C.; Schiøtz,~J.; Schütt,~O.; Strange,~M.; Thygesen,~K.~S.;
  Vegge,~T.; Vilhelmsen,~L.; Walter,~M.; Zeng,~Z.; Jacobsen,~K.~W. The atomic
  simulation environment—a {Python} library for working with atoms.
  \emph{Journal of Physics: Condensed Matter} \textbf{2017}, \emph{29},
  273002\relax
\mciteBstWouldAddEndPuncttrue
\mciteSetBstMidEndSepPunct{\mcitedefaultmidpunct}
{\mcitedefaultendpunct}{\mcitedefaultseppunct}\relax
\EndOfBibitem
\bibitem[Bahn and Jacobsen(2002)Bahn, and Jacobsen]{ase-paper2}
Bahn,~S.~R.; Jacobsen,~K.~W. An object-oriented scripting interface to a legacy
  electronic structure code. \emph{Computing in Science \& Engineering}
  \textbf{2002}, \emph{4}, 56--66\relax
\mciteBstWouldAddEndPuncttrue
\mciteSetBstMidEndSepPunct{\mcitedefaultmidpunct}
{\mcitedefaultendpunct}{\mcitedefaultseppunct}\relax
\EndOfBibitem
\bibitem[Monkhorst and Pack(1976)Monkhorst, and Pack]{PhysRevB.13.5188}
Monkhorst,~H.~J.; Pack,~J.~D. Special points for {Brillouin}-zone integrations.
  \emph{Physical Review B} \textbf{1976}, \emph{13}, 5188--5192\relax
\mciteBstWouldAddEndPuncttrue
\mciteSetBstMidEndSepPunct{\mcitedefaultmidpunct}
{\mcitedefaultendpunct}{\mcitedefaultseppunct}\relax
\EndOfBibitem
\bibitem[Perdew and Zunger(1981)Perdew, and Zunger]{Perdew_LDA}
Perdew,~J.~P.; Zunger,~A. Self-interaction correction to density-functional
  approximations for many-electron systems. \emph{Physical Review B}
  \textbf{1981}, \emph{23}, 5048--5079\relax
\mciteBstWouldAddEndPuncttrue
\mciteSetBstMidEndSepPunct{\mcitedefaultmidpunct}
{\mcitedefaultendpunct}{\mcitedefaultseppunct}\relax
\EndOfBibitem
\bibitem[Perdew \latin{et~al.}(1996)Perdew, Burke, and Ernzerhof]{PBE}
Perdew,~J.~P.; Burke,~K.; Ernzerhof,~M. Generalized Gradient Approximation Made
  Simple. \emph{Phys. Rev. Lett.} \textbf{1996}, \emph{77}, 3865--3868\relax
\mciteBstWouldAddEndPuncttrue
\mciteSetBstMidEndSepPunct{\mcitedefaultmidpunct}
{\mcitedefaultendpunct}{\mcitedefaultseppunct}\relax
\EndOfBibitem
\bibitem[Perdew \latin{et~al.}(2008)Perdew, Ruzsinszky, Csonka, Vydrov,
  Scuseria, Constantin, Zhou, and Burke]{PBEsol}
Perdew,~J.~P.; Ruzsinszky,~A.; Csonka,~G.~I.; Vydrov,~O.~A.; Scuseria,~G.~E.;
  Constantin,~L.~A.; Zhou,~X.; Burke,~K. Restoring the Density-Gradient
  Expansion for Exchange in Solids and Surfaces. \emph{Phys. Rev. Lett.}
  \textbf{2008}, \emph{100}, 136406\relax
\mciteBstWouldAddEndPuncttrue
\mciteSetBstMidEndSepPunct{\mcitedefaultmidpunct}
{\mcitedefaultendpunct}{\mcitedefaultseppunct}\relax
\EndOfBibitem
\bibitem[Armiento and Mattsson(2005)Armiento, and Mattsson]{PhysRevB.72.085108}
Armiento,~R.; Mattsson,~A.~E. Functional designed to include surface effects in
  self-consistent density functional theory. \emph{Physical Review B}
  \textbf{2005}, \emph{72}, 085108\relax
\mciteBstWouldAddEndPuncttrue
\mciteSetBstMidEndSepPunct{\mcitedefaultmidpunct}
{\mcitedefaultendpunct}{\mcitedefaultseppunct}\relax
\EndOfBibitem
\bibitem[Sun \latin{et~al.}(2015)Sun, Ruzsinszky, and Perdew]{SCAN}
Sun,~J.; Ruzsinszky,~A.; Perdew,~J.~P. Strongly Constrained and Appropriately
  Normed Semilocal Density Functional. \emph{Phys. Rev. Lett.} \textbf{2015},
  \emph{115}, 036402\relax
\mciteBstWouldAddEndPuncttrue
\mciteSetBstMidEndSepPunct{\mcitedefaultmidpunct}
{\mcitedefaultendpunct}{\mcitedefaultseppunct}\relax
\EndOfBibitem
\bibitem[Moldabekov \latin{et~al.}(2021)Moldabekov, Dornheim, Böhme,
  Vorberger, and Cangi]{moldabekov_jcp21}
Moldabekov,~Z.; Dornheim,~T.; Böhme,~M.; Vorberger,~J.; Cangi,~A. The
  relevance of electronic perturbations in the warm dense electron gas.
  \emph{The Journal of Chemical Physics} \textbf{2021}, \emph{155},
  124116\relax
\mciteBstWouldAddEndPuncttrue
\mciteSetBstMidEndSepPunct{\mcitedefaultmidpunct}
{\mcitedefaultendpunct}{\mcitedefaultseppunct}\relax
\EndOfBibitem
\bibitem[Moldabekov \latin{et~al.}(2021)Moldabekov, Dornheim, Vorberger, and
  Cangi]{moldabekov2021benchmarking}
Moldabekov,~Z.; Dornheim,~T.; Vorberger,~J.; Cangi,~A. Benchmarking
  Exchange-Correlation Functionals in the Spin-Polarized Inhomogeneous Electron
  Gas under Warm Dense Conditions. 2021\relax
\mciteBstWouldAddEndPuncttrue
\mciteSetBstMidEndSepPunct{\mcitedefaultmidpunct}
{\mcitedefaultendpunct}{\mcitedefaultseppunct}\relax
\EndOfBibitem
\bibitem[Zastrau \latin{et~al.}(2014)Zastrau, Sperling, Harmand, Becker,
  Bornath, Bredow, Dziarzhytski, Fennel, Fletcher, F{"o}rster, G{"o}de,
  Gregori, Hilbert, Hochhaus, Holst, Laarmann, Lee, Ma, Mithen, Mitzner,
  Murphy, Nakatsutsumi, Neumayer, Przystawik, Roling, Schulz, Siemer,
  Skruszewicz, Tiggesb{"a}umker, Toleikis, Tschentscher, White, W{"o}stmann,
  Zacharias, D{"o}ppner, Glenzer, and Redmer]{Zastrau}
Zastrau,~U.; Sperling,~P.; Harmand,~M.; Becker,~A.; Bornath,~T.; Bredow,~R.;
  Dziarzhytski,~S.; Fennel,~T.; Fletcher,~L.~B.; F{"o}rster,~E.; G{"o}de,~S.;
  Gregori,~G.; Hilbert,~V.; Hochhaus,~D.; Holst,~B.; Laarmann,~T.; Lee,~H.~J.;
  Ma,~T.; Mithen,~J.~P.; Mitzner,~R.; Murphy,~C.~D.; Nakatsutsumi,~M.;
  Neumayer,~P.; Przystawik,~A.; Roling,~S.; Schulz,~M.; Siemer,~B.;
  Skruszewicz,~S.; Tiggesb{"a}umker,~J.; Toleikis,~S.; Tschentscher,~T.;
  White,~T.; W{"o}stmann,~M.; Zacharias,~H.; D{"o}ppner,~T.; Glenzer,~S.~H.;
  Redmer,~R. Resolving ultrafast heating of dense cryogenic hydrogen.
  \emph{Phys. Rev. Lett} \textbf{2014}, \emph{112}, 105002\relax
\mciteBstWouldAddEndPuncttrue
\mciteSetBstMidEndSepPunct{\mcitedefaultmidpunct}
{\mcitedefaultendpunct}{\mcitedefaultseppunct}\relax
\EndOfBibitem
\bibitem[Dornheim and Vorberger(2021)Dornheim, and
  Vorberger]{Dornheim_JCP_2021}
Dornheim,~T.; Vorberger,~J. Overcoming finite-size effects in electronic
  structure simulations at extreme conditions. \emph{The Journal of Chemical
  Physics} \textbf{2021}, \emph{154}, 144103\relax
\mciteBstWouldAddEndPuncttrue
\mciteSetBstMidEndSepPunct{\mcitedefaultmidpunct}
{\mcitedefaultendpunct}{\mcitedefaultseppunct}\relax
\EndOfBibitem
\bibitem[Dornheim \latin{et~al.}(2021)Dornheim, Moldabekov, and
  Tolias]{PhysRevB.103.165102}
Dornheim,~T.; Moldabekov,~Z.~A.; Tolias,~P. Analytical representation of the
  local field correction of the uniform electron gas within the effective
  static approximation. \emph{Phys. Rev. B} \textbf{2021}, \emph{103},
  165102\relax
\mciteBstWouldAddEndPuncttrue
\mciteSetBstMidEndSepPunct{\mcitedefaultmidpunct}
{\mcitedefaultendpunct}{\mcitedefaultseppunct}\relax
\EndOfBibitem
\end{mcitethebibliography}

\end{document}